\documentclass[a4paper,11pt]{article}

\usepackage{jheppub}
\usepackage{float,caption}
\usepackage{slashed}
\usepackage{graphicx}
\usepackage{amsmath}

\DeclareMathOperator{\arcsinh}{arcsinh}
\DeclareMathOperator{\arctanh}{arctanh}

\usepackage{bm}
\usepackage{physics}
  \usepackage{hyperref}
\usepackage[utf8]{inputenc}
\usepackage{upgreek}

\allowdisplaybreaks
\allowbreak

\usepackage{IEEEtrantools}
\newcommand{\beq}{\begin{equation}}
\newcommand{\eeq}{\end{equation}}

\newcommand{\be}{\beta}

\def\th{\theta}

\def\be{\begin{equation}}
\def\ee{\end{equation}}

\title{ \huge The first law of differential entropy and holographic complexity  }
\author[a]{Debajyoti Sarkar}
\emailAdd{dsarkar@iiti.ac.in}
\affiliation[a]{Indian Institute of Technology Indore,\\ 
Khandwa Road, Simrol 453552 Indore, India.\\}

\author[b]{and Manus Visser}
\emailAdd{manus.visser@unige.ch}
\affiliation[b]{University of Geneva, Department of Theoretical Physics,\\ 
24 quai Ernest-Ansermet, 1211 Geneve 4, Switzerland.\\}

\vspace{3mm}


\abstract{
We construct the CFT  dual of the first law of spherical  causal diamonds in three-dimensional AdS spacetime. A spherically symmetric causal diamond  in AdS$_3$ is    the domain of dependence of a spatial circular disk   with vanishing extrinsic curvature. The bulk first law   relates the variations of the  area of the boundary of the disk, the spatial volume of the disk, the cosmological constant and   the matter Hamiltonian. In this paper we  specialize to first-order metric variations  from pure AdS to the conical defect spacetime, and   the bulk first law is derived following a coordinate based approach. The AdS/CFT dictionary    connects the area of the boundary of the disk   to  the differential entropy in~CFT$_2$, and assuming the `complexity=volume' conjecture,   the volume of the disk  is considered to be dual to the    complexity of a cutoff CFT. On the CFT side we explicitly compute the differential entropy and holographic complexity   for the vacuum state  and the excited state dual to conical AdS using the kinematic space formalism. As a result, the boundary dual of the bulk first law   relates the first-order variations of   differential entropy and   complexity to the variation of the scaling dimension of the excited state, which corresponds to the matter Hamiltonian variation in the bulk. We also include the variation of the central charge with     associated chemical potential  in the boundary first law. Finally, we comment on the boundary dual of the first law for the Wheeler-deWitt patch of AdS,   and we propose an extension of our CFT first law    to higher dimensions. 
}
\begin{document}

\maketitle

\section{Introduction}\label{sec:intro}

Deriving     gravitational thermodynamics of black holes \cite{Bekenstein:1973ur,Bardeen:1973gs,Hawking:1974sw} from a microscopic perspective remains one of the guiding principles in the quest for quantum gravity. 
The microscopic state counting  of black hole entropy \cite{Strominger:1996sh}  is   considered to be one of the major successes of string theory. Later, this microscopic derivation of black hole entropy was reinterpreted~\cite{Strominger:1997eq} in terms of the     Anti-de Sitter (AdS)/ Conformal Field Theory (CFT) correspondence~\cite{Maldacena:1997re}, where the entropy of  three-dimensional AdS black holes \cite{Banados:1992wn,Banados:1992gq} matches with the thermodynamic entropy in two-dimensional CFTs \cite{Cardy:1986ie}. In higher dimensions, it has also been argued  that the mass, entropy and temperature of   AdS black holes  can be identified with the energy, entropy and temperature of  a   thermal state in the dual CFT at high temperature~\cite{Witten:1998zw}.  
 
 Furthermore, the correspondence between gravitational entropy and CFT entropy can be extended to the entanglement entropy of subregions on the conformal boundary of AdS. The   Ryu-Takayagani (RT) formula \cite{Ryu:2006bv,Ryu:2006ef}  states that the entanglement entropy of a subregion~$\mathcal R$ in the CFT is, to leading order in Newton's constant, dual to the Bekenstein-Hawking entropy $A/(4G)$ of the minimal bulk surface which intersects the conformal boundary at $\partial \mathcal R$. The entanglement entropy satisfies a first law-like relation, which is the quantum generalization of the first law of thermodynamics~\cite{Blanco:2013joa,Wong:2013gua}. 
 An important result in AdS/CFT shows that the  linearized gravitational dynamics in the bulk emerge  from the RT formula and the first law of entanglement  on the boundary \cite{Faulkner:2013ica}. 
  
 More recently, the area of non-extremal codimension-two surfaces in three-dimensional AdS spacetime, which are not necessarily homologous to the boundary,  was related to  the notion of \emph{differential entropy} in $2d$ CFTs, via  equation \eqref{eq:defde} \cite{Balasubramanian:2013rqa,Balasubramanian:2013lsa}. The authors discovered that  closed  curves in a spatial slice of AdS$_3$ can be reconstructed by adding and subtracting boundary-anchored geodesics tangent to the   curve. Since RT surfaces in AdS$_3$ are   boundary-anchored geodesics, they were able to express the length (`area') of the closed curve in terms of  an integral over entanglement entropies, associated to the boundary intervals subtended by the geodesics, which they dubbed `differential entropy'.  
This new field theoretic quantity can be qualitatively interpreted as the uncertainty about the global state  for local  observers who make measurements for a finite time in the CFT, because the exterior of a  bulk closed curve is naturally associated to a time strip in the dual CFT. The formalism of differential entropy   was extended to higher dimensions \cite{Myers:2014jia,Czech:2014wka, Balasubramanian:2018uus},   covariant set-ups \cite{Hubeny:2014qwa, Headrick:2014eia}, bulk curves near horizons or singularities \cite{Balasubramanian:2014sra}, bulk points and distances \cite{Czech:2014ppa}, the Poincar\'{e} and Rindler wedges of AdS \cite{Espindola:2017jil, Espindola:2018ozt}, and it was reinterpreted in terms of   kinematic space in~\cite{Czech:2015qta}, reviewed in section \ref{subsec:holeography}.   In the present work, in similarity to   the first law of entanglement,  we derive a   first law of differential entropy for a~holographic~CFT$_2$. 

To construct the first law of differential entropy we find inspiration from the bulk side,  where  gravitational thermodynamics has been extended to  spherical causal diamonds in  maximally symmetric spacetimes (hence including in AdS) \cite{Jacobson:2015hqa,Jacobson:2018ahi}. Spherically symmetric causal diamonds are defined as  the future and past domain of dependence of spherical, codimension-two, spatial regions with vanishing extrinsic curvature (see figure \ref{fig:bulkckv}). These  spherical regions in AdS are relevant for our purposes, since their boundary area is dual to differential entropy in the CFT. 
In general, maximally symmetric causal diamonds  admit only a  conformal Killing vector $\zeta$,  instead of a true Killing vector  like  for   stationary black holes, although in certain limits $\zeta$ becomes a true Killing vector (e.g. for Rindler spacetime and the static patch of de Sitter spacetime). 
Hence, generic maximally symmetric  diamonds are only `conformally stationary', but this seems to be sufficient for them  to behave as thermodynamic equilibrium states under gravitational perturbations. The variational relation to nearby solutions of these diamonds in Einstein gravity    is given by~\cite{Jacobson:2015hqa,Jacobson:2018ahi}  
\begin{equation}\label{eq:bulkfirstlawintro}
\delta H^{\text{mat}}_\zeta  =  \frac{1}{8 \pi G} \left (  - \kappa  \delta A  + \kappa k \delta V  - V_\zeta \delta \Lambda  \right) . 
\end{equation}
This  is the so-called    \emph{first law of causal diamonds}.  Let us briefly explain the notation: $H_{\zeta}^{\text{mat}}$~is the matter Hamiltonian  generating the evolution of classical matter fields along the conformal Killing flow, $A$ is the area of the edge of the diamond, $V$ is the volume of the maximal slice, $k$ is the trace of the extrinsic curvature of the edge  as embedded in the maximal slice, $\kappa$~is the surface gravity associated to   $\zeta$, and $V_\zeta$ is the `thermodynamic volume' of the maximal slice conjugate to the variation of the cosmological constant $\Lambda$.   

 In this paper we restrict to causal diamonds associated to circular disks in AdS$_3$. The main goal   is to   derive a dual   first law in a  CFT$_2$ with a large central charge.  
 For  simplicity,  we consider   excited states in the CFT dual to   a conical defect in AdS, which arises due to the presence of a  classical  point particle \cite{Deser:1983tn,Deser:1983nh}. For this setting we prove the first law of causal diamonds by fixing the global  coordinates of AdS$_3$ and   changing the metric and classical matter fields from pure AdS$_3$ to conical AdS$_3$ (see section~\ref{subsec:firstlawbulkall}). We compute the variation of the bulk area, volume and matter Hamiltonian due to changes in the boundary interval size (associated to geodesics tangent to the boundary of the disk), the conical defect parameter  and  the cosmological constant. By combining these variations in a particular way we find that the term proportional to the variation of the boundary interval size drops out of   the first law  and we reproduce~\eqref{eq:bulkfirstlawintro}.
The main difference compared to \cite{Jacobson:2018ahi} is that we derive the first law using a fixed coordinate approach, rather than Wald's covariant phase space formalism~\cite{Wald:1993nt,Iyer:1994ys}. The latter approach is more general since it holds for     arbitrary variations to nearby solutions, whereas here we   consider only  metric perturbations to conical AdS. The advantage of our approach is, however,   that it provides   a  controlled setting to compare variations in AdS and in  the~CFT.

The   boundary dual to the   first law of causal diamonds can be derived in a similar fashion.  Two important ingredients in our boundary first law are differential entropy~$S_{\text{diff}}$ and a version of \emph{holographic complexity} $\mathcal{C}$   based on the `complexity=volume' proposal and the volume formula for finite bulk regions   in  \cite{Abt:2017pmf,Abt:2018ywl}. Both notions can be formulated in    the kinematic space formalism, and  are defined in terms of   entanglement entropies, cf. \eqref{eq:defde} and \eqref{eq:ourCVrel1}. The holographic dictionary used in this paper reads (with $L$ the AdS~radius) 
\begin{equation} \label{eq:dictionaryfirst}
	S_{\text{diff}} = \frac{A}{4G} \qquad \text{and} \qquad \mathcal C  =  \frac{V}{4GL} .
\end{equation} 
 We compute the variations of $ 	S_{\text{diff}}$ and $\mathcal C$ with respect to the subregion size $\alpha$, the scaling dimension $\Delta$ and the   central charge $c$. The scaling dimension is associated to the (twist) operator acting on the vacuum state, and the central charge is varied in the space of CFTs. We assume   $\Delta \sim c \gg 1$ such that the CFT excited state  is dual to a classical   geometry in the bulk. 
 Varying  
 $c$ corresponds to changing the coupling constants $G$ and $\Lambda$ in the bulk.   The combination of    the  variations of $ 	S_{\text{diff}}$ and $\mathcal C$  yields  the following CFT first law  
\begin{equation} \label{eq:boundaryfirstlaw1}
	\delta E = T\delta S_{\text{diff}} + \nu \delta \mathcal C + \mu \delta c.
\end{equation}
We call this the \emph{first law of differential entropy}.   Here $E$ is a rescaled energy in the CFT,  whose variation is given by 
\begin{equation} 
	\delta E=\kappa f(\alpha) \delta \Delta \qquad \text{with} \qquad f(\alpha)=  \frac{1}{\cos \alpha} -\frac{ \sin \alpha }{\cos \alpha}  ,
\end{equation}
  where $\kappa$ is an arbitrary normalization which  could depend on $\alpha$ and corresponds in the bulk to the surface gravity of the diamond. The   function $f(\alpha)$ is positive in the range $\alpha \in [0, \pi/2]$ and is related to the norm  of the bulk conformal Killing vector $\zeta$ evaluated at the center of the  diamond, via $\sqrt{-\zeta\cdot \zeta} \big |_{O} = \kappa L f(\alpha)$.  Further, the   boundary energy $E$ is dual to the bulk matter Hamiltonian $H_\zeta^{\text{mat}}$   (see section \ref{subsec:bulkmattervar}). The  conjugate quantities in the boundary  first law    depend on the normalization and subregion size  as follows
\begin{equation}
  T =   - \frac{\kappa}{2\pi},   \qquad \nu  = \frac{\kappa}{2\pi} \frac{1}{\cos \alpha}, \qquad \text{and} \qquad   \mu = \frac{1}{c}\big( \!-T S_{\text{diff}}^{\text{vac}} - \nu  \mathcal C_{\text{vac}} \big)= \frac{\kappa}{2\pi} \frac{\pi}{3} f(\alpha) \, .  
\end{equation}  
In the paper we set $\kappa=2\pi.$ Here, $\mu$ is a chemical potential to changing the number of field degrees of freedom  in the CFT, and $\nu$ is the energy cost of changing the   complexity. The formal `temperature' $T$ is negative, in line with the gravitational thermodynamics of  causal diamonds \cite{Jacobson:2018ahi}.  In  section \ref{subsec:limits} we study two limiting cases of the boundary first law:  large and small boundary subregions. The zero   subregion  size limit ($\alpha\to 0$),  cf.~\eqref{eq:delawwdw},   is dual to the   first law for the `Wheeler-deWitt' (WdW) patch of pure AdS, which is a limiting case of the first law of causal diamonds \cite{Jacobson:2018ahi}. In related work,  a similar WdW first law was derived for coherent states in the bulk and on the boundary, without the area variation, and argued to be dual to the `first law of complexity' \cite{Bernamonti:2019zyy,Bernamonti:2020bcf} or to the boundary symplectic form  \cite{Belin:2018fxe,Belin:2018bpg}. Hence, our first law~\eqref{eq:boundaryfirstlaw1} can be viewed as   an extension of the first law of complexity which includes the variation of   differential entropy and central charge, and which depends on the boundary subregion size $\alpha$ (corresponding to finite bulk regions).


The plan of the paper is as follows. In section~\ref{sec:bdryfirstlaw}  we derive the first law of differential entropy and holographic complexity. Section \ref{sec:bulkderiv} is   devoted to   the  first law of causal diamonds applied to the present geometric setting.  We match the boundary first law and bulk first law in section \ref{sec:matching}. We first show how the former follows from the latter, and afterwards we discuss a possible higher dimensional generalization of the boundary first law.  
 We end with      concluding remarks and an outlook in section \ref{sec:conclude}. 
 
 Finally, we have a total of four appendices. Appendix \ref{app:embed}  discusses the embedding formalism and several coordinate systems for pure AdS$_3$ and conical AdS$_3$. In appendix~\ref{app:embed_chord_crofton} we compute the geodesic equation  and the   chord length of finite geodesic arcs in conical AdS. Further, in  appendix \ref{app:ckvs}   we derive the boundary  conformal Killing vector  of a causal diamond on the cylinder, both from the generators of the conformal group on the cylinder and from the boundary limit of the boost Killing vector of AdS-Rindler space. Appendix~\ref{app:couplings} studies the contributions from the variation  of $G$ and $\Lambda$  in the first law of causal diamonds, using the covariant phase space formalism, and   shows that the term proportional to the variation of Newton's constant vanishes in     the first law.

\section{A   first law in CFT$_2$}\label{sec:bdryfirstlaw}

We are interested in studying the physics of bounded     regions in the bulk from a   field theory perspective, in the context of the AdS/CFT correspondence. For simplicity, we restrict to  AdS$_3$/CFT$_2$ and we focus on the example of a circular disk $D$ of coordinate radius~$R$ inside a time slice of AdS. A~gravitational first law \eqref{eq:bulkfirstlawintro} has recently been    derived for metric perturbations of such disks  in pure AdS  which satisfy the linearized  Einstein equation~\cite{Jacobson:2018ahi}. For a gravitational theory with a boundary dual field theory, it is   a natural question whether a CFT version of such a gravitational first law  exists. The CFT first law is   an unexplored subject within the AdS/CFT literature, and in what follows  we will   derive a  non-trivial variational relation between various boundary quantities that is dual to the bulk first law. This establishes a new relational entry in the AdS/CFT dictionary.

There are two terms in the gravitational first law   which allow for an immediate holographic interpretation in AdS$_3$/CFT$_2$: the area variation  of the boundary of the disk  and the volume variation of the disk.  First, there is   a fair amount of literature that investigates the CFT dual of the area of an arbitrary differentiable curve  on a spatial slice of AdS$_3$  \cite{Balasubramanian:2013rqa,Balasubramanian:2013lsa,Czech:2014ppa, Myers:2014jia,Headrick:2014eia,Espindola:2017jil}. This goes by the name of \emph{differential entropy}, which  is a derived quantity from entanglement entropy and    is related to the area of any closed, differentiable bulk curve  in a broad class of gravitational backgrounds. Second, we interpret the volume of the disk as \emph{holographic complexity}, following the `complexity=volume' conjecture \cite{Susskind:2014rva,Stanford:2014jda}. Although the disk is a finite bulk region, instead of an entire bulk time slice,  we can still relate it to complexity because such a region corresponds to a CFT at a UV cutoff  according to the well-known UV/IR correspondence \cite{Susskind:1998dq,Peet:1998wn}.  We   use the  volume formula of \cite{Abt:2017pmf,Abt:2018ywl} to express the volume as a pure CFT quantity,  an integral involving entanglement entropies    analogous to differential entropy. An important technicality is that the volume formula only applies to   quotients of pure AdS, which is sufficient for our purposes, since we take the perturbed geometry in the bulk first law to be AdS$_3$ with a conical singularity. Both   differential entropy and  the volume formula  can be formulated in terms of the formalism of integral geometry and \emph{kinematic space} \cite{Czech:2015qta}, which we review  shortly  below.

 Our   setup is as follows. We work with Einstein gravity in locally   AdS$_3$ spacetimes  in global coordinates, and we mostly specialize to pure AdS and AdS with a conical singularity. The   metric of the latter spacetime is  
 \begin{equation} \label{eq:conicalmetricfirst}
 ds^2 = -\left ( \gamma^2 +  \frac{r^2}{L^2} \right) dt^2 +\left ( \gamma^2 +  \frac{r^2}{L^2} \right)^{-1} dr^2 + r^2 d\phi^2,
 \end{equation}
 where $\phi \in [0, 2\pi)$ and $\gamma \in (0,1)$ parametrizes the departure away from pure AdS ($\gamma =1$). 
 The   dual CFT$_2$ lives on the conformal boundary, which is a  Lorentzian cylinder. We fix the conformal frame on the boundary    such that the CFT time is the same as global AdS time~$t$, i.e. $ds^2_{\text{bndy}} =\lim_{r\to \infty} \frac{L^2}{r^2} ds^2 $, and we   distinguish   the boundary angular coordinate $\theta$ from the bulk angular coordinate $\phi$  by shifting the origin. Thus, the  boundary metric is
  \begin{equation} \label{eq:bndmetricfirst}
  ds^2_{\text{bndy}} = -dt^2 + L^2 d \theta^2.
  \end{equation}
  Note that the radius   $L$ of the cylinder   is equal to the AdS curvature radius in this   frame.

\subsection{Review of kinematic space}\label{subsec:holeography}

\begin{figure}
\begin{center}
\includegraphics[width=\textwidth]{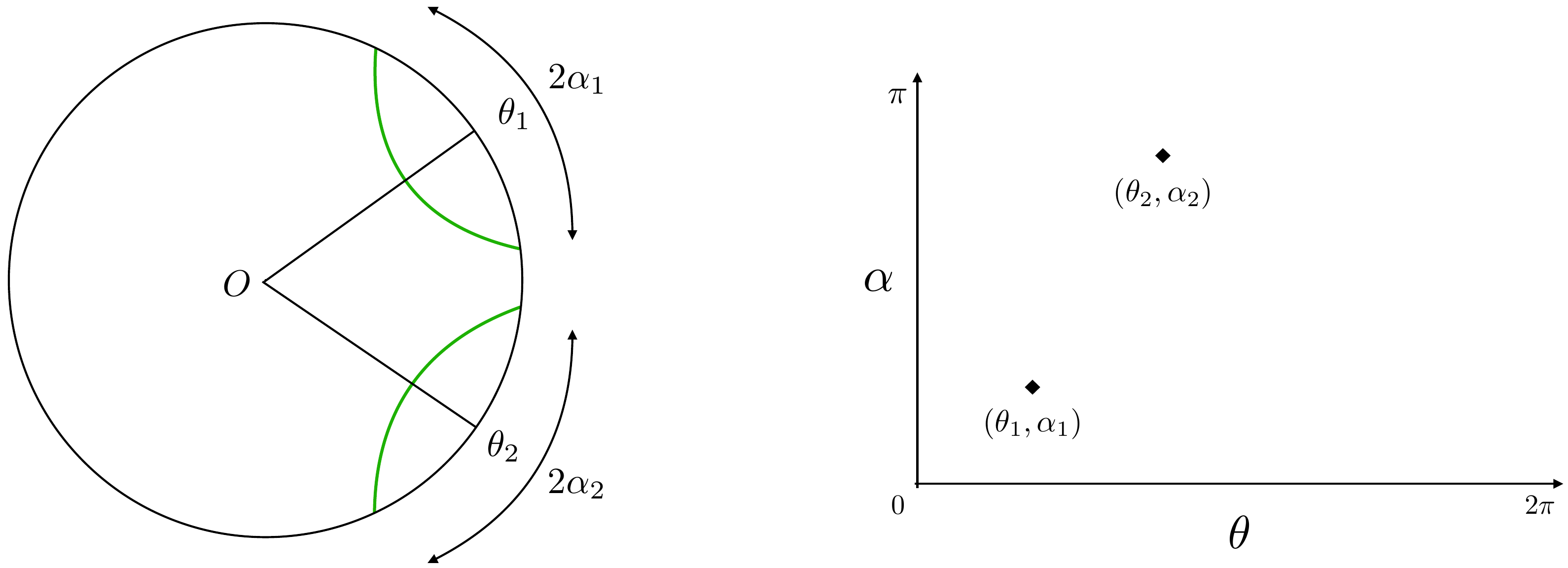}
\end{center}
\caption{\emph{Left diagram:}  a     time slice of pure AdS$_3$ containing two boundary anchored geodesics (in green) parametrized by $(\theta_1,  \alpha_1)$ and $(\theta_2, \alpha_2)$. \emph{Right diagram:} The associated  two points in kinematic space with coordinate system $(\theta,  \alpha)$, where $\theta$  denotes the midpoint  of the boundary subregion  in   angular coordinates  and $ \alpha$  is the angular radius. The orientation  of each  geodesic  is reversed by the transformation  $(\theta,\alpha)\to (\pi+\theta ,   \pi- {\alpha})$, which exchanges the boundary subregion with its complement.\label{fig:physical_kinematic_space}}  
\end{figure}

 Kinematic space   is   the space of oriented spacelike  geodesics in the bulk which are anchored on the boundary.  In this article we restrict to static, locally  AdS$_3$ geometries and to geodesics inside a time slice of those geometries. For vacuum AdS$_3$   kinematic space is the space of RT   surfaces \cite{Ryu:2006bv,Ryu:2006ef} passing through a time slice. An equivalent parametrization of  kinematic space  is via   the boundary subregion that the geodesics subtend: for a given pair $(\theta, \alpha)$ on the boundary, with    $\theta$   the midpoint and $ \alpha$ the opening angle of the subregion, there exists a unique oriented geodesic in the bulk  (see figure~\ref{fig:physical_kinematic_space}). For the conical defect and BTZ geometry this is no longer the case: several  geodesics  (i.e. minimal and non-minimal geodesics) can be associated to a given  boundary interval \cite{Balasubramanian:2014sra}. Kinematic space  for the conical defect spacetime can still be defined though as the space of oriented geodesics,   thereby also taking into account non-minimal geodesics,  but it cannot be defined as the space of boundary intervals (see \cite{Cresswell:2017mbk} though for a   CFT definition in terms of   OPE blocks).  
 

\subsubsection{Differential entropy}\label{sec:diffentropy}

 The main idea behind differential entropy  is to trace out every point of  a closed bulk curve by {unique} boundary anchored geodesics of opening angle $\alpha(\theta)$ which are    {tangent} to the bulk curve at that point. For a central bulk circle these geodesics are just  the RT surfaces corresponding to    subregions of a  fixed, constant angular size $2\alpha$ for every angle $\theta$.   
 
\emph{Differential entropy} is   defined as the $\theta$  integral over the derivative of the entanglement entropy $S(\alpha)$ with respect to $\alpha$ \cite{Balasubramanian:2013rqa,Balasubramanian:2013lsa,Czech:2014ppa} 
\begin{equation}\label{eq:defde}
	S_{\text{diff}}
	=\frac{1}{2}\int_{0}^{2\pi}d\theta\,\frac{dS(\alpha)}{d\alpha}\Bigg{|}_{\alpha=\alpha(\theta)}  \qquad \text{(boundary)}. 
\end{equation}
Using the Ryu-Takayanagi formula $S= \ell /(4G)$, where $\ell$ is the length of the geodesic which is anchored at the boundary coordinates $\theta -\alpha$ and $\theta + \alpha$, the differential entropy can be expressed in terms of bulk quantities. It turns out that   differential entropy is dual to the Bekenstein-Hawking entropy \cite{Bekenstein:1973ur,Hawking:1974sw} of the   closed    curve corresponding to the function $\alpha(\theta)$
\begin{equation}\label{eq:defde3}
	S_{\text{diff}}
	=\frac{1}{8 G}\int_0^{2\pi}  d \theta\frac{d \ell (\alpha) }{d\alpha}\Bigg{|}_{\alpha=\alpha(\theta)}\!=\,\,\frac{A}{4G} \qquad  \text{(bulk)}. 
\end{equation}
Here $A$ is the area (i.e. circumference) of the bulk curve and $G$ is the three-dimensional Newton  constant. For example, for a CFT in the vacuum state on the cylinder, the  entanglement entropy of a subregion  of size $2\alpha$  with its complement is \cite{Calabrese:2004eu,Holzhey:1994we}
\begin{equation}\label{eq:ee}
	S^{\text{vac}}(\alpha)=\frac{c}{3}\log\left(\frac{2L}{\mu}\sin\alpha\right),
\end{equation}
 where $c$ is the central charge of the boundary CFT and $\mu$ is the UV cutoff scale. If we restrict  the closed curve in the bulk to be a central circle, centered at the origin in global coordinates, 
then $\alpha$ is independent of $\theta$ for every point on the bulk curve, and 
the   differential entropy is simply
 \begin{equation} \label{eq:vacdiffentropy1}
 	S_{\text{diff}}^{\text{vac}} (\alpha) = \frac{\pi c}{3} \cot \alpha.
 \end{equation}
 Note that   the    two scales $L$ and $\mu$  drop out in   the differential entropy.  Using the dictionary between the bulk radius and the boundary opening angle in pure AdS, given by $R = L \cot \alpha$ with $L$ the curvature radius of AdS,\footnote{In appendix \ref{app:geoeq} we provide   a derivation of this equation, see  \eqref{eq:radiuscon2} with $\gamma=1$.}   and the dictionary for   the central charge  $c = 3 L/ (2 G)$ \cite{Brown:1986nw},  we find that the differential entropy is indeed equal to the circumference of the circle divided by $4G$
 \begin{equation}
 	S_{\text{diff}}^{\text{vac}}(R) = \frac{2\pi R}{4 G}.
 \end{equation}
 This is only a simple example of the equality between differential entropy and the Bekenstein-Hawking entropy --  which is nonetheless relevant for this paper --  but the equality has been proven more generally  for     any closed,  piecewise differentiable     curve  on a spatial slice of AdS$_3$ in \cite{Balasubramanian:2013lsa}, and for time varying curves on   arbitrary holographic backgrounds which possess  a generalized planar symmetry in~\cite{Headrick:2014eia}.
In what follows, the holographic dictionary between differential entropy and bulk area plays an important role in our boundary interpretation of the bulk first law.

There are several proposals in the literature for the   physical interpretation of differential entropy. In the original paper \cite{Balasubramanian:2013rqa}   it has been conjectured that it signifies the amount of entanglement between quantum gravitational degrees of freedom associated to the interior and exterior of the bulk subregion.\footnote{Note that this is the leading order quantity in a $1/c$ expansion in the dual CFT, i.e. it is of order $\mathcal O(c)$. It is not to be confused with the subleading quantum correction due to the entanglement of   bulk fields.} This was immediately challenged in the follow-up paper \cite{Balasubramanian:2013lsa}, where it was suggested that the Hilbert space of quantum gravity does not factorize between the inside and outside of the bulk curve. This is because the exterior of the bulk curve is holographically dual to a finite time strip on the boundary cylinder and the density matrix on such a region  still acts on the full Hilbert space of the CFT and not on a tensor factor.  
Instead, a separate interpretation was proposed    based on the idea that observers who make measurements for a finite   duration in time  only have access to   local CFT data, and not to the global state. 
 As a result, the authors of \cite{Balasubramanian:2013lsa} suggested that differential entropy measures the uncertainty in reconstructing the global quantum state from the local data  collected by all observers   in the finite time strip.  
However, this interpretation was contested in \cite{Swingle:2014nla} since the global ground state cannot always be reconstructed with arbitrary high accuracy from local data. This is the case if, for example, there is a degeneracy of locally indistinguishable ground states.  Therefore, the maximal global (`reconstruction') entropy of the global ground state does not always admit a precise bulk geometric interpretation.

Another interesting perspective   was provided by \cite{Czech:2017zfq}, which interprets differential entropy as the Wilson loop of the boundary modular Berry connection in kinematic space. This Berry connection relates the eigenspaces of modular Hamiltonians of different subsystems in the CFT. In the bulk the modular Berry connection   ties two infinitesimally separated geodesics  under the action of  the bulk modular Hamiltonian, or more precisely the modular translation operator, which translates   geodesics along a spatial direction of a fixed time slice. As mentioned above, the bulk disk is indeed mapped by a collection of such geodesics, so it is quite natural that the integrand in differential entropy serves as a   connection in   kinematic space. Finally, from a slightly different viewpoint, differential entropy also finds a quantum information theoretic definition in  \cite{Czech:2014tva}. In this language, the length and shape of the bulk curve is expressed   in terms of a  communication protocol called `constrained state merging'. The differential entropy is then the `entanglement cost' of sending the state of the boundary subregion from one party to another, modulo   locality constraints on the operations.

\subsubsection{Volume formula}

Next, we move to the   term in the bulk first law    proportional to  the change in volume of the bulk subregion, which we interpret in terms of the change in  holographic complexity. The volume of a maximal slice anchored at a boundary time slice in the eternal black hole spacetime  has   been   conjectured to be dual to the complexity of the state on the boundary time slice in the CFT \cite{Susskind:2014moa,Brown:2015bva}. This `complexity=volume' conjecture has been extended to the   volume  of the extremal bulk region bounded by a boundary subregion and the RT surface for this subregion,  which is supposed to be dual to the complexity of the mixed state associated to the boundary subregion \cite{Alishahiha:2015rta,Carmi:2016wjl}.\footnote{Note that if the boundary subregion spans the entire   boundary time slice, then the scenario is the same as when our bulk disk has infinite radius on a   time slice of AdS$_3$. The corresponding   causal diamond in the bulk is called the `Wheeler-deWitt' (WdW) patch of pure AdS, which has been a topic of   interest due to the `complexity=action' conjecture \cite{Brown:2015bva,Brown:2015lvg}. 
 In section \ref{subsec:limits}  we also explore  the CFT dual of the first law for   the WdW patch of pure AdS.} These conjectures  have not been proven~yet, due to a lack of understanding of complexity in interacting quantum field theories at strong coupling.   
    
    However, some progress in this subject has been made for the CFT dual of the volume  of bulk subregions in (quotients of) pure AdS$_3$~\cite{Abt:2017pmf,Abt:2018ywl,Huang:2019ajv}.  The authors of \cite{Abt:2017pmf} have proven a  `volume formula'  which  expresses the volume of a bulk subregion as an integral over kinematic space, where the integrand  can be interpreted in terms of pure CFT quantities. Their original motivation was to find a CFT definition of subregion complexity using the kinematic space formalism, but their proposal also holds for bulk regions which are not anchored on the asymptotic boundary (such as a   disk in AdS). In essence, the calculation of the bulk volume amounts to counting the total number of boundary anchored geodesics that pass through the bulk subregion and   integrating   the corresponding \emph{chord lengths}~$\lambda$  in kinematic space, which are the lengths of the intersection of the geodesics with the subregion. In the following we will explain the volume formula and the necessary kinematic space concepts in more detail.  

\begin{figure}
\begin{center}
\includegraphics[width=\textwidth]{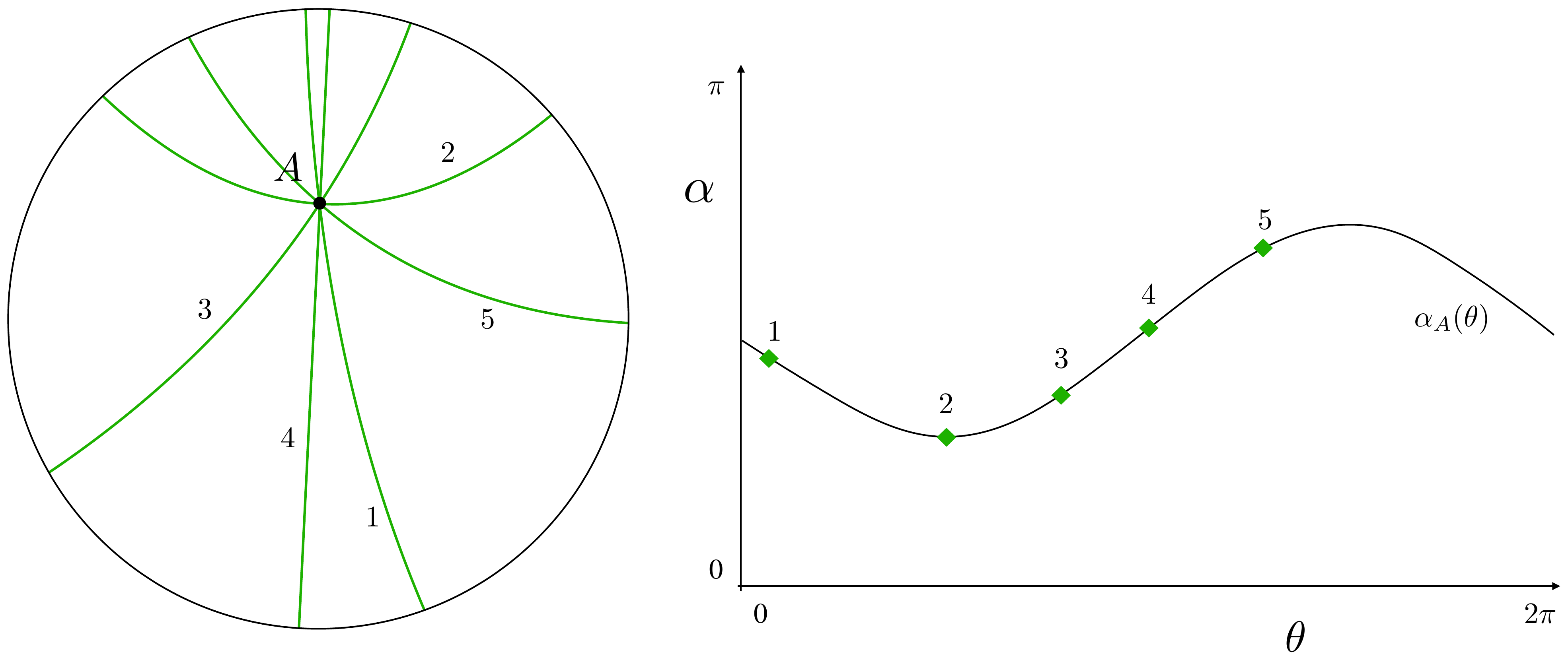}
\end{center}
\caption{\emph{Left diagram}: A   point $A$ on a time slice of AdS, and five geodesics (in green) that intersect $A$.  \emph{Right diagram:} A point curve $\alpha_A(\theta)$    in kinematic space  which represents all geodesics that pass through the bulk point $A$.  The five   diamonds (in green) on the point curve in the right diagram correspond to the five geodesics  in the left~diagram.    \label{fig:pcurve}  } 
 \end{figure}

The computation of the  total number of RT geodesics passing through a given bulk region is facilitated by the so-called \emph{Crofton form}  $\omega$, which is the volume form on kinematic space.
For the kinematic space of the hyperbolic plane, i.e. a time slice of pure AdS, the Crofton form     depends only on   $\alpha$ (and not on~$\theta$) \cite{Czech:2015qta} 
\begin{equation}\label{eq:cf1}
	\omega =\frac{c}{6\sin^2\alpha}d\theta\wedge d\alpha,
\end{equation}
where we haven chosen a normalization that is convenient for AdS$_3$/CFT$_2$. Using~\eqref{eq:ee} we find that  the Crofton form can be written in terms of the second derivative  of the entanglement entropy\footnote{For time slices of non-static geometries there is an additional derivative in the Crofton form   with respect to  the location $\theta$ of the boundary subregion, i.e. $\omega = \frac{1}{2} ( \partial_\theta^2 - \partial_\alpha^2 ) S (\theta, \alpha) d \theta \wedge d\alpha$ \cite{Czech:2015qta, Abt:2018ywl}.}    
\begin{equation}\label{eq:cf2}
	\omega 
	=-\frac{1}{2}\partial_{\alpha}^2S(\alpha)d\theta\wedge d\alpha.
\end{equation} 
Since the Crofton form  characterizes the density of geodesics, the length  of a bulk curve  can now be computed by integrating the Crofton form over the region in kinematic space consisting of all geodesics that intersect the curve.   For instance, the geodesic distance or {chord length} between two   points $A$ and $B$ on a bulk time slice   is given by the so-called \emph{Crofton formula} in integral geometry \cite{santalo_kac_2004,Czech:2015qta} 
\begin{equation} \label{eq:originalcroftonformula}
	\frac{\lambda (A,B)}{4G}=\frac{1}{4}\int_{\Delta_{AB}}\omega(\theta,\alpha).
\end{equation}
We have normalized the Crofton form  appropriately   such that its integral   yields the Bekenstein-Hawking entropy. For convex curves it can be easily verified using Stokes' theorem that the Crofton formula reproduces the differential entropy formula \eqref{eq:defde} if the Crofton form is given by \eqref{eq:cf2}.\footnote{The factor of $1/4$ in \eqref{eq:originalcroftonformula} is cancelled by two factors of 2, one due to the orientation  and one  due to the intersection number of a geodesic with the convex curve \cite{Czech:2015qta}.} The integration region  $\Delta_{AB}$   in kinematic space is given by the region bounded by the two so-called point curves $\alpha_A(\theta)$ and $\alpha_B(\theta)$. The point curve of a given point $A$   is   the collection of all   geodesics that pass through the point $A$, which in kinematic space is a single line $\alpha_A (\theta)$ (see figure \ref{fig:pcurve}). Thus, the region $\Delta_{AB}$    corresponds in AdS to the set of all geodesics (or RT surfaces) that intersect the geodesic arc between $A$ and~$B$ (see figure \ref{fig:twofigappB2} in the appendices).  

\begin{figure}
\begin{center}
\includegraphics[width=.5\textwidth]{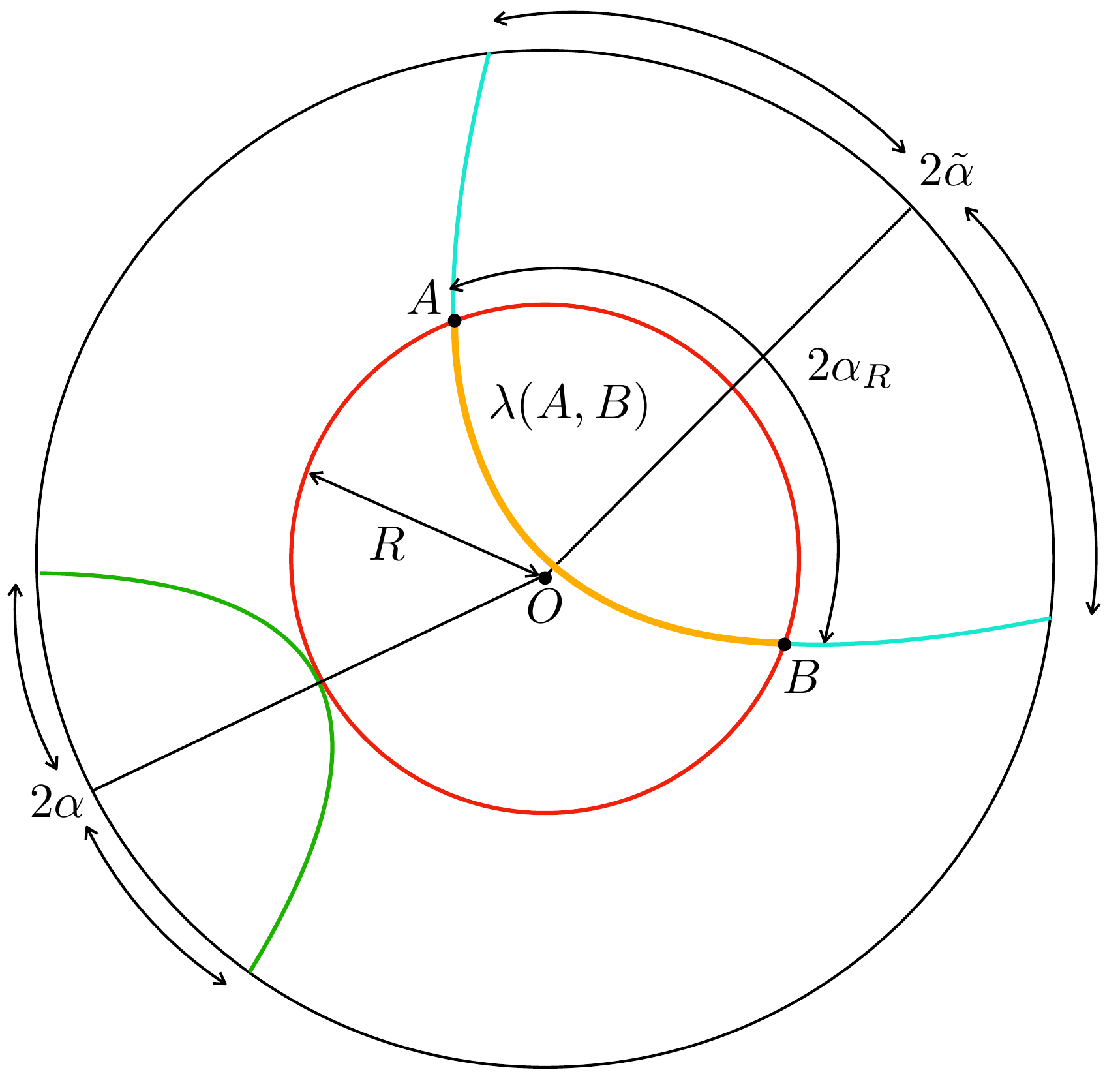}
\end{center}
\caption{A central circle (in red) of coordinate radius $R$ on a time slice of pure  AdS$_3$.  A boundary anchored geodesic (in turquoise) associated to a boundary subregion  of size~$2\tilde \alpha$ intersects the   circle   at two points $A$ and $B$. The  geodesic arc (in orange) between $A$ and~$B$ has a bulk angular size $2 \alpha_R$  and its chord length is denoted by  $\lambda (A,B)$. Geodesics (in green) whose turning point is   tangent to the circle, satisfy $\alpha_R=0$ and $\tilde \alpha=\alpha$. The chord length can be formulated as an integral over all geodesics intersecting the arc between $A$ and $B$, and the proper volume of the disk is an integral over    geodesics between $\alpha\le \tilde \alpha \le\pi-\alpha$. 
\label{fig:genpic}}
\end{figure}

Surprisingly, an explicit derivation of the   chord length for pure AdS from   the   Crofton formula   seems to be absent  in the literature. For completeness, we have provided this computation   in appendix \ref{app:geoembed}, for the more general case of AdS with a conical defect (which reduces to pure AdS by setting~$\gamma=1$). 
As a result, in   vacuum AdS$_3$  the   chord length between two points $A$ and $B$  which lie on a circle of radius $R$  is~\cite{Abt:2017pmf,Abt:2018ywl}  
\begin{equation}\label{eq:adschord}
	\lambda_{\text{vac}}(\alpha_R)=L \, \text{arccosh} \! \left[1+2 (R/L)^2\sin^2(\alpha_R)\right].
\end{equation}
Here, $2\alpha_R$   is the bulk    angle between the points $A$ and $B$ on the circle  (see figure~\ref{fig:genpic}). These two points lie on a geodesic which is anchored on the asymptotic boundary at the angular coordinates $\theta - \tilde \alpha$ and $\theta + \tilde \alpha$. The geodesic equation which relates the bulk and boundary opening angles $\alpha_R$ and $\tilde \alpha$, respectively,  takes the form 
\begin{equation}\label{eq:modRrrel}
	\frac{R}{\sqrt{R^2+L^2}}\cos\alpha_R=\cos\tilde{\alpha}.
\end{equation}
 We give a derivation of this geodesic equation  in appendix \ref{app:geoeq}, i.e. it follows from the second equation in \eqref{eq:finalgeodesic} by setting $r=R$ and $\gamma=1$. One can think of $\alpha_R$ as the difference between  the bulk angular coordinate $\phi$ and the boundary angular coordinate~$\theta$, i.e. $\alpha_R = \phi - \theta$.
For geodesics  which are tangent to the circle we have $ \alpha_R=0$, and  we denote the value of the boundary opening angle by $ \alpha$ for such geodesics (see again figure~\ref{fig:genpic}). Hence the geodesic equation can also be written as 
\begin{equation}
\begin{aligned}\label{eq:modRrrelanother}
\cos \alpha \, \cos\alpha_R &=\cos\tilde{\alpha}, \\    \text{with} \quad  \cos\alpha  &= \frac{R}{\sqrt{R^2+L^2}}.
\end{aligned}
\end{equation}
The chord length   \eqref{eq:adschord}  vanishes, of course, for geodesics tangent to the circle, since  $\alpha_R=0$, and      is by definition only non-vanishing   for  $\tilde \alpha \in (\alpha,\pi-\alpha)$. In the rest of the paper we denote   generic boundary opening angles by  $\tilde \alpha$ and we reserve the notation $\alpha$ for CFT intervals whose RT surfaces are tangent to the boundary of a given bulk codimension-one region (like in differential entropy).

Now we can  write the bulk volume in terms of the chord length and Crofton form.  
Recalling that the Crofton form can be interpreted as the density of geodesics, one would expect that the   volume of a bulk subregion  is proportional to the integral of the chord length times the Crofton form, with  an integration region in kinematic space that corresponds to all geodesics intersecting the bulk subregion. By reinstating appropriate normalizations the \emph{volume formula} in integral geometry   reads  \cite{santalo_kac_2004,Abt:2017pmf,Abt:2018ywl} (see also \cite{Huang:2019ajv} for a similar expression)
\begin{equation} \label{eq:volumeformula1}
	\frac{V}{4G} =\frac{1}{2\pi} \int_K\lambda(\theta',\tilde{\alpha}')\,\omega(\theta,\tilde{\alpha}),
\end{equation}
where $K$ is the set of geodesics that intersect the bulk subregion, and $\lambda$ is the chord length of the intersection of those geodesics and the bulk region (see figure \ref{fig:genpic}).

 To gain some intuition for the volume formula, we now compute it explicitly for a  circular disk   $D$ of coordinate radius $R$  inside a time slice of pure AdS, following \cite{Abt:2017pmf}. 
Using equation~\eqref{eq:cf2} for the Crofton form in pure AdS and the Ryu-Takayangi formula $S = \ell/(4G),$ we can   write the volume formula for a disk  as 
\begin{equation}\label{eq:abtetalvol2}
	 V_{\text{vac}} =-\frac{1}{4\pi}\int_{0}^{2\pi}d\theta\int_{\alpha}^{\pi-\alpha}d\tilde{\alpha}\,\lambda_{\text{vac}} (\theta', \tilde \alpha')\,\partial_{\tilde{\alpha}}^2 \ell_{\text{vac}}(\tilde{\alpha}),
\end{equation}
where the subscript `$\text{vac}$' signifies  that the  chord length $\lambda$ of a geodesic arc and the length~$\ell$ of a boundary anchored geodesic are evaluated in vacuum AdS. 
For computational purposes it is convenient  to replace the integral over $\tilde \alpha$ by an integral over $\alpha_R$, based on the identity
\begin{equation}\label{eq:SDEderchord}
	\partial^2_{\tilde{\alpha}}\ell_{\text{vac}}(\tilde{\alpha})\,d\tilde{\alpha}= \partial^2_{\alpha_R}\lambda_{\text{vac}} (\alpha_R)\,d\alpha_R\,,
\end{equation}
which   can be   checked using the equations \eqref{eq:ee} and \eqref{eq:adschord}. 
The volume formula then becomes 
\begin{equation}
\begin{aligned}\label{eq:volchordgeo}
	V_{\text{vac}} &=-\frac{1}{4\pi}\int_{0}^{2\pi}d\theta\int_{0}^{\pi }d\alpha_R\,\lambda_{\text{vac}}  \,\partial^2_{\alpha_R}\lambda_{\text{vac}} =\frac{1}{2}\int_{0}^{\pi}d\alpha_R\left(\partial_{\alpha_R}\lambda_{\text{vac}}\right)^2  \\
	&= \int_0^\pi d \alpha_R  \frac{2 R^2 \cos^2 (\alpha_R)}{1 + (R/L)^2 \sin^2(\alpha_R)} = 2 \pi L^2 \left (\sqrt{1+ (R/L)^2} -1 \right).
\end{aligned}
\end{equation}
This reproduces the proper volume of a disk in pure AdS.   In the second equality on the first  line we performed the trivial integral over $\theta$ and we partially   integrated, noting that  the boundary term vanishes since    $\lambda_{\text{vac}}=0$  at  $\alpha_R=0,\pi$.

 The volume formula~\eqref{eq:volumeformula1} actually holds for an arbitrary bulk subregion in pure AdS, as  shown in~\cite{Abt:2018ywl}. Similarly, it applies to bulk subregions  in quotient spaces of pure~AdS$_3$, since the kinematic space for these geometries can be obtained from  quotients of the kinematic space for pure~AdS \cite{Abt:2018ywl,Cresswell:2017mbk}. The Crofton form follows from the quotient procedure and   still takes the form \eqref{eq:cf2} for time slices of   static quotient spaces.
 Next we repeat the computation above for a  disk  in the quotient space of AdS$_3$     with a conical defect.

The volume formula for a disk in the conical AdS spacetime  is similar to the expression~\eqref{eq:abtetalvol2} for pure AdS, except that the integration region now depends on the   defect parameter $\gamma$\footnote{We emphasize again that the chord length of both minimal and non-minimal geodesics should be taken into account in the volume formula, i.e. kinematic space for conical AdS is defined here as the space of all spatial, boundary anchored geodesics (and is not restricted to only minimal geodesics) \cite{Abt:2018ywl,Cresswell:2017mbk}.} 
\begin{equation} \label{eq:volumeformulaconfirst}
V_{\text{con}} = -\frac{1}{4\pi}	\int_0^{2\pi} d\theta \int_{\alpha}^{\pi /\gamma- \alpha} d \tilde \alpha \, \lambda_{\text{con}} (\theta',\tilde \alpha') \partial_{\tilde \alpha}^2 \ell_{\text{con}} (\tilde \alpha) .
\end{equation}
The  chord length for the conical defect spacetime \eqref{eq:conicalmetricfirst} is computed in appendix \ref{app:geoembed} in two different ways, using the embedding space formalism and   the kinematic space formalism. The result is
 \begin{equation} \label{eq:conchordlength}
	  \lambda_{\text{con}} (\alpha_R)= L \text{arccosh} \left[1+2  R^2 /(\gamma L)^2 \sin^2(\gamma\alpha_R)\right],
\end{equation}
where $\alpha_R$ is the bulk opening angle between two points on a circle of radius $R$ (see figure~\ref{fig:genpic}). These two points lie on a boundary anchored  geodesic, with boundary opening angle $\tilde \alpha$, for which the geodesic equation reads (see appendix \ref{app:geoeq} for a derivation)
\begin{equation}
\begin{aligned}  \label{eq:geoequationconical}
\cos(\gamma \alpha) \, \cos( \gamma \alpha_R) &=\cos(\gamma \tilde{\alpha}). 
\end{aligned}
\end{equation}
Here, $\alpha$ is the value of the boundary opening angle $\tilde \alpha$ for which the boundary anchored (Ryu-Takayanagi) geodesic is tangent to the circle of radius $R$, i.e. $\alpha_R=0$, satisfying
\begin{equation} \label{eq:radiusconalpha}
 \cos (\gamma \alpha ) = \frac{R}{\sqrt{R^2 + \gamma^2 L^2}} \qquad \text{or} \qquad R = L \gamma \cot (\gamma \alpha).
\end{equation}
We now compute the volume of a disk in conical AdS using the chord length. 
In equation \eqref{eq:volumeformulaconfirst}
we can replace the integral over $\tilde \alpha$ by an integral over $\alpha_R$   using
\begin{equation} \label{eq:congeodesicequation}
		\partial^2_{\tilde{\alpha}}\ell_{\text{con}}(\tilde{\alpha})\,d\tilde{\alpha}= \partial^2_{\alpha_R}\lambda_{\text{con}} (\alpha_R)\,d\alpha_R\,.
\end{equation}
This can be seen geometrically from figure~\ref{fig:genpic}, but it can also be explicitly checked from \eqref{eq:conchordlength} and \eqref{eq:conicalee}.  After inserting this and performing the trivial integral over $\theta$, the volume formula reduces to a single integral over $\alpha_R$
\begin{equation} 
\begin{aligned}
V_{\text{con}} 
&=-\frac{1}{2} \int_0^{\pi/ \gamma} d \alpha_R \, \lambda_{\text{con}} \partial^2_{\alpha_R}  \lambda_{\text{con}} = \frac{1}{2} \int_0^{\pi /\gamma} d \alpha_R (\partial_{\alpha_R}  \lambda_{\text{con}})^2 \\
&= \int_0^{\pi/\gamma} d \alpha_R\, \frac{2R^2   \cos^2 (\gamma \alpha_R)}{ 1 + R^2 /(\gamma L )^2 \sin^2 (\gamma\alpha_R) } =2\pi L^2\left ( \sqrt{\gamma^2 +(R/L)^2 }  - \gamma   \right).  \label{eq:volumeconicalnew}
\end{aligned}
\end{equation}
In the second equality we integrated by parts and removed the boundary term, since $\lambda_{\text{con}}=0$ at $\alpha_R = 0, \pi/\gamma.$ The final expression is indeed the volume of a disk in conical~AdS.

\subsubsection{Boundary dual of finite bulk volume}
\label{sec:boundarydualbulkvolume}

In this section we discuss the CFT$_2$ dual of the   volume of a     subregion inside a time slice of pure AdS$_3$ (or a static quotient space of AdS$_3$). The   volume formula~\eqref{eq:volumeformula1} can be expressed in terms of entanglement entropies, using  the Crofton formula \eqref{eq:originalcroftonformula} and equation~\eqref{eq:cf2}     for the Crofton form of the hyperbolic plane,
\begin{equation}\label{eq:abtetalvol}
	V =\frac{G^2}{2\pi}\int_{K}d\theta\,d\tilde{\alpha}\int_{\Delta_{AB}}d\theta'\,d\tilde{\alpha}'\,\partial_{\tilde{\alpha}}^2\,S(\tilde{\alpha})\,\partial_{\tilde{\alpha}'}^2\,S(\tilde{\alpha}').
\end{equation}
Clearly, the right-hand side is not a CFT quantity, as it still involves   Newton's constant~$G$. However,   we can  define a manifestly field theoretic quantity by dividing the volume by an appropriate dimensionful factor. Following the   `complexity=volume' proposal \cite{Susskind:2014moa,Brown:2015bva} we divide the volume by $GL$, where $L$ is the AdS radius, and we call the resulting dimensionless    quantity \emph{holographic complexity}
\begin{equation}\label{eq:ourCVrel}
	\mathcal C = \frac{V }{4 G L}.
\end{equation}
The factor $1/4$ is conveniently chosen since   the same factor   appears in differential entropy.
The   {dimensionful} proportionality factor $1/(GL)$  has been used in earlier definitions of holographic complexity   for boundary thermofield double  states \cite{Susskind:2014moa} and for boundary subregion density matrices \cite{Alishahiha:2015rta,Carmi:2016wjl}. This is also the same factor that connects boundary Fisher information and bulk volume \cite{Banerjee:2017qti,Sarkar:2017pjp}. As a result, we find  the following definition   of the   boundary dual of the bulk volume 
\begin{equation}\label{eq:ourCVrel1}
	\mathcal C=\frac{3}{16\pi c}\int_{K}d\theta\,d\tilde{\alpha}\int_{\Delta_{AB}}d\theta'\,d\tilde{\alpha}'\,\partial_{\tilde{\alpha}}^2\,S(\tilde{\alpha})\,\partial_{\tilde{\alpha}'}^2\,S(\tilde{\alpha}'),
\end{equation}
where we employed  $c=3L/(2G).$  
This is a pure CFT quantity, since the regions $K$ and $\Delta_{AB}$ in kinematic space can be defined in terms of   boundary coordinates $(\theta, \tilde \alpha)$ (see also section \ref{sec:varycomplexity}). The information theoretic interpretation of this expression is not   clear to us, but at least it provides a precise dictionary between the bulk volume and a boundary integral over entanglement entropies. This dictionary is our second input for the boundary interpretation of the bulk first law (differential entropy being the first input).
 
Regarding the complexity interpretation of the bulk volume, the CFT quantity could be defined as the complexity of a global   state on a time slice in the CFT, where a UV cutoff has been implemented in the theory. The cutoff scale is related to the boundary of a  bulk subregion in a   time slice of AdS  through the UV/IR correspondence \cite{Susskind:1998dq,Peet:1998wn} (the CFT time slice   coincides with the asymptotic boundary of the bulk time slice). If the bulk subregion is a central disk of a fixed radius, then   the CFT lives at a radial cutoff in AdS.  It would be interesting to make this proposal for \emph{cutoff complexity} more precise, see for example the recent paper \cite{Chen:2020nlj}.
 
We should be careful in distinguishing this notion of cutoff complexity from the usual notion  of \emph{subregion complexity} \cite{Alishahiha:2015rta,Carmi:2016wjl}. The latter is argued to be the  complexity of a reduced density   matrix associated to a boundary subregion,   dual to    the volume of the extremal bulk codimension-one region bounded by   the boundary subregion and the associated RT surface (or  dual to the action of the   Wheeler-deWitt patch of the bulk region).   
Cutoff complexity depends on the global state of a time slice of the CFT, or on the reduced state associated to a time strip, whereas subregion complexity is a property of a reduced     density matrix associated to a subregion. 
The two definitions are only equivalent in the limit where the boundary subregion coincides with    the entire   time slice in the CFT. The cutoff complexity and subregion complexity are in that case dual to the  volume of an extremal time slice of AdS,   which can be regularized by choosing an IR cutoff in the bulk which matches the UV cutoff on the boundary. We discuss this limit   further in section \ref{subsec:limits}.
 
Note that the complexity=volume proposal \eqref{eq:ourCVrel} differs from the holographic dictionary  in   \cite{Abt:2017pmf,Abt:2018ywl}. In particular, their definition of \emph{topological complexity} $\mathcal{C}_{\text{top}}$ for a bulk subregion $\Sigma$ of constant intrinsic scalar curvature $\mathcal{R}$ is given by 
\begin{equation} \label{eq:topcomplexity1}
 \mathcal{C}_{\text{top}}=-\frac{1}{2} \int_\Sigma   dV \mathcal{R} =\frac{V_\Sigma }{L^2} .
\end{equation}
In the last step, we inserted  the expression $\mathcal{R}=-2/L^2$, which holds for time slices of AdS$_3$.  
In terms of entanglement entropy the topological complexity reads
\begin{equation} \label{eq:topcomplexity2}
 \mathcal{C}_{\text{top}}=\frac{9}{8\pi c^2}\int_{K}d\theta\,d\tilde{\alpha}\int_{\Delta_{AB}}d\theta'\,d\tilde{\alpha}'\,\partial_{\tilde{\alpha}}^2\,S(\tilde{\alpha})\,\partial_{\tilde{\alpha}'}^2\,S(\tilde{\alpha}').	
\end{equation}
Note that $C_{\text{top}} $ does not depend on the central charge, because the $1/c^2$  cancels against the central charges in the two factors of the   entanglement entropy   $S \sim c$ (at least for the vacuum).   The region $\Sigma$ is often taken to be the codimension-one region bounded by the RT surface and a boundary interval, but the   expression above applies to any bulk subregion (since the volume formula applies to arbitrary regions).
The motivation for considering this    definition  comes from the Gauss-Bonnet theorem, where the integral in \eqref{eq:topcomplexity1}  term appears as being associated with the volume of the region. Through the Gauss-Bonnet theorem the resulting complexity is related to the Euler number of the bulk subregion, cf. equation~\eqref{eq:gaussbonnet},  manifesting the topological nature of the definition. Both proposals for holographic complexity \eqref{eq:ourCVrel} and \eqref{eq:topcomplexity1} are   valid dimensionless CFT quantities, but we work with the first  proposal in the rest of the paper for three reasons: a)  it is more widely used in the literature, b) it is proportional to the central charge  like differential entropy,  and c) it  is well defined in higher dimensions (see the comment below equation \eqref{eq:topocomplfirstlaw}).

\subsection{First law of differential entropy and holographic complexity}\label{subsec:boundarylaw}

In this section  we   derive a CFT$_2$ counterpart of the   first law of causal diamonds in~AdS$_3$. The   CFT first law involves a variation of the differential entropy and holographic complexity, which are respectively dual to (the variation of) the area and volume of a disk in~AdS. We proceed in deriving the boundary   relation by first computing   the   variations of $S_{\text{diff}}$ and $\mathcal C$ separately in the CFT, and then combining them into one variational identity. 
We consider three independent types of variations in the CFT:   1) a variation of the boundary subregion size $\alpha$ in a fixed coordinate system, 2)   a variation of the scaling dimension $\Delta$ of  operators  acting on the vacuum state, and 3)  a variation of the central charge $c$ in the space of CFTs. 
While studying one particular variation, we     keep the other two quantities fixed. Both  differential entropy and holographic complexity change under the   variations of $(\alpha, \Delta, c)$. A particular  combination of   $\delta S_{\text{diff}}$ and $\delta \mathcal C$ yields     a new variational relation, which  for brevity   we call  the `first law of differential entropy'. 
One can think of this as a   dynamical constraint that any  $2d$ holographic field theory must satisfy.  

For the most part, we     assume the central charge to be   large such that the holographic dual has a  classical geometry. Further, we mostly consider those state variations in the CFT which are dual  to the creation of a conical defect in AdS. In other words,  we take the perturbed geometry in the bulk (after a metric perturbation of pure AdS$_3$)  to be the  classical  spacetime  which is locally identical to AdS$_3$ but globally  has   an angular deficit (see section \ref{subsec:conicalreview}). A conical defect spacetime with angular periodicity $2\pi/N$ corresponds to  the quotient space AdS$_3/\mathbb Z_N$,  with $N$ a positive integer. A conical defect in AdS is dual to an excited state  created, via the state/operator correspondence,  by a heavy operator in the CFT \cite{Balasubramanian:1999zv, Balasubramanian:2000rt,Krasnov:2000ia,Lunin:2002bj}. There is substantial  evidence that the  CFT state dual to   AdS$_3/\mathbb Z_N$ is excited by an   operator    which, at large $c$, has   scaling dimension   \cite{Benjamin:2020mfz,Asplund:2014coa,Cresswell:2018mpj}  
\begin{equation}\label{eq:dimtwist}
	 \Delta = \frac{c}{12}\left(1-\frac{1}{N^2}\right) + \mathcal O(c^0).
\end{equation}
The leading-order term is the scaling dimension of a twist operator \cite{Dixon:1986qv,Knizhnik, Calabrese:2009qy}. At the orbifold point of the D1-D5 CFT the dual of AdS$_3/\mathbb Z_N$ has indeed been identified as the state created by acting with a twist field on the vacuum \cite{Lunin:2000yv,Martinec:2001cf,Balasubramanian:2014sra}. Taking subleading  corrections into account in $\Delta$   is equivalent to including perturbative quantum corrections in the bulk stress-energy tensor, due to the presence of quantum fields in a fixed AdS background. In the following we neglect these subleading $1/c$ corrections and we only consider CFT excited states with large $\Delta$ dual to a classical geometry with a point particle.

 In terms of the conical defect parameters $\gamma = 1/N$ and $\epsilon = 1- \gamma$, which we often use,   the scaling dimension to leading order reads 
\begin{equation}\label{eq:Delta-delta-gamma}
	\Delta = \frac{c}{12} \left ( 1 - \gamma^2 \right) = \frac{c}{6}  \epsilon - \frac{c}{12}\epsilon^2 .
\end{equation}
For first-order variations   around the vacuum  state we thus have
\begin{equation}\label{eq:epsdelrel}
	\delta \Delta = \frac{c}{6} \delta \epsilon. 
\end{equation}
Note that  $\delta\Delta=\Delta$ and $\delta \epsilon = \epsilon$, as the vacuum state corresponds to $\Delta=0$ and $\epsilon=0$.  

\subsubsection{Varying differential entropy} 
\label{sec:vardiffentropy}

The change in    differential entropy  under the variation of the  subregion size $\alpha$, scaling dimension $\Delta$ and central charge $c$ is 
\begin{equation}\label{eq:genvarde}
	\delta S_{\text{diff}}=\delta_{\alpha}S_{\text{diff}}\big{|}_{\Delta,c}+\delta_{\Delta}S_{\text{diff}}\big{|}_{\alpha,c}+\delta_{c}S_{\text{diff}}\big{|}_{\alpha,\Delta}.
\end{equation}
Note that if $\Delta$ is kept fixed, we should evaluate the differential entropy   in the ground state of CFT. The variation $\delta_\Delta$ in the second term denotes a state variation  induced by acting with an operator of dimension $\Delta$ on the vacuum. We     start with computing the  second term   using  two different methods, and afterwards we discuss the other terms.

\emph{Method 1)} Since differential entropy \eqref{eq:defde} is expressed in terms of the entanglement entropy~$S(\alpha)$, we can employ the first law of entanglement to calculate the change in differential entropy under a  state variation.    Recall that the   reduced density matrix associated to a subregion  can be expressed as 
\begin{equation}
	\rho  = {e^{- H_\text{mod}} \over Z},
\end{equation}
where $H_\text{mod}$ is the so-called modular Hamiltonian.
Under a variation of the   state of the system, it follows that the  variation of the entanglement entropy is equal to the variation of the expectation value of $H_\text{mod}$ \cite{Blanco:2013joa,Wong:2013gua}   
\begin{equation}
\delta_{\Delta} S = \delta_\Delta \!\left< H_\text{mod} \right>.	
\end{equation}
This is the first law of entanglement. In order to derive a first law for differential entropy, we differentiate  with respect to $\alpha$ on both sides and then take the integral over $\theta$:
\begin{equation}
\label{eq:firstlawdiffentropy1}
	\frac{1}{2}\int^{2\pi}_{0} d\theta \,\frac{d }{d \alpha} \delta_{\Delta} S(\alpha) \Big|_{\alpha(\theta)} =
\frac{1}{2}\int^{2\pi}_{0} d\theta\, \frac{d }{d \alpha}\,  \delta_{\Delta}\! \left< H_\text{mod} \right>\Big|_{\alpha(\theta)}.	
\end{equation}
The left-hand side is, of course, the state variation of the differential entropy. The challenge lies in understanding the right-hand side of this equation. In order to do so, we use the fact that the modular Hamiltonian for the reduced density matrix of the CFT global vacuum state restricted to a ball-shaped region $B$  can be interpreted as a conserved charge \cite{Casini:2011kv}
\begin{equation}\label{eq:HAformula}
	H_\text{mod} = \int_B  d \Sigma^\mu \, \xi^\nu\, T_{\mu \nu}  = L\int_{\th - \alpha}^{\th+ \alpha} d\bar{\th} \, \xi^{t }_{(\theta,\alpha)}(\bar{\th}) \, T_{tt} (\bar{\theta}) \Big |_{t=0}.
\end{equation}
Here $\Sigma^\mu$ and $T_{\mu\nu}$   are, respectively, the volume-form on the subregion  and the CFT stress-energy tensor. The curvature radius $L$ appears due to the square root of the determinant of the metric \eqref{eq:bndmetricfirst} on the boundary cylinder. Further,   $\xi^{t}_{(\theta,\alpha)} $ is the time component of the  conformal Killing vector that  generates a flow which remains inside the past and future domain of dependence (a.k.a. the causal diamond) of the ball-shaped subregion. Since   the two-dimensional CFT lives on a cylinder in our setup,   the spherical subregion is an interval  with angular size $2\alpha$ and center  at $\bar \theta = \theta$, and we assume it lies inside the $t=0$ time slice.  In appendix~\ref{app:ckvs} we provide a derivation of the conformal Killing vector $\xi$ generating the conformal isometry that preserves a causal diamond on the   cylinder, both from a boundary and a bulk perspective.\footnote{We   only need the time component of $\xi$ here evaluated at $t=0$, which was already obtained in~\cite{Blanco:2013joa}, since they find   the modular Hamiltonian of a spatial interval in the CFT$_2$ vacuum on the cylinder.} 

By plugging in the expression above for the modular Hamiltonian into (\ref{eq:firstlawdiffentropy1}) and taking the derivative with respect to $\alpha$ inside the $\bar{\theta}$-integral -- this is allowed since $\xi_{(\theta,\alpha)} =0$ at the boundary values of the integral, i.e. at $t=0$ and $\bar \theta = \th \pm \alpha$, which are the   edges of the diamond -- one finds
\begin{equation}
	\delta_{\Delta} S_{\text{diff}}\big{|}_{\alpha,c} = \frac{1}{2}L\int^{2\pi}_{0} d\theta \int_{\th - \alpha}^{\th + \alpha} d \bar{\theta}\, \frac{d }{d \alpha}  \, \xi^t_{(\theta,\alpha)}(\bar{\th})  \delta_{\Delta}\! \left < T_{tt} (\bar{\theta}) \right>  \Big |_{t=0}.
\end{equation}
We can interchange the order of the integrals by imposing $|\bar{\theta}-\theta|\leq\alpha$. This gives
\begin{equation}
	\delta_{\Delta} S_{\text{diff}}\big{|}_{\alpha,c} = \frac{1}{2}L\int^{2\pi}_{0} d\bar{\theta} \int_{\bar{\th} - \alpha}^{\bar{\th} + \alpha} d \theta\, \frac{d }{d \alpha}  \, \xi^t_{(\theta,\alpha)}(\bar{\th})  \delta_{\Delta} \!\left < T_{tt} (\bar{\theta}) \right>  \Big |_{t=0}.
\end{equation}
Lastly, we insert the   conformal Killing vector whose flow preserves a causal diamond on the boundary cylinder and has unit surface gravity\footnote{Here, we use a slightly different notation compared to equation \eqref{eq:ckvcylinder2}:  we employ a dimensionful   time coordinate $t = \tau L$,  the angular coordinate is denoted by $\bar \theta$ and the center of the causal diamond is    located   at~$\bar \theta =  \theta$, instead of at $\bar \theta =0$.\label{fn:ckv}}   
\begin{equation}
	\xi_{(\theta,\alpha)}(\bar{\th}) =\frac{1}{\sin\alpha}\left[\left(\cos (t/L)\cos( \bar{\theta}-\theta )-\cos\alpha\right)\!L\partial_t-\sin (t/L)\sin(\bar{\theta}-\theta)\partial_{\bar {\theta}}\right].
\end{equation}
This vector field vanishes at the edges of the diamond $\{t=0, \bar \theta = \theta \pm \alpha \} $ and the past and future vertices $\{t/ L = \pm \alpha, \bar \theta = \theta \}$, and acts as a null flow on the  null boundaries  of the diamond $\{   t/L + \bar \theta =  \theta \pm \alpha;  t/L - \bar \theta =- \theta \pm \alpha \}$.
The derivative of the time component of this vector with respect to $\alpha$, evaluated at $t=0$,~is
\begin{equation}
	\frac{d}{d \alpha} \xi^t_{( {{\theta}},\alpha)} \Big|_{t=0} =L\left [ 1- \frac{\cos \alpha}{\sin^2 \alpha} \left ( \cos (\bar \theta -  \theta) - \cos \alpha \right) \right].
\end{equation}
After evaluating the inner ${\theta}$-integral we find
\begin{equation}
\begin{aligned}
\label{eq:differentialenergy}
	\delta_{\Delta} S_{\text{diff}}\big{|}_{\alpha,c} &= L^2 \int_0^{2 \pi} d \bar{\theta} \left ( -\frac{ \cos \alpha}{\sin \alpha} + \frac{\alpha}{\sin^2 \alpha} \right) \delta_{\Delta} \!\left<  T_{tt} (\bar{\theta}) \right> \\
	&=2\pi\left ( -\frac{ \cos \alpha}{\sin \alpha} + \frac{\alpha}{\sin^2 \alpha} \right)\delta \Delta ,
\end{aligned}
\end{equation}
where in the last line we assumed $\alpha \neq \alpha (\bar \theta)$, which corresponds to a spherical region in AdS$_3$, and we inserted the relationship between the stress-energy tensor expectation value and the scaling dimension
\begin{equation}\label{eq:TDeltaref}
	L^2 \int_0^{2\pi}d\bar \theta\,\delta_{\Delta}\langle T_{tt} (\bar{\theta}) \rangle=2\pi\delta\Delta.
\end{equation}
\noindent \emph{Method 2)}  Note that   the   derivation above holds for   arbitrary state perturbations and for arbitrary boundary subregions $\alpha(\bar \theta)$, until the first line of \eqref{eq:differentialenergy}, corresponding to arbitrary bulk regions.  However,   if the perturbed state is dual to a conical defect in~AdS$_3$,  and if $\alpha$ is constant, there is an alternate method of obtaining the state variation of differential entropy. In particular,  we can   compute the state variation by subtracting the differential entropy for the vacuum state from the differential entropy for the excited state dual to the conical defect spacetime. To do so, we need the entanglement entropy for the excited state dual to a conical defect spacetime~\cite{Balasubramanian:2014sra,Asplund:2014coa,Visser2} 
\begin{equation}\label{eq:conicalee}
 S^{\text{con}} (\alpha) = \frac{c}{3} \log \left [ \frac{2L}{\mu \gamma} \sin (\gamma \alpha) \right].	
\end{equation}
\noindent The variation in differential  entropy is now equal to the difference between the differential entropy for the excited state dual to the conical defect space and the differential entropy for the vacuum state
\begin{equation}
\begin{aligned}\label{eq:areavarbysubtract}
\delta_\Delta S_{\text{diff}}\big{|}_{\alpha,c}&=\frac12\int_0^{2\pi} d\theta \frac{d}{d\alpha} S^{\text{con}}( \alpha)-\frac12\int_0^{2\pi} d\theta \frac{d}{d\alpha} S^{\text{vac}}( \alpha)    \\
&=\frac12\int_0^{2\pi} d\theta\, \frac{c}{3}\,\left[(1-\epsilon)\, \frac{\cos\left[(1-\epsilon)\alpha\right]}{\sin\left[(1-\epsilon)\alpha\right]}-\frac{\cos \alpha}{\sin \alpha}\right]  \\
&= 2\pi   \left ( -\frac{\cos \alpha}{\sin \alpha} + \frac{\alpha }{\sin^2 \alpha}\right) \frac{c}{6}\, \epsilon+\mathcal O (\epsilon^2).
\end{aligned}
\end{equation}
\noindent This is the same result as (\ref{eq:differentialenergy}), since $\delta \Delta = (c/6) \epsilon$ up to first order in $\epsilon$, cf. equation~\eqref{eq:epsdelrel}.

After dealing with the state variation, we can now concentrate on the first and third term in~\eqref{eq:genvarde}, i.e. the variation of the  subregion size and central charge, respectively. The change in differential entropy under an $\alpha$ variation, at fixed $\Delta$ and $c$, is to first order
\begin{equation}\label{eq:alphavarde}
	\delta_{\alpha}S_{\text{diff}}\big{|}_{\Delta,c}=\frac{\partial S_{\text{diff}}^{\text{vac}}}{\partial \alpha}\delta\alpha=-\frac{ \pi c}{3} \frac{1}{\sin^2\alpha}\delta\alpha,
\end{equation}
where we   used the expression \eqref{eq:vacdiffentropy1} for the vacuum differential entropy with $\alpha \neq \alpha (\theta)$. Similarly, the differential entropy variation  due to a variation of the central charge, at fixed $\alpha$ and $\Delta$, is  
\begin{equation}\label{eq:SDEcvar}
	\delta_c S_{\text{diff}}\big{|}_{\alpha,\Delta}=\frac{\partial S_{\text{diff}}^{\text{vac}} }{\partial c}\, \delta c = \frac{1}{c}\, S_{\text{diff}}^{\text{vac}} \, \delta c=\frac{\pi}{3} \frac{\cos \alpha}{\sin \alpha}     \delta c.
\end{equation}
Finally, combining the variations \eqref{eq:differentialenergy}, \eqref{eq:alphavarde} and \eqref{eq:SDEcvar}, we find that the total differential entropy variation is to first order given by
\begin{align}\label{eq:totaldevar}
\delta S_{\text{diff}}
&=   \frac{\pi}{3} \frac{\cos \alpha}{\sin \alpha}     \delta c- \frac{\pi c }{3} \frac{1}{\sin^2 \alpha} \delta \alpha +  2 \pi     \left ( -\frac{\cos \alpha}{\sin \alpha} +   \frac{\alpha }{\sin^2 \alpha} \right)\delta \Delta.
\end{align}

\subsubsection{Varying  holographic complexity}
\label{sec:varycomplexity}

Our next task is to study the change in holographic complexity \eqref{eq:ourCVrel}  due to $\alpha$, $\Delta$,~and~$c$ variations 
\begin{equation}\label{eq:genvarvol}
	\delta \mathcal C=\delta_{\alpha}\,\mathcal C\big{|}_{\Delta,c}+\delta_{\Delta}\,\mathcal C\big{|}_{\alpha,c}+\delta_{c}\,\mathcal C\big{|}_{\alpha,\Delta}.
\end{equation}
We start with computing the second term, i.e. the state variation of holographic complexity at fixed $\alpha$ and $c$. Unfortunately, unlike with the differential entropy variation, we  cannot use the first law of entanglement in this case, since the volume formula \eqref{eq:volumeformula1}, on which our holographic complexity is based, applies only to locally AdS$_3$ spacetimes. This means that the holographic complexity formula \eqref{eq:ourCVrel1} holds   for CFT states which are dual to quotients of AdS$_3$, such as the conical defect spacetime and the BTZ black hole, but   does not apply to arbitrary variations of the vacuum state. In particular, it does not hold for small deviations from the vacuum state, i.e. with $\Delta \ll c$,  dual to geometries with small local variations away from pure AdS$_3$ \cite{Abt:2018ywl}. 

Therefore, we use the second method in the previous section to determine the complexity variation. That means we define the complexity state variation as  the difference between the complexity for the excited state dual to conical AdS and the   complexity for the vacuum  
\begin{align}\label{eq:Cdiffs}
	&\delta_{\Delta}\mathcal C\big{|}_{{\alpha},c}:=\left ( \mathcal C_{\text{con}}-\mathcal C_{\text{vac}} \right) \Big |_{\alpha,c} .
\end{align}
The holographic complexity for both states could in principle be computed by inserting the   entanglement entropies for the vacuum and excited state, i.e.   \eqref{eq:ee} and \eqref{eq:conicalee} respectively, into the definition   \eqref{eq:ourCVrel1} (see \cite{Abt:2018ywl} for similar computations).
However,   we find it computationally easier to express  the complexity first in terms of the chord length, instead of the entanglement entropy.  Following the steps in \eqref{eq:volchordgeo} we can write the complexity   for the vacuum state as    
\begin{equation}\label{eq:complexviachord}
	\mathcal C_{\text{vac}}=\frac{c}{12}\int_{0}^{\pi}d\alpha_R\,(\partial_{\alpha_R}\tilde \lambda_{\text{vac}})^2.
\end{equation}
Here $\tilde \lambda_{\text{vac}}=\lambda_{\text{vac}} /L$ is the dimensionless chord length  for the vacuum state, given in \eqref{eq:adschord}. Although the chord length is   a bulk quantity, defined as the length of a geodesic arc, it can also be interpreted as a boundary quantity. By writing the bulk opening angle $\alpha_R$ and the radius $R$ in terms of the boundary   opening angles $\alpha$ and $\tilde{\alpha}$, using \eqref{eq:modRrrel} and \eqref{eq:modRrrelanother}, we find a boundary expression for the vacuum chord length
\begin{equation}\label{eq:chordasbdry}
	\tilde \lambda_{\text{vac}} (  \tilde \alpha) =\text{arccosh} \left[1+2\frac{  \cos^2\alpha-\cos^2\tilde{\alpha} }{\sin^2\alpha}\right].
\end{equation}
Notice that this is only a function of $\tilde \alpha$,  the angle $\alpha$ determines the size of the disk in the bulk and is hence fixed in the computation of the volume formula. 
The vacuum complexity can be derived by plugging this expression for the chord length into~\eqref{eq:complexviachord}  and, after a change of variables, performing the integral over~$\tilde \alpha$ (or,~equivalently, inserting equation~\eqref{eq:adschord} for the chord length and integrating over $\alpha_R$)
\begin{equation}\label{eq:vacCCFT}
	\mathcal C_{\text{vac}}=\frac{\pi c}{3}\left (  \frac{1}{\sin \alpha} -1 \right).
\end{equation}
Note that, unlike entanglement entropy \eqref{eq:ee}, the complexity and differential entropy of the vacuum state are independent of the UV cutoff $\mu$ of the CFT.

Next, we   turn to the complexity of the CFT state dual to conical AdS (see also \cite{Ageev:2018nye}). We compute the holographic complexity through the volume formula, which can be expressed entirely       in terms of   the   chord length, cf. equation \eqref{eq:volumeconicalnew}, in a similar way as for pure AdS,  
\begin{equation}\label{eq:cdiffchord}
 \mathcal C_{\text{con}} =\frac{c}{12} \int_{0}^{\pi/\gamma}d\alpha_R\,(\partial_{\alpha_R}\tilde \lambda_{\text{con}})^2.
\end{equation}
By changing variables we   replace the integral over $\alpha_R$   by an integral over  $\tilde \alpha$,  and  we   write  the dimensionless chord length   $\tilde \lambda_{\text{con}}=\lambda_{\text{con}} /L  $ as a boundary quantity, using \eqref{eq:geoequationconical}~and~\eqref{eq:radiusconalpha},
\begin{equation}
	\tilde  \lambda_{\text{con}} (  \tilde \alpha)=  \text{arccosh} \left[1+2\frac{ \cos^2(\gamma \alpha)-\cos^2(\gamma \tilde{\alpha})}{\sin^2(\gamma \alpha)}\right].
\end{equation}
This means that the complexity formula \eqref{eq:cdiffchord} can be  written purely in terms of CFT quantities.  
Performing the remaining integral over $\tilde \alpha$ (or, equivalently, over $\alpha_R$),  the complexity  for      CFT states dual to conical AdS turns out to be
\begin{equation} 
	\mathcal C_{\text{con}}= \frac{\pi c}{3}\gamma \left (  \frac{1}{\sin(\gamma  \alpha)} -1 \right).
\end{equation}
Expanded to first order in $\epsilon= 1- \gamma$,   the complexity is given by
\begin{equation}
	\mathcal C_{\text{con}}=\frac{\pi c}{3}\left [  \frac{1}{\sin \alpha} -1 + \left (  \frac{\alpha \cos \alpha}{\sin^2 \alpha} - \frac{1}{\sin \alpha}  +1\right) \epsilon + \mathcal O(\epsilon^2)\right].
\end{equation}
Thus, the state variation of the complexity is to first order
\begin{equation}\label{eq:Cvarstate}
	\delta_{\Delta}\mathcal C\big{|}_{\alpha,c}=2 \pi    \left (  \frac{\alpha \cos \alpha}{\sin^2 \alpha} - \frac{1}{\sin \alpha}  +1 \right)   \delta \Delta,
\end{equation}
where we have used \eqref{eq:epsdelrel} to replace $\delta \epsilon$ by $\delta \Delta$.

Finally, we turn to the   $\alpha$ and $c$ variations of holographic complexity. As the state is unchanged in both variations we can directly use equation \eqref{eq:vacCCFT} for the vacuum   complexity and take partial derivatives with respect to $\alpha$ and $c$.  
We end up with the following first-order variations
\begin{equation}\label{eq:Cvaralpha}
	\delta_{\alpha}\mathcal C\big{|}_{\Delta,c}=\frac{\partial \mathcal C_{\text{vac}}}{\partial \alpha}\delta \alpha=- \frac{  \pi c}{3} \frac{\cos \alpha}{\sin^2 \alpha} \delta \alpha,
	\end{equation}
and 
\begin{equation}\label{eq:cvarofC}
	\delta_c\,\mathcal{C}\big{|}_{\alpha, \Delta}= \frac{\partial \mathcal C_{\text{vac}}}{\partial c}\delta c =\frac{1}{c}\,\mathcal{C}_{\text{vac}}\, \delta c =\frac{\pi  }{3}\left (  \frac{1}{\sin \alpha} -1 \right) \delta c .
\end{equation}
Thus, the total   complexity variation is  
 \begin{equation}
\delta \mathcal C =  \frac{\pi}{3}   \left (  \frac{1}{\sin \alpha}  - 1 \right) \delta c - \frac{  \pi c}{3} \frac{\cos \alpha}{\sin^2 \alpha} \delta \alpha +2 \pi    \left (  \frac{\alpha \cos \alpha}{\sin^2 \alpha} - \frac{1}{\sin \alpha}  +1 \right)   \delta \Delta .
\end{equation}

\subsubsection{Combining the variations} \label{subsec:combining1}

At this point, we are ready to write down a   first law in CFT$_2$ involving the change in differential entropy and complexity under the variations of $\alpha$, $\Delta$ and $c$. We proceed by  computing a particular combination of variations of   differential entropy and complexity,
\begin{equation}\label{eq:varconj}
	\delta S_{\text{diff}}- \left(\frac{\partial S_{\text{diff}}}{\partial\mathcal{C}} \right )_c\,\delta\mathcal{C}, 
\end{equation}
since such a combination typically appears in  a  thermodynamic relation. To be specific, if the internal energy depends on the three equilibrium state variables  $S_{\text{diff}}$, $\mathcal C$ and $c$, i.e. $E = E(S_{\text{diff}}, \mathcal C,c)$, then its variation is by definition equal to 
\begin{equation} \label{eq:thermodynamiclaw}
	\delta E =\left ( \frac{\partial E}{\partial S_{\text{diff}}} \right)_{  \mathcal C,c}\!\! \delta S_{\text{diff}} +\left ( \frac{\partial E}{\partial \mathcal C} \right)_{S_{\text{diff}}, c} \!\!\delta \mathcal C + \left ( \frac{\partial E}{\partial c} \right)_{  S_{\text{diff}}, \mathcal C} \!\!\delta c.
\end{equation}
This can be written in a different form using   Maxwell's relation
\begin{equation}
		\delta E =\left ( \frac{\partial E}{\partial S_{\text{diff}}} \right)_{  \mathcal C,c}\!\!  \left ( \delta S_{\text{diff}} - \left ( \frac{\partial S_{\text{diff}}}{\partial \mathcal C} \right)_{E, c} \!\!\delta\mathcal C \right) + \left ( \frac{\partial E}{\partial c} \right)_{  S_{\text{diff}}, \mathcal C} \!\!\delta c.
\end{equation}
We find that the combination of variations \eqref{eq:varconj} indeed  appears in the variation of the internal energy. 
 
Another motivation for studying the particular combination of variations  \eqref{eq:varconj} is that it corresponds to $\delta (A/4G) - k L \delta (V/4GL)$ in the dual AdS spacetime, where $k$ is the trace of the extrinsic curvature of the boundary of the disk as embedded in the disk. The first law of causal diamonds  can be expressed  in terms of this combination of area and volume variations, as we will show in section \ref{sec:matching}. Hence, we expect that the combination of variations \eqref{eq:varconj}  appears in the dual boundary first law.
The partial derivative of $S_{\text{diff}}$ with respect to~$\mathcal C$  should be evaluated in the vacuum,  at fixed central charge,  and can be easily calculated as
\begin{equation}\label{eq:varconjlhs}
 \left(\frac{\partial S_{\text{diff}}^{\text{vac}}}{\partial\mathcal{C}_{\text{vac}}} \right)_c = \frac{\partial S_{\text{diff}}^{\text{vac}}}{\partial\alpha} \left( \frac{\partial \mathcal{C}_{\text{vac}}}{\partial \alpha} \right)^{-1} = -\frac{\pi c}{3}\frac{1}{\sin^2 \alpha} \left (- \frac{\pi c}{3} \frac{\cos \alpha}{\sin^2 \alpha} \right)^{-1} = \frac{1}{\cos \alpha}.
\end{equation} 
In the bulk this is dual to the product of the extrinsic trace $k$ and the AdS radius $L$, i.e.~$k L = 1/\cos \alpha$, cf. equation \eqref{eq:combvar}. 

We can now compute  the combination  of variations of differential entropy and complexity separately for $\alpha, \Delta$ and~$c$ induced variations.
Firstly, for   variations which change the subregion size $\alpha$, but keep  $\Delta$ and $c$ fixed, the combination  vanishes  to first order 
\begin{equation}\label{eq:onlyalphavar}
	\left ( \delta_\alpha S_{\text{diff}}-\frac{1}{\cos\alpha}\,\delta_\alpha \mathcal{C} \right)_{\Delta, c}=0.
\end{equation}
This follows directly from the $\alpha$ variations of differential entropy and complexity,  \eqref{eq:alphavarde} and \eqref{eq:Cvaralpha} respectively. Note  that the choice of relative coefficient \eqref{eq:varconjlhs}   is crucial for
 the cancellations between the two variations.  The $\alpha$ variation is an example of a global conformal transformation in the CFT, in particular a  dilatation, hence it  is dual to a bulk isometry transformation. In \cite{Jacobson:2018ahi} it has been shown that the combination $\delta_\chi A- k \delta_\chi V$ vanishes for variations  of ball-shaped regions in AdS with vanishing extrinsic curvature induced by arbitrary diffeomorphisms $\chi$, in particular it is zero for variations induced by isometries. Hence, from the AdS/CFT duality we   expect that \eqref{eq:varconj} vanishes for variations induced by  \emph{arbitrary} global conformal transformations (since they are dual to isometries in the bulk), not just for $\alpha$ variations, but we do not have a proof of this   in the CFT.  

Secondly, under a small change in   $\Delta$, at fixed $\alpha$ and $c$, the combination of variations becomes  
\begin{equation} 
		\left ( \delta_\Delta   S_{\text{diff}} - \frac{1}{\cos \alpha}	\delta_\Delta \mathcal C \right)_{\alpha, c} =- 2 \pi \left (  \frac{1}{\cos \alpha} -\frac{ \sin \alpha }{\cos \alpha} \right) \delta \Delta.
\end{equation}
This can be derived by inserting   the state variations of differential entropy and complexity. The state variation of   differential entropy \eqref{eq:differentialenergy} is valid for arbitrary perturbations, but  the variation of    complexity \eqref{eq:Cvarstate}  is specific for first-order variations from the vacuum state to the excited state dual to conical AdS. It would be interesting to see whether the equality above holds more generally for arbitrary first order perturbations of the vacuum state.

Furthermore, under  the variation of the central charge, at fixed $\alpha$ and $\Delta$, the combination of variations takes a very simple form
\begin{equation}\label{eq:onlycvar}
	\left ( \delta_c  S_{\text{diff}} - \frac{1}{\cos \alpha}	\delta_c\mathcal C \right)_{\alpha, \Delta} =\left(S_{\text{diff}}^{\text{vac}} - \frac{1}{\cos \alpha}   \mathcal C_{\text{vac}}\right)\,\frac{\delta c}{c} =    \frac{\pi}{3\cos \alpha}(  1- \sin \alpha) \delta c .
\end{equation}
The variation of   differential entropy and complexity under changing $c$  is simply given by    
$\delta_c S_{\text{diff}} = (S_{\text{diff}}^{\text{vac}}/c)  \delta c$  and $ \delta_c\, \mathcal C = (\mathcal C_{\text{vac}}/c)\,  \delta c$, cf.   \eqref{eq:SDEcvar} and \eqref{eq:cvarofC}.
Hence, the first equality above    follows from inserting these central charge variations, and the second equality follows from   the explicit expressions for $S_{\text{diff}}^{\text{vac}}$ an~$\mathcal C_{\text{vac}}$ as a function of $\alpha,$ cf.   \eqref{eq:vacdiffentropy1} and \eqref{eq:vacCCFT}.

Thus, combining the $\alpha, \Delta$ and $c$ variations above yields
\begin{equation}\label{eq:defirstlaw2}
 \delta S_{\text{diff}} - \frac{1}{\cos \alpha} \delta \mathcal C=   \left ( S_{\text{diff}}^{\text{vac}} - \frac{1}{\cos \alpha} \mathcal C_{\text{vac}} \right) \frac{\delta c}{c} - 2 \pi \left (  \frac{1}{\cos \alpha} -\frac{ \sin \alpha }{\cos \alpha} \right) \delta \Delta. 
\end{equation}
This is our proposal for the    CFT$_2$ dual of the first law of causal diamonds, which we call the \emph{first law of differential entropy}. Note that the variational relation above is written purely in terms of CFT quantities:   the subregion size $\alpha$,  scaling dimension~$\Delta$, central charge $c$, holographic complexity $\mathcal{C}$ and differential entropy $S_{\text{diff}}$. In section \ref{sec:matching} we prove explicitly that 
this boundary first law is dual to the bulk first law for disks in AdS$_3$.

 Let us now write the first law of differential entropy as a   `thermodynamic' first law. We emphasize, however, that the CFT   is not in a standard thermal state, hence this might just be a formal analogy. The first law of differential entropy   should probably rather be interpreted  as a   variational relation in a quantum theory,   just like the first law of entanglement. 
Nevertheless, we can associate the $\Delta$ variation   with the change in internal energy of the CFT. 
Formally, we   can define a   positive, $\alpha$ dependent  energy in the~CFT up to first order in $\Delta/c$ (or $\epsilon$), via its variation    
\begin{equation}\label{eq:rescaledE}
\delta E =   2 \pi   f(\alpha) \delta   \Delta \qquad \text{with} \qquad f(\alpha) =   \frac{1}{\cos \alpha} (1-\sin \alpha)   . 
\end{equation}
One can   think of this as a finite $\alpha$ modification of the CFT energy. Note that $f(\alpha)$ is positive in the range $\alpha \in[0,\pi/2]$. As we will see in section~\ref{subsec:bulkmattervar}, the  boundary energy variation   corresponds   to  the variation of the matter Hamiltonian    in the bulk, which generates evolution along the flow of the conformal Killing vector. The function $f(\alpha)$ is proportional  to the norm of the bulk conformal Killing vector $\zeta$ evaluated at the center of the causal diamond,   their precise relation is  given by  equation~\eqref{eq:normckv} 
\begin{equation}
	\sqrt{-\zeta \cdot \zeta} \big |_{O}= \kappa L f(\alpha),
\end{equation} 
with $\kappa$   the   surface gravity.  It would be interesting to understand the $\alpha$ dependent factor in \eqref{eq:rescaledE} from the CFT side as well, and to find a covariant definition of this energy in the dual field theory. 

In terms of the energy defined above the    first law of differential entropy  takes the   form 
\begin{equation} \label{eq:finalcftfirstlaw}
\delta E = T \delta S_{\text{diff}} + \nu \delta \mathcal C + \mu \delta c\,,
\end{equation}
where the conjugate quantities are defined in equation \eqref{eq:thermodynamiclaw} and given by
\begin{equation}
\begin{aligned} \label{eq:conju1}
T&:=\left ( \frac{\partial E}{\partial S_{\text{diff}}} \right)_{  \mathcal C,c}\!\!=-1, \\
 \nu  &:=\left ( \frac{\partial E}{\partial \mathcal C} \right)_{S_{\text{diff}}, c} \!\! = -T \left( \frac{\partial S_{\text{diff}}}{\partial\mathcal{C}} \right)_{E,c}   = \frac{1}{\cos \alpha},  \\
  \mu  &:=\left ( \frac{\partial E}{\partial c} \right)_{  S_{\text{diff}}, \mathcal C} \!\!\!=- T \left (  \frac{ \partial  S_{\text{diff}}}{\partial c} \right)_{E}   - \,  \nu \left (  \frac{ \partial \mathcal C }{\partial c}  \right)_{E}  = \frac{\pi}{ 3}  f(\alpha) \,.
\end{aligned}
\end{equation}
The   function $\mu$ is a chemical potential for  the change in   the number of field degrees of freedom in the CFT, and its dependence on $\alpha$ follows from equation \eqref{eq:onlycvar}. The fact that both $  E$ and $\mu$ are proportional to $f(\alpha)$ is a peculiarity for two-dimensional CFTs and does not generalize to higher dimensions, cf. equation \eqref{eq:verygeneralfirstlaw}. 
The conjugate quantity  $\nu$  can be interpreted as the energy cost of a unit change in complexity, at fixed $S_{\text{diff}}$ and $c$, and we have computed this in equation \eqref{eq:varconjlhs}. 
Furthermore, in \cite{Jacobson:2018ahi,Jacobson:2019gco}  it was argued  that negative temperature is a property of causal diamonds in maximally symmetric spacetimes, and hence it is natural that we find a negative `temperature'  $T$ in the dual field theory as well. 
The formal definition of the temperature on the boundary is in terms of the partial derivative of differential entropy with respect to the $\alpha$ dependent energy,  $1/T := (\partial S_{\text{diff}}/ \partial E) \big |_{\mathcal C,  c}$. However, this  does not seem to be  a standard     temperature as the differential entropy and energy are not   thermodynamic quantities. It is clear from the definition and from \eqref{eq:defirstlaw2} that $T$ is negative since the entropy $S_{\text{diff}}$ decreases as the energy $E$ increases.   The normalization of the temperature is related to the normalization of   the   conformal Killing vector in the bulk. Here, we have normalized the conformal Killing vector such that the surface gravity is~$\kappa=2\pi$, hence $T=-\kappa/2\pi = -1$  (see the discussion below~\eqref{eq:ckv}).

Another way to organize   the  first law of differential entropy is in terms of the variation of the standard  dimensionless  CFT energy,  $\delta \bar E  = 2\pi \delta \Delta$, without the factor $f(\alpha)$.\footnote{We thank Ted Jacobson for this suggestion.} Then the  first law  becomes 
\begin{equation}
	\delta \bar  E = \bar T \delta S_{\text{diff}} + \bar \nu \delta \mathcal C + \bar \mu \delta c,
\end{equation}
and the associated conjugate quantities change into 
\begin{equation} \label{eq:newconjugate}
	\bar T = - \frac{\cos \alpha}{1 - \sin \alpha },   \qquad \bar\nu  = \frac{1}{1- \sin \alpha}, \qquad \text{and} \qquad \bar \mu = \frac{\pi}{3} \, . 
\end{equation}
Notice that in this case the chemical potential is constant and the temperature depends on $\alpha$, whereas in the previous form of the first law the temperature was constant and the chemical potential depended on $\alpha$. 
 The choice of energy $\bar E$ (or temperature $\bar T$) corresponds in the bulk to a different normalization of    the conformal Killing vector. In particular, the normalization of the conformal Killing vector is such that the surface gravity is given by  $\kappa = 2\pi/  f(\alpha)$, hence $\bar T = - \kappa / 2\pi = -1/f(\alpha)$, and the norm of the conformal Killing vector evaluated at the center of the disk is now independent of $\alpha$, i.e. $\sqrt{-\zeta \cdot \zeta} \big|_O= 2\pi L.$
  
For an arbitrary normalization of the bulk conformal Killing vector,  the conjugate quantities in the boundary first law depend  on the surface gravity and on the subregion size as  follows
 \begin{equation}  \label{eq:blablabla}
	  T = - \frac{\kappa}{2\pi},   \qquad \nu  =\frac{\kappa}{2\pi} \frac{1}{\cos \alpha}, \qquad \text{and} \qquad   \mu =\frac{\kappa}{2\pi} \frac{\pi}{3} f(\alpha) \, ,
\end{equation}
and the boundary energy variation is given by 
 \begin{equation} \label{eq:blblabla}
 	 \delta E  =\kappa f(\alpha) \delta \Delta ,
 \end{equation}  
which is dual to the matter Hamiltonian variation in the bulk, cf. equation \eqref{eq:hamvarcft}.  Thus,  different normalizations  of the surface gravity   correspond to different forms of the first law of differential entropy. Notice though that all `thermodynamic' quantities in \eqref{eq:blablabla} and \eqref{eq:blblabla} are proportional to the surface gravity, so $\kappa$ is   just some arbitrary normalization of the first law which  can also    be left out (corresponding  to the choice $\kappa=1$). Finally, we would like to mention that the first law of differential entropy can be formulated in a $\kappa$~independent way, by multiplying \eqref{eq:finalcftfirstlaw} with the inverse temperature $\beta =1/ T$,
\begin{equation} \label{eq:newblalbla}
	\beta \delta E = \delta S_{\text{diff}} + \tilde \nu \delta \mathcal C + \tilde \mu \delta c.
\end{equation}
The  dimensionless product    $\beta \delta E$ and the conjugate quantities $\tilde \nu = \beta \nu$ and $\tilde \mu = \beta \mu$    now do not depend on $\kappa. $ This formulation of the   first law of differential entropy is similar to  the first law of entanglement, $\delta \langle K_\xi \rangle = \delta S$, applied to thermodynamic     systems which admit a (conformal) Killing vector $\xi$ (see section \ref{subsec:comparefirstlaws}). For those systems    the modular Hamiltonian  is equal to product of the inverse temperature and the (conformal) Killing Hamiltonian, $  K_\xi   = \beta H_\xi$. The modular Hamiltonian does not depend on the normalization of $\xi$, but the inverse temperature and Killing Hamiltonian do, just like $\beta$ and $E$  in \eqref{eq:newblalbla} depend on an arbitrary normalization but their product does not. However, 
for the first law of differential entropy we do not have a physical   interpretation of the   product of the inverse temperature and the energy, like   the modular Hamiltonian. 
Therefore, 
   in the rest of the paper we adhere to the formulation  \eqref{eq:finalcftfirstlaw} of the first law, and  we choose   $\kappa=2\pi$ for convenience.
 
 %
 
\subsection{Limiting cases: small  and large boundary intervals}\label{subsec:limits}

Our proposed first law of differential entropy is an explicit function of the subregion size~$\alpha$, that uniquely specifies the size of the disk in the bulk. It is interesting and straightforward to study  two limiting cases of the first law, i.e.
 \begin{equation}
\begin{aligned}
&\text{small interval:} \quad \alpha \to 0, \qquad \text{and} \qquad    \text{large interval:} \quad \alpha \to \pi/2 .
\end{aligned}
\end{equation}
These limits correspond in the bulk, respectively, to a   disk of infinite size   (whose causal diamond is called the   Wheeler-deWitt    patch in AdS) and a   disk of zero size (a point in AdS).  The first limit is relevant for the standard definition of state complexity, which applies to a global  state, whereas the second limit might shed   light on the holographic description of flat space, as it probes    scales much smaller than the AdS radius.

\emph{Small intervals}: In the limit $\alpha \to 0$, our boundary first law \eqref{eq:defirstlaw2} reduces to a much simpler form 
\begin{equation}\label{eq:defirstlawwdw}
	2 \pi (1-\alpha)\, \delta \Delta \overset{(\alpha \approx 0)}= -\delta S_{\text{diff}} +\delta \mathcal C+ \frac{\pi}{3} (1-\alpha)  \delta c,
\end{equation}
where we   kept terms up to first order in $\alpha$. Note that at fixed $\Delta$ and $c$ the differential entropy variation and complexity variation are equal. This is reminiscent of the dual interpretation of the eternal AdS black hole, for which complexity and thermal entropy are   proportional \cite{Susskind:2014rva}. Thus, in the strict $\alpha=0$ limit, we end up~with 
\begin{equation}\label{eq:delawwdw}
	 2 \pi   \delta \Delta\overset{(\alpha = 0)}=-\delta S_{\text{diff}} +\delta \mathcal C+\frac{\pi}{3}  \delta c.
\end{equation}
The left-hand side can be interpreted  as the integral of the variation of the stress-energy expectation value over the entire boundary circle, as in  \eqref{eq:TDeltaref}, and the factor $2\pi$ here appears due to the  size of the   circle. Hence,  in the zero size   limit  the $\alpha$ dependent energy~$E$ becomes   equal to the standard  dimensionless  energy $\bar E = 2\pi \Delta$ of the global CFT state, which is independent of $\alpha$. Further, notice that the chemical potential for the central charge simplifies to $\mu_{\alpha=0}=\frac{\pi}{3}$ for  intervals of zero size, like in equation \eqref{eq:newconjugate}. 

The bulk dual of the boundary first law for the zero interval size is the first law of the  WdW patch, which was studied in  \cite{Jacobson:2018ahi} as a limiting case of the first law of causal diamonds in AdS.  
 Further, in   \cite{Bernamonti:2020bcf} a   first law for WdW was   derived in the context of the `complexity=volume' conjecture, by perturbing AdS with coherent state excitations of a free scalar field. The CFT dual relation was dubbed the `first law of complexity'~\cite{Bernamonti:2019zyy}, which can be independently obtained from Nielsen’s geometric approach to   circuit complexity by perturbing the target state and keeping the reference state fixed.\footnote{The change in complexity was also considered in the context of the   `second law of complexity'  \cite{Brown:2017jil}. In that context the complexity variation was related to the notion of  uncomplexity, i.e. the difference between the   complexity of the system and the maximum value  it can attain, which  in turn  is related to the  entropy of an auxiliary system.} In the context of the `complexity=action' conjecture a    bulk first law was also proposed in \cite{Bernamonti:2019zyy,Bernamonti:2020bcf} for the action of the WdW patch, and extended in \cite{Hashemi:2019aop} to arbitrary perturbations and backgrounds. 
Alternatively, similar variational relations for complexity were studied by interpreting the change in volume as the boundary symplectic form \cite{Belin:2018fxe,Belin:2018bpg}, and by performing  local conformal transformations on the AdS vacuum \cite{Flory:2018akz,Flory:2019kah}. These works are different from our results in the sense that they only study   the complexity variation, and not   the differential entropy (or area) variation in the first law (i.e. they keep the   volume of a spatial slice of the CFT fixed). In this sense, our first law of differential entropy generalizes their results in a non-trivial way both for the WdW patch and, perhaps more importantly,  away from this limit. It is satisfying, however, that the authors of~\cite{Bernamonti:2020bcf} showed that their WdW first law, sans the area variation, precisely coincides with the first laws studied in \cite{Jacobson:2018ahi} and~\cite{Belin:2018bpg}.  

\emph{Large intervals}: For the $\alpha \to \pi/2 $ limit, we expect the differential entropy and complexity variations to vanish, as the area and volume of the disk go to zero. 
 For small variations away from $\beta:=\alpha -\pi/2=0$, the CFT first law is   to first order in $\beta$ equal to
 \begin{equation}\label{eq:delawflat}
\beta	\delta S_{\text{diff}} + \delta \mathcal C \overset{(\beta \approx 0)}=   0.
 \end{equation}
Hence, in the strict  $\beta=0$ limit, we find a trivial version of the first law
\begin{equation}
	\delta\mathcal{C}\overset{(\beta = 0)}=0.
\end{equation}
There is no longer any energy associated to this variation, and there is also no contribution from the variation of the central charge in the bulk point limit.    
Since the first-order variation  gives a trivial result, we also study the     second-order variation of complexity, which is nonvanishing in the   $\beta = 0$ limit
 \begin{equation}
 	\delta^2 \mathcal C  = \frac{1}{2}   \frac{ \partial^2 \mathcal C_{\text{vac}}}{\partial \alpha^2}  \delta  \alpha^2 +   \frac{1}{2}   \frac{ \partial^2 \mathcal C_{\text{con}}}{\partial \epsilon^2} \delta  \epsilon^2 \overset{(\beta = 0)}= \frac{\pi c}{6} \delta  \alpha^2 + \frac{\pi^3 c}{24} \delta \epsilon^2.
 \end{equation}
 Note that the final term can be written in terms of the second-order variation of the scaling dimension \eqref{eq:Delta-delta-gamma} of a twist operator, since $\delta^2 \Delta = - (c/12) \delta \epsilon^2$, which is dual to the negative  gravitational binding energy  of the conical defect spacetime, cf. \eqref{eq:gravbindingenergy}.
 
\section{A   first law in  AdS$_3$}\label{sec:bulkderiv}

In this section we   provide an explicit   derivation of the   first law of causal diamonds in~AdS$_3$, which matches the   CFT$_2$  first law of the previous section. In \cite{Jacobson:2018ahi} the first law was derived using the covariant phase space method \cite{Wald:1993nt,Iyer:1994ys}, whereas in this section we follow a   coordinate based approach. Moreover,  in contrast to \cite{Jacobson:2018ahi} we   specialize to variations from vacuum~AdS to a locally AdS$_3$ spacetime  with a conical singularity, which is a three-dimensional static solution to the Einstein equation with a classical stress-energy tensor for a point particle \cite{osti4101075,Gott:1982qg,Deser:1983nh,Deser:1983tn}. 
Three-dimensional spacetimes with constant negative curvature can be constructed as    quotients  of pure AdS$_3$ by a discrete subgroup of the isometry group \cite{Mess2007LorentzSO,Banados:1992gq}. In particular, a spatial section of the conical defect spacetime follows from quotienting two-dimensional hyperbolic space by the conjugacy class of its isometry group that is generated by an elliptic  element, which depends on a single parameter \cite{Carlip:1994gc,Steif:1995pq,Martinec:1998wm,Cresswell:2018mpj}. The conical defect metric therefore differs from that of pure AdS by a single parameter $\gamma$ or $\epsilon$, which simplifies the derivation (and applicability) of the first law considerably.

\begin{figure}
\begin{center}
\includegraphics[width=65mm,scale=0.5]{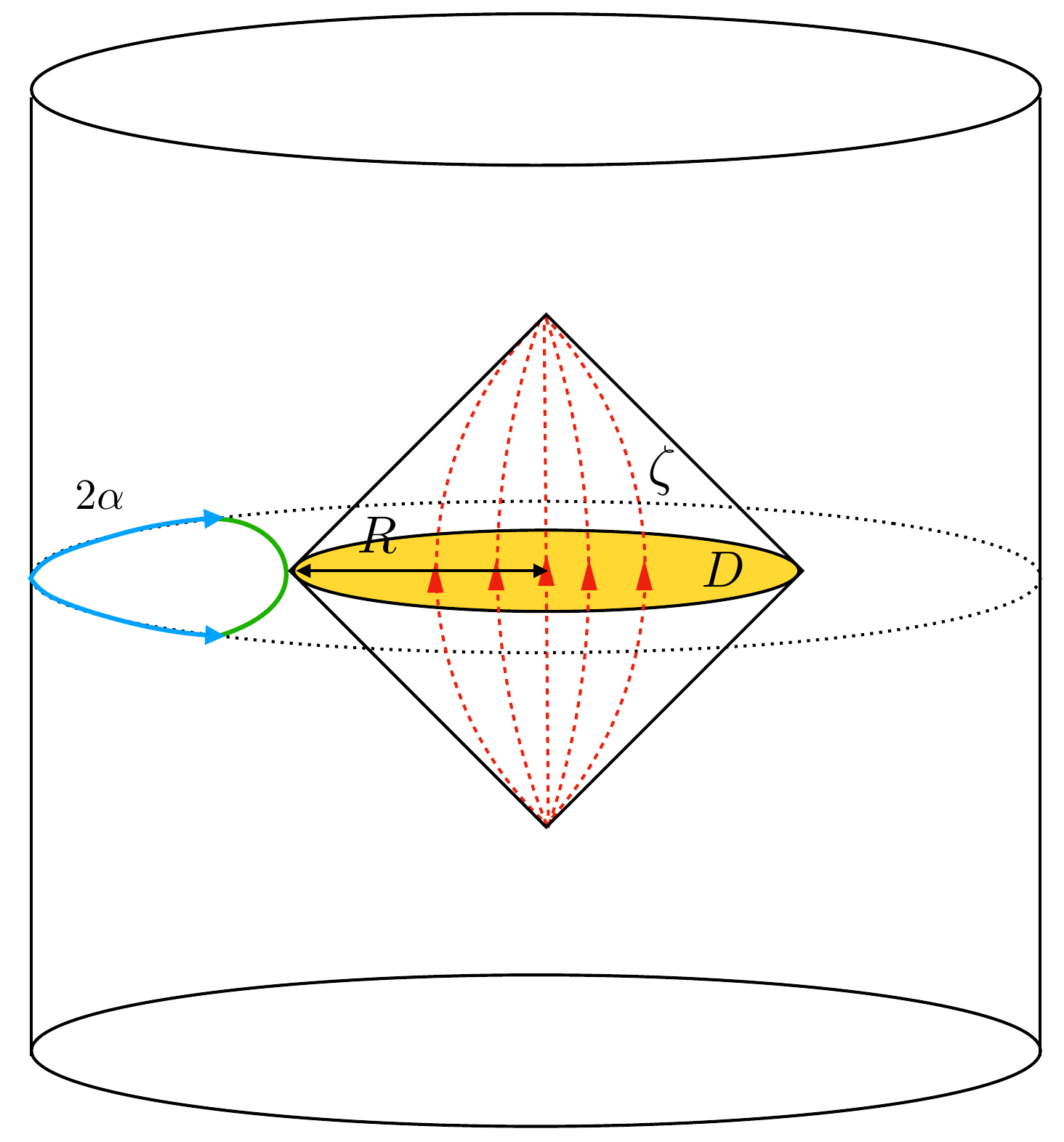}
\end{center}
\caption{A causal diamond in pure AdS$_3$ associated to a disk $D$ (yellow region) of radius~$R$. The bulk conformal Killing vector $\zeta$ generates a   flow (in red)   within the causal diamond,   which sends the boundary of the diamond into itself, and leaves fixed the vertices and the circle $\partial D$. The minimal geodesic  or  Ryu-Takayanagi surface (in green), which is tangent to the circle, subtends an angle $2\alpha$  at the asymptotic boundary (in blue).   Note that a larger boundary subregion corresponds to a smaller bulk disk and vice-versa.\label{fig:bulkckv}}
\end{figure}

To begin with, let us   summarize the first law of causal diamonds in general relativity, derived in   \cite{Jacobson:2015hqa} for flat space and extended in \cite{Jacobson:2018ahi} to (A)dS space.  The causal diamonds  under consideration are defined as the domain of dependence of  (codimension-1)  ball-shaped  regions  of any size in  maximally symmetric spacetimes. There exists a unique conformal isometry that  preserves these causal  diamonds, which is generated by the conformal Killing vector $\zeta$. Figure \ref{fig:bulkckv} illustrates the flow of $\zeta$ for a causal diamond in AdS. The flow is tangent to the null boundary of the diamond, and leaves fixed the future and past vertices and the boundary of the ball. The first law  applies to  arbitrary first-order variations of these maximally symmetric causal  diamonds to nearby solutions -- i.e. the variations   satisfy the linearized Einstein (constraint) equations -- and it reads  
\begin{equation} \label{eq:firstlawcd}
\delta H^{\text{mat}}_\zeta  =  \frac{1}{8 \pi G} \left (  - \kappa  \delta A  + \kappa k \delta V  - V_\zeta \delta \Lambda  \right) . 
\end{equation}
Here $\delta H^{\text{mat}}_\zeta$ is the variation of the bulk matter Hamiltonian which generates evolution along the flow of  $\zeta$, $\kappa$ is the (positive) surface gravity associated to $\zeta$, $A$ is the area of the boundary   of the ball, $k$ is the trace of the extrinsic curvature of the boundary as embedded inside the ball,   $V$ is the proper volume of the ball, $V_\zeta$ is the thermodynamic volume defined as the proper volume locally weighted by the norm of $\zeta$, and $\Lambda$ is the cosmological constant. In the upcoming sections we compute these quantities explicitly for disks, denoted by $D$, in locally AdS$_3$ spacetimes.

The variations we consider in this section are of three different types, in analogy with the three CFT variations of the previous section.\footnote{In the   covariant phase space approach followed in \cite{Jacobson:2018ahi},  the variations were defined as   variations of  the metric and matter fields, while holding the manifold, the vector field $\zeta$ and the disk $D$ of the unperturbed diamond fixed. In the present coordinate based approach we fix the position of the diamond in a global coordinate system of pure AdS and vary the metric within the diamond.} The first variation involves changing the boundary interval size $\alpha$ of a RT surface in a fixed global  coordinate system for pure AdS. The metric of pure AdS remains the same under this variation, but  the radius~$R$ of the disk  decreases as $\alpha$ increases (see figure \ref{fig:bulkckv}). The second variation is a first order   variation of the metric and matter fields from pure AdS to a point mass in AdS, while  again keeping the coordinate system fixed. This variation is parametrized by the conical defect parameter~$\epsilon$, and is related to varying the scaling dimension $\Delta$ in the dual~CFT. Thirdly, we consider variations of   the coupling constants of the gravitational theory, i.e.   the cosmological constant~$\Lambda$ and Newton's constant $G$. Assuming the sign of the cosmological constant does not change, a variation of $\Lambda$ can also be interpreted as a metric perturbation due to changing only the   curvature scale $L$ of the AdS background, since $\Lambda = -1/L^2$ in~AdS$_3$.   
Our motivation for varying $\Lambda$ (or $L$) and $G$ is not so much that they are separate equilibrium state variables or that the quantity  $-\Lambda/8\pi G$ is the pressure of a perfect fluid in the bulk \cite{Kastor:2009wy},  but rather that   they can be combined to   form a single  quantity in the dual CFT, the  central charge   $c =3 L /(2G)$, which can be varied in the space of CFTs~\cite{Kastor:2014dra}. We will show that the variations of $L$ and~$G$    appear in the first law in the particular combination $\delta (L/G)$, which is   dual to varying the central charge  of the holographic~CFT. Another reason for varying $G$ is that it can be   combined with  the boundary area  of the disk to form      the differential  entropy.
 
We start the rest of this section with a   review of locally AdS$_3$ spacetimes with a conical defect. This   enables us to compute the first-order variations of   the area, volume and matter Hamiltonian for a disk in   empty AdS$_3$.  The area and volume variations  match precisely with the variations of the differential entropy and holographic complexity, respectively,  of the previous section. Finally, the area, volume and matter Hamiltonian variations can be combined into a single variational relation, which is   the first law of causal diamonds applied to the present geometric setting.

\subsection{AdS$_3$ with a  conical defect}\label{subsec:conicalreview}

The AdS$_3$ geometry with a conical defect is locally identical to pure AdS$_3$ -- except for one singular point -- but it has a different global structure.  The static geometry is  constructed by cutting out a wedge  of two-dimensional hyperbolic space (the spatial sections of AdS) along two spatial geodesics, and identifying the edges to form a cone. The tip of the cone is a singular point,   called the `conical singularity'.  This is a naked singularity, since it is not shielded by an event horizon. The metric takes the same form as that of pure AdS$_3$ 
\begin{equation} \label{eq:conicalmetric1}
ds^2=-\left(1+\frac{r^{\prime2}}{L^2}\right)dt^{\prime2}+\left(1+\frac{r^{\prime2}}{L^2}\right)^{-1}dr^{\prime2}+r'^2 d \phi^{\prime2}, 
\end{equation}
where $\phi$ is the bulk angular coordinate with    periodicity: $\phi' \sim \phi' + 2\pi \gamma$  where   $0 < \gamma < 1$. The value $\gamma=1$ corresponds to vacuum AdS$_3$, and  for $\gamma=0$ the geometry is identical to that of  a massless BTZ black hole. If $\gamma =1/N$ with $N$  a positive integer,  then the conical defect space corresponds to the quotient space AdS$_3/\mathbb{Z}_N$.\footnote{As shown in \cite{Lunin:2002iz,Alday:2006nd}, non-integer values of $1/\gamma$ are ruled out for supersymmetric conical  metrics,  i.e.~solutions to six-dimensional supergravity theories which reduce to the $3d$ conical defects upon compactification~\cite{Balasubramanian:2000rt,Maldacena:2000dr}. Moreover,    the dual description of a conical defect spacetime in terms of a twist operator, discussed in \cite{Benjamin:2020mfz}, applies  only for integer $1/\gamma$. Therefore, in the context of AdS/CFT (and hence also in the present paper) the conical defect parameter $
1/\gamma$   should  most likely be taken to be integer.\label{ft:integer}}  
 The conical singularity is located at $r'=0$, and  
the   deficit angle of the geometry is   
\begin{equation}
\delta \phi' = 2 \pi (1 - \gamma) \equiv 2 \pi \epsilon,
\end{equation}
with $0<\epsilon <1.$ The angular periodicity  can be modified  by rescaling the coordinates 
\begin{equation} \label{eq:rescalingcoord}
t =t^{\prime} /\gamma,\,\,\,\,\,\,\,\,\,\,r =r^{\prime}\gamma,\,\,\,\,\,\,\,\,\,\phi=\phi^{\prime}/\gamma \, . 
\end{equation}
Under this coordinate transformation the metric becomes
\begin{equation} \label{eq:conicalmetric2}
ds^2=-\left(\gamma^2+\frac{r^2}{L^2}\right)dt^2+\left(\gamma^2+\frac{r^2}{L^2}\right)^{-1}dr^2+r^2 d\phi^2 ,
\end{equation}
where $\phi$ now ranges from $0$ to $2\pi$. We   mainly use this coordinate system in the following sections. In   appendix \ref{app:embed} we derive this metric   from the embedding formalism, and we present some other metrics for conical AdS, one of which is the  line element  in the original paper by Deser-Jackiw  \cite{Deser:1983nh}  (cf. equation \eqref{eq:DJsystem}). 

The conical defect geometry is not a solution to the vacuum Einstein equation, like pure AdS. 
It is a   solution to the   Einstein equation in $2+1$ dimensions with a negative cosmological constant, $G_{\mu \nu } + \Lambda g_{\mu \nu} = 8 \pi G T_{\mu \nu},$ and a  point particle stress-energy tensor
\begin{equation} \label{eq:pointstresstensor1}
\sqrt{g^{(2)}} T_{\mu \nu} u^\mu u^\nu = m \delta^{(2)} (r) .
\end{equation}
Obviously, the location of the classical point mass, $r=0$, coincides with the location of the  conical singularity.  Here we have chosen a time slice,  $t=0$ for convenience, for which  $g^{(2)}$ denotes the determinant of the induced metric  and $u^{\mu}$ is the future directed  unit normal, given by $\sqrt{g^{(2)}} =r \left ( \gamma^2 + (r/L)^2 \right)^{-1/2}$ and  $u^\mu \partial_\mu = \left ( \gamma^2 + (r/L)^2 \right)^{-1/2} \partial_t$. 

Next we determine the relation between the mass $m$ and the conical~defect~parameter~$\gamma$ (see \cite{Klemm:2002ir} for a similar computation). We   derive the stress-energy tensor for the metric~\eqref{eq:conicalmetric2} from the  Hamiltonian constraint  
\begin{equation} \label{eq:hamiltonianconstraint}
\mathcal R - 2 \Lambda  + K^2 - K_{\mu\nu}K^{\mu \nu}= 16 \pi G T_{\mu \nu} u^\mu u^\nu ,
\end{equation}
where $\mathcal{R}$ is the intrinsic curvature scalar of the two-dimensional  spacelike hypersurface.
The extrinsic curvature of a constant $t$ hypersurface vanishes, i.e. $K_{\mu \nu}=0$, because the hypersurface is time symmetric.  For vacuum AdS   the spatial curvature scalar is thus  constant:
$
\mathcal{R}_{\text{vac}}=2 \Lambda.  
$
The intrinsic curvature scalar  of conical AdS only differs from that of vacuum AdS at the conical singularity. Therefore, it is equal to the sum of the vacuum AdS curvature scalar and      a singular part: 
$
\mathcal{R}_{\text{con}}= 2 \Lambda + \mathcal{R}_{\text{sing}}.
$
The singular part of the  curvature scalar can be derived from the
Gauss-Bonnet theorem applied to a disk  of radius~$R$, within the $t=0$ hypersurface and centered at $r=0$,  
\begin{equation} \label{eq:gaussbonnet}
  \chi_{e} = \frac{1}{4\pi} \int_D dV \,  \mathcal{R}  +\frac{1}{2 \pi}  \int_{\partial D} d A \,  k .
\end{equation}
Here $\chi_e$ denotes the Euler number, which is equal to one for a disk. The proper volume element of the disk is $dV = \sqrt{g^{(2)}} d\phi dr $ and the line element of the boundary circle is $dA = r d\phi$. Further,  $k$~denotes the trace of the extrinsic curvature of $\partial D$ as embedded in the disk, which for  the current set-up is    given by  
\begin{equation} \label{eq:extrinsictrace}
k =\frac{1}{\sqrt{g_{rr}}} \partial_r \log \sqrt{g_{\phi \phi}} \Big |_{r=R}= \frac{1}{R} \sqrt{\gamma^2 + \frac{R^2}{L^2}} \,  .
\end{equation}
After performing some simple integrals, we obtain  the relation
\begin{equation}
\int_D dV  \, \mathcal{R}_{\text{sing}} = 4 \pi (1 - \gamma) . 
\end{equation}
Therefore, the singular part of the two-dimensional Ricci scalar  is 
\begin{equation}
\mathcal{R}_{\text{sing}} = \frac{4\pi}{\sqrt{g^{(2)}}} \left (1 - \gamma   \right) \delta^{(2)} (  r) .
\end{equation}
Inserting this back into the Hamiltonian constraint \eqref{eq:hamiltonianconstraint} yields the following result for  the stress-energy tensor of conical AdS$_3$  
\begin{equation} \label{eq:pointstresstensor2}
\sqrt{g^{(2)}} T_{\mu \nu} u^{\mu } u^\nu =  \frac{1- \gamma}{4 G}  \delta^{(2)} ( {r})    . 
\end{equation}
This is the only non-zero component of the stress-energy tensor. 
Comparing with \eqref{eq:pointstresstensor1}, we find that the mass of the point particle is related to the defect parameter through 
\begin{equation} \label{eq:djhmass}
m = \frac{1- \gamma}{4 G} = \frac{\epsilon}{4 G}\,  .
\end{equation}
This is the expression for the mass of a point particle   in flat space~\cite{Deser:1983tn}  and  in AdS~\cite{Deser:1983nh}. We therefore call it the Deser-Jackiw-'t Hooft (DJH) mass. It is equal to the `proper' mass of the conical defect spacetime, defined as $m=\int dV \rho(r) $     where $dV$ is the `proper' volume element  and $\rho$ is the energy density,  but it differs from the total mass of the spacetime.  

We can compute the total mass of the conical defect spacetime through the ADM formula
\begin{equation} \label{eq:admmass}
M = - \frac{1}{8 \pi G}\lim_{\partial D \to \infty} \int_{\partial D} dA  \, N(k - k_{\gamma=1}) \, . 
\end{equation}
Note that we have subtracted the    value of the ADM mass in the   pure AdS background (with $\gamma=1$) in order to cancel the divergences at spatial infinity. This is sufficient for the present context,   because   the first law only features the energy difference  between conical AdS and pure AdS.\footnote{The absolute value of the   mass can be obtained through holographic renormalization \cite{Balasubramanian:1999re}. The result for global pure AdS$_3$ is $M_{\text{vac}} = - 1/8G$ and for AdS$_3$ with a conical defect  $M_{\text{con}}= - \gamma^2 / 8 G$ \cite{Balasubramanian:2000rt}.} The extrinsic trace  $k$ is given by \eqref{eq:extrinsictrace} and the lapse function $N$ is
\begin{equation}
N = \sqrt{|g_{tt}|} \Big |_{r=R}= \sqrt{\gamma^2 + \frac{R^2}{L^2}} \, . 
\end{equation}
Inserting this into the ADM formula and taking the limit $R \to \infty$ yields
\begin{equation}  \label{eq:admmasscon}
M = \frac{1-\gamma^2}{8 G} = \frac{\epsilon}{4 G} - \frac{\epsilon^2}{8G} \,  . 
\end{equation}
Note that for small deficit angles, i.e.   $\epsilon \ll1$, the ADM mass agrees with the DJH mass. This implies that we can use both masses interchangeably in the first law of causal diamonds, since  the first law only holds for first order perturbations away from pure AdS. For large deficit angles, however, the ADM mass contains an extra $O(\epsilon^2)$ term compared to the DJH mass. This term could   be related to  the binding energy of the point mass to the gravitational background. 
The   (negative)  gravitational  binding energy is defined as  the difference between the  total mass and the proper mass\footnote{See for instance p.~126 in~\cite{Wald:1984rg}, where however the gravitational binding energy is defined to be positive, i.e. $E_B = m-M$.}
\begin{equation} \label{eq:gravbindingenergy}
E_B =  M -m =  - \frac{\epsilon^2}{8 G} \,  .
\end{equation}
The negative binding energy might be interpreted as a reduction in the gravitational energy due to the fact that the conical defect cuts out part of the spacetime \cite{Balasubramanian:2001nb}.
Because we are interested only in small point masses, we   neglect this   binding energy  in the remainder of the~paper. 

Finally, we   would like to mention a simple relation between the DJH and ADM mass
\begin{equation} \label{eq:djhadmmass}
m = \frac{1}{4G} \left ( 1 - \sqrt{1- 8 G M} \right) , 
\end{equation}
and  we remark that   the    ADM mass (not the DJH mass) is related to the scaling dimension in the  $2d$ CFT, via $\Delta = ML.$ This can be seen, for example, by comparing the ADM mass~\eqref{eq:admmasscon} and the conformal dimension of a twist operator \eqref{eq:Delta-delta-gamma}, where one should restrict to integer values  for the conical defect parameter, i.e. $1/\gamma = N \in \mathbb N$ (see footnote~\ref{ft:integer}).

\subsection{First law of causal diamonds in AdS$_3$}
\label{subsec:firstlawbulkall}
\subsubsection{Area variation}\label{subsec:areavarbulk}

In this section we compute the   variation of the boundary area of a circular disk, centered at the point $r=0$ in a constant time slice of pure AdS$_3$. 
 In order to compare the area variation with the differential entropy variation in the CFT, we express the radius  $R$ of the disk in terms of the boundary opening angle $\alpha$. By construction, the spacelike geodesic anchored at the endpoints of the boundary interval of size $2\alpha$ is tangent to the boundary circle  of the disk (see figure \ref{fig:bulkckv}). If the AdS spacetime contains a conical singularity, the coordinate radius of the disk is given by\footnote{See  appendix \ref{app:geoeq} for a derivation,  i.e.   equation 
\eqref{eq:radiuscon}  is identical to \eqref{eq:radiuscon2} for  $\tilde R = R$ and $\tilde \alpha = \alpha$.}  
\begin{equation} \label{eq:radiuscon}
R = L \gamma \cot (\gamma {\alpha}) \, . 
\end{equation}

\noindent The area of the disk   in   conical AdS$_3$   is therefore
 \begin{equation}
\begin{aligned} \label{eq:areacon}
A_{\text{con}}  = 2 \pi R 
 = 2 \pi L  \left [\frac{\cos \alpha}{\sin \alpha} +\left ( - \frac{\cos \alpha}{\sin \alpha} + \frac{\alpha }{\sin^2 \alpha} \right)\epsilon + \mathcal O \left(\epsilon^2 \right)\right] .
\end{aligned}
\end{equation}
We see that the area in conical AdS depends on three variables, $A_{\text{con}} = A_{\text{con}} ( \alpha, \epsilon,L)$. The area in pure AdS depends only on two variables, $A_{\text{vac}} = A_{\text{vac}} (  \alpha,L)$, since the defect parameter is fixed to be $\epsilon_{\text{vac}} = 0.$ We therefore consider three types of first-order variations  of the pure AdS background:  1)~variations of $\alpha$, i.e.   rescaling   the size of the boundary interval,   2)~variations of $\epsilon$, i.e.   metric perturbations due to the presence of a  conical singularity in AdS$_3$, and 3)~variations of $L$, i.e.  changing   the cosmological constant of the background (in the differential entropy variation below we also include variations of the gravitational constant). The total change in area under $\alpha$, $\epsilon$ and $L$ variations is   defined as
\begin{equation} \label{eq:areavar1}
\delta A   =\delta_{\alpha} A \Big |_{ \epsilon,L} + \delta_\epsilon A  \Big |_{ \alpha,L}+ \delta_L A  \Big |_{\alpha, \epsilon}.
\end{equation}

\noindent First, the change in area under an $\alpha$  variation,  at fixed $\epsilon$ and $L$, is to first order
\begin{equation}  \label{eq:areachangealpha}
\delta_{\alpha} A \Big |_{\epsilon,L} = \frac{\partial A_{\text{vac}}}{\partial \alpha}   \delta \alpha = -  2 \pi L \frac{1}{\sin^2 \alpha} \delta \alpha .
\end{equation}
Here we used an explicit expression for the area in pure AdS, 
\begin{equation}
	A_{\text{vac}} =2\pi L \cot \alpha,
\end{equation}
 which follows from \eqref{eq:areacon} by setting $\epsilon=0$.  Thus, the area decreases  if the boundary interval becomes larger.  This can be easily verified from  figure \ref{fig:bulkckv}.  

Further,  from \eqref{eq:areacon} we see that  the area change due to an infinitesimal variation of the conical defect parameter is, keeping $\alpha$ and $L$ fixed,
\begin{equation} \label{eq:areavarepsilon}
\delta_\epsilon A \Big |_{\alpha, L} =  2 \pi L   \left ( -\frac{\cos \alpha}{\sin \alpha} + \frac{\alpha }{\sin^2 \alpha} \right)\delta \epsilon ,
\end{equation}
where we   introduced $\delta \epsilon = \epsilon_{\text{con}} -  \epsilon_{\text{vac}} = \epsilon$, since $\epsilon_{\text{vac}}=0$. 
The function of $\alpha$ between brackets is positive   since $\alpha \ge 0$, and we have $\delta \epsilon >0$ since the perturbed geometry has a conical defect.\footnote{Note that $\delta \epsilon$ would   be negative for perturbations to  geometries with a conical excess, but we discard these rather unphysical solutions, since their energy spectrum is unbounded from below.} Therefore, the area of the disk increases due to the presence of a conical defect, if the  background curvature and boundary opening angle are kept fixed. 

Finally, if we vary the AdS radius, but keep $\alpha$ and $\epsilon$ fixed, the area change is simply
\begin{equation}  \label{eq:areavarads}
\delta_L A \Big |_{\alpha, \epsilon} = \frac{\partial A_{\text{vac}}}{\partial L}  \delta L  =  2 \pi \frac{\cos \alpha}{\sin \alpha}  \delta L .
\end{equation}
Thus, the full area variation is to first order given by
\begin{equation} \label{eq:fullareavar}
\delta A =  2 \pi \frac{\cos \alpha}{\sin \alpha}  \delta L - 2\pi L \frac{1}{\sin^2 \alpha} \delta \alpha + 2 \pi L   \left ( -\frac{\cos \alpha}{\sin \alpha} + \frac{\alpha }{\sin^2 \alpha} \right)\delta \epsilon.
\end{equation}
In order to compare the area variation of a disk in pure AdS$_3$ with the variation of the differential entropy, we need the holographic dictionary between differential entropy and bulk area \cite{Balasubramanian:2013rqa,Balasubramanian:2013lsa}
\begin{equation} \label{eq:diffentropyarea}
S_{\text{diff}} = \frac{A}{4G}.
\end{equation}
If we allow for variations of Newton's constant~$G$, the   variation of the differential entropy is, written in terms of    bulk quantities,
\begin{equation} \label{eq:vardiffentropybulk}
\delta S_{\text{diff}}=  \frac{\pi}{2}  \frac{\cos \alpha}{\sin \alpha}  \delta \left ( \frac{L}{G} \right) - \frac{\pi L}{2 G} \frac{1}{\sin^2 \alpha} \delta \alpha +\frac{\pi L}{2 G}     \left ( -\frac{\cos \alpha}{\sin \alpha} + \frac{\alpha }{\sin^2 \alpha} \right)\delta \epsilon.
\end{equation}
 Note that the variation of the ratio of the AdS radius and Newton's constant appears in this formula, since the (vacuum) differential entropy is proportional to the same fraction: $S_{\text{vac}}^{\text{DE}} \sim L/G$. 
We can now translate  the right-hand side of the equation above   in terms of  pure boundary quantities, using the standard holographic dictionary for   AdS$_3$/CFT$_2$  
\begin{equation} \label{eq:holographicdictionary}
c= \frac{3L}{2 G} \qquad \text{and} \qquad   \Delta =    M  L=  \frac{c}{12} \left ( 1-\gamma^2  \right) =  \frac{c}{6} \epsilon + \mathcal  O(\epsilon^2) . 
\end{equation} 
The first equation is the   Brown-Henneaux formula for the central charge \cite{Brown:1986nw}, and the second equation is the holographic relation between the scaling dimension of an operator in the dual CFT and the ADM mass of the (conical) geometry.  
Importantly, the first term on the right-hand side of \eqref{eq:vardiffentropybulk} can be written solely in terms of the variation of the central charge, due to the appearance of   $\delta (L/G)$. 
Furthermore, the final term in \eqref{eq:vardiffentropybulk} is proportional to  the variation of the scaling dimension, since $\delta \Delta = (c/6) \delta \epsilon$ \eqref{eq:epsdelrel}.    Therefore, based on  the holographic dictionary  we arrive at
\begin{align}
\delta S_{\text{diff}}
&=   \frac{\pi}{3} \frac{\cos \alpha}{\sin \alpha}     \delta c- \frac{\pi c }{3} \frac{1}{\sin^2 \alpha} \delta \alpha +  2 \pi     \left ( -\frac{\cos \alpha}{\sin \alpha} +   \frac{\alpha }{\sin^2 \alpha} \right)\delta \Delta.
\end{align}
This is identical to the differential entropy variation derived in section \ref{sec:vardiffentropy}.

\subsubsection{Volume variation}\label{subsec:volvarbulkdisk}

In this section we vary the proper volume of a disk in pure AdS$_3$. We consider again the three ($ \alpha, \epsilon,L$) variations of the previous section, under which the volume changes to first order as
\begin{equation} \label{eq:volumevardef}
\delta V   =  \delta_{\alpha} V \Big |_{ \epsilon,L} + \delta_\epsilon V  \Big |_{ \alpha,L} +\delta_L V  \Big |_{\alpha, \epsilon} .
\end{equation}
The $\alpha$ and $L$  variations can be derived from   the proper volume   in pure AdS
\begin{equation}
V_{\text{vac}}= \int_{0}^R \frac{2\pi r d  r}{\sqrt{1 +( r/L)^2}} = 2 \pi L^2 \left ( \sqrt{1 + (R/ L)^2}   - 1   \right) =2 \pi L^2   \left ( \frac{1}{\sin \alpha  } -1 \right).
\end{equation}
The change of volume under a variation of $\alpha$, at fixed $\epsilon$ and $L$, is given by 
\begin{equation} \label{eq:volalphavar}
\delta_\alpha V \Big |_{\epsilon,L} = \frac{\partial V_{\text{vac}}}{\partial \alpha}   \delta \alpha = - 2 \pi L^2 \frac{\cos \alpha}{\sin^2 \alpha} \delta \alpha . 
\end{equation}
The minus sign indicates that in pure AdS the volume of a disk decreases if the opening angle on the boundary increases (and $L$ is kept fixed). Further, because of the quadratic scaling of the volume with the AdS radius,   the   change  in volume under a variation of $L$, at fixed $\alpha$ and $\epsilon$, is
\begin{equation} \label{eq:volumevarads}
 \delta_L V \Big |_{\alpha, \epsilon} = \frac{\partial V_{\text{vac}}}{\partial L}   \delta L=4 \pi L \left (  \frac{1}{\sin \alpha}  - 1 \right) \delta L .
 \end{equation}
 We simply find that  in pure AdS the volume of a disk increases if the curvature radius increases (and $\alpha$ is kept fixed).

The $\epsilon$ variation in \eqref{eq:volumevardef} is equal to the first order change in the volume due to the presence of a conical defect in AdS. In order to derive this we need the proper volume of a disk in   AdS  with a conical defect, which in terms of the coordinates \eqref{eq:conicalmetric2} is given by
\begin{equation}\label{eq:convol}
V_{\text{con}}= \int_{0}^R \frac{2\pi r d  r}{\sqrt{\gamma^2 + r^2/L^2}} = 2 \pi L^2 \left ( \sqrt{\gamma^2 + R^2 / L^2}   - \gamma   \right) .
\end{equation}
Although we use the same notation $R$  in conical AdS  and  in pure~AdS for the coordinate radius of a disk, we emphasize that its relation to, for instance, the geodesic radius of the disk\footnote{The geodesic radius is given in terms of the coordinate radius   by: $\ell = 2 L \arctanh\left [ \frac{-\gamma + \sqrt{\gamma^2 + (R/L)^2}}{R/L} \right].$ } or to the boundary opening angle $\alpha$, as in     \eqref{eq:radiuscon}, is different for conical and pure AdS, since  their metrics   differ. In terms of the boundary opening angle, the proper volume reads
\begin{equation}
\begin{aligned}
V_{\text{con}} 
&=  2 \pi L^2 \gamma \left ( \frac{1}{\sin (\alpha \gamma)} -1 \right) \\
&= 2 \pi L^2 \left [  \frac{1}{\sin \alpha} -1 + \left (  \frac{\alpha \cos \alpha}{\sin^2 \alpha} - \frac{1}{\sin \alpha}  +1\right) \epsilon + \mathcal O \left(\epsilon^2 \right) \right] .
\end{aligned}
\end{equation}
In the last equation we expanded the proper volume around $\epsilon_{\text{vac}}=0.$
The leading-order  term in the expansion is, of course, the proper volume   in pure AdS. The subleading-order term defines the first order volume change  under a variation of $\epsilon$, at fixed $\alpha$ and $L$,
\begin{align} \label{eq:volumevarepsilon}
\delta_\epsilon V  \Big |_{\alpha,L}&= 2 \pi L^2  \left (  \frac{\alpha \cos \alpha}{\sin^2 \alpha} - \frac{1}{\sin \alpha}  +1 \right) \delta \epsilon  .
\end{align}
  The term between brackets is positive    for the entire range of $\alpha$, i.e. for $0 \le \alpha \le \pi/2$. Hence, just like the area, the proper volume of a disk increases due to the presence of a point particle in AdS, if $\alpha$ and $L$ are kept fixed.
In total, the   volume variation is given by
\begin{equation}
\delta V =  4 \pi L \left (  \frac{1}{\sin \alpha}  - 1 \right) \delta L - 2 \pi L^2 \frac{\cos \alpha}{\sin^2 \alpha} \delta \alpha + 2 \pi L^2  \left (  \frac{\alpha \cos \alpha}{\sin^2 \alpha} - \frac{1}{\sin \alpha}  +1 \right)  \delta \epsilon .
\end{equation}
Next, we want to convert this bulk variational identity for the volume into a boundary variational identity for  holographic complexity.  We  employ  the `complexity=volume'  conjecture \cite{Susskind:2014rva,Stanford:2014jda}, which can be extended to a correspondence between the   complexity of the vacuum state in a cutoff CFT and the volume of a ball-shaped  region  in the bulk, whose radius is related to the dual energy cutoff scale in the CFT  (see section \ref{sec:boundarydualbulkvolume})
\begin{equation} \label{eq:complexityequalsvolume}
\mathcal C = \frac{V}{4 G L}  \, .
\end{equation}
Allowing for variations of both Newton's constant  and the AdS radius, we thus find the following variational relation for holographic complexity    in terms of   bulk quantities  
\begin{equation}
\delta \mathcal C =   \frac{\pi}{2}   \left (  \frac{1}{\sin \alpha}  - 1 \right) \delta \left (  \frac{L}{G } \right)  - \frac{  \pi L}{2G} \frac{\cos \alpha}{\sin^2 \alpha} \delta \alpha + \frac{  \pi L}{2G}  \left (  \frac{\alpha \cos \alpha}{\sin^2 \alpha} - \frac{1}{\sin \alpha}  +1 \right)  \delta \epsilon .
\end{equation}
The   variation $\delta (L/G)$ appears again, just like in the differential entropy variation \eqref{eq:vardiffentropybulk}, because the  vacuum  holographic complexity scales with the same fraction:   $  \mathcal C_{\text{vac}} \sim L/G$. This term is proportional to the variation of the central charge  \eqref{eq:holographicdictionary}.  Note that with a different normalization of holographic complexity, such as    topological complexity  $\mathcal C_{\text{top}} = V/L^2$ \cite{Abt:2017pmf,Abt:2018ywl}, the complexity variation would not involve a term   proportional to  the variation of the central charge, since $C_{\text{top}}^{\text{vac}}$ does not depend on the central charge, cf. equation \eqref{eq:topcomplexity2}. With our choice of normalization, however,  the  complexity variation can be written  in terms of  the variation of the three  CFT variables  ($c, \alpha, \Delta$), by employing the    holographic dictionary for~AdS$_3$/CFT$_2$, 
 \begin{equation}
\delta \mathcal C =  \frac{\pi}{3}   \left (  \frac{1}{\sin \alpha}  - 1 \right) \delta c - \frac{  \pi c}{3} \frac{\cos \alpha}{\sin^2 \alpha} \delta \alpha +2 \pi    \left (  \frac{\alpha \cos \alpha}{\sin^2 \alpha} - \frac{1}{\sin \alpha}  +1 \right)   \delta \Delta .
\end{equation}
As expected, this agrees   with the result derived in section \ref{sec:varycomplexity}.

\subsubsection{Bulk matter Hamiltonian variation}\label{subsec:bulkmattervar}

In this section we compute the variation of the bulk matter Hamiltonian, $\delta H_\zeta^{\text{mat}}$, featuring on the left-hand side of the first law of causal diamonds \eqref{eq:firstlawcd}. The   matter Hamiltonian is the generator of evolution of  matter fields  along the  flow generated by the conformal Killing vector $\zeta$. Its first-order  variation takes the form\footnote{Ref. \cite{Jacobson:2018ahi} distinguished between the perfect  fluid matter corresponding to the cosmological constant and other matter fields with a fluid description.  In the present paper we treat the cosmological constant as a coupling constant of the theory and therefore it does not contribute to the stress tensor in \eqref{eq:matterHamvar}. The notation $\delta H_\zeta^{\text{mat}}$  used here corresponds to $\delta H_\zeta^{\text{\~m}}$ in  \cite{Jacobson:2018ahi}.} \cite{Jacobson:2018ahi}  
\begin{equation} \label{eq:matterHamvar}
\delta H_\zeta^{\text{mat}}  = \int_D \delta   ( {T_\mu}^\nu   )   \zeta^\mu u_\nu dV . 
\end{equation}
Recall that $u$ is the future pointing unit normal to the disk $D$. Moreover,   $ {T_\mu}^\nu    = g^{\nu \alpha} T_{\mu \alpha}$ denotes the Hilbert stress-energy tensor (with one index raised) associated to   matter fields  in the bulk. Since the stress-energy tensor for a typical field theory action is quadratic in the matter fields, and the matter fields vanish in the AdS background, the first-order variation  of the stress tensor away from pure AdS vanishes. The stress-energy associated to fluid matter, however, can contribute to the first-order variation of the Hamiltonian, and thus to the first law of causal diamonds.\footnote{The fact that variations of the  stress tensor for perfect fluids can contribute to the    first law of black hole mechanics was already pointed out in the original paper by Bardeen, Carter and Hawking \cite{Bardeen:1973gs}. }

For the present field content, the perfect fluid consists of a point mass in AdS and its stress-energy indeed contributes to the matter Hamiltonian variation. 
Inserting the point particle stress-energy tensor  \eqref{eq:pointstresstensor1}   into the matter Hamiltonian variation \eqref{eq:matterHamvar} yields 
 \begin{equation} \label{eq:conicalhamvar}
 \delta H_\zeta^{\text{mat}} = \int_D d \phi dr \delta^{(2)} (r)  (-\zeta \cdot u ) \delta m=  \sqrt{- \zeta \cdot \zeta}  \Big |_{  O} \delta m,
 \end{equation}
where $\delta m$ is the variation of the DJH mass and $O$ is the location of the point particle, i.e.~$O=\{t=0,r=0\}$ for the current set-up.  In the first equality we used that $u$ has unit norm, $u \cdot u = -1$, and in the last equality we wrote  $u$ as the   velocity vector of the conformal Killing flow, $u^\mu = \zeta^\mu / \sqrt{- \zeta \cdot \zeta}$, which also defines the extension of $u$ off of   $D$. This relation between $u$ and $\zeta$ holds in particular at $D$ because the conformal Killing vector   is normal to the disk. We thus find that the Hamiltonian generating the evolution of a point mass along the conformal Killing flow   is equal to  the DJH mass times a `redshift factor'    
  \begin{equation} \label{eq:hammassrel}
  H_\zeta^{\text{mat}} =  m  \sqrt{- \zeta \cdot \zeta} \Big |_{O} = m \frac{\partial \tau}{\partial s}  \Big |_{O} . 
  \end{equation}
 This result was to be expected, since $H_\zeta^{\text{mat}}$ and $m$ are defined with respect to different time variables, called $s$ and $\tau$ respectively.  The conformal Killing time $s$ satisfies $\zeta \cdot d s = 1$, with the initial condition $s=0$ at $D$, and the time variable $\tau$ is the proper time along the flow lines of $\zeta$, which is 
 similarly defined through $u \cdot d \tau = 1$ and the condition $\tau =0$ at~$D$. However, the function $s$ is  not uniquely defined,\footnote{The ambiguity is to add to $\zeta=\partial_s$ any vector $v$ in the codimension-1 subspace satisfying $v \cdot d s =0.$ (We thank Ted Jacobson for this point.)} and is only meaningful relative to a complete coordinate system. In appendix B of \cite{Jacobson:2018ahi} a complete coordinate chart $(x,s)$ was constructed for   a maximally symmetric diamond (suppressing the angular coordinates). Here $x\in[0,\infty)$ is a radial coordinate, satisfying  $|dx|=|ds|$, $x=0$ at $r=0$, and $\zeta \cdot d x =0$. This implies in particular that $\zeta = \partial_s$ and $\tau = \tau (s,x).$\footnote{The proper time can be computed by the integral $ \int_0^s |\zeta| ds'$, where $|\zeta|:= \sqrt{-\zeta\cdot\zeta}$. For the conformal Killing vector of an AdS causal diamond,  which has   surface gravity $\kappa$, the norm is
\begin{equation} \label{eq:norm3}
|\zeta| (s,x)= \frac{ \kappa R }{   \sqrt{1+(R/L)^2}\cosh s + \cosh x   }=\frac{  \kappa  L \cos \alpha}{  \cosh s + \sin \alpha \cosh x }.
\end{equation} 
This follows from equation (B.5) in \cite{Jacobson:2018ahi}, since $|\zeta| = C$ in their notation, by first replacing $L \to i L$ and then setting $R_* = L \arctan (R/L)$.   The proper time along the flow lines of $\zeta$ is
\begin{equation}
	\tau (s,x)= \frac{2 \kappa L \cos \alpha}{\sqrt{\sin^2 \alpha \cosh^2 x - 1}} \arctanh \left [ \tanh(s/2) \sqrt{\frac{\sin \alpha \cosh x -1}{\sin \alpha \cosh x + 1 }}\right],
\end{equation}
which   vanishes at $s=0$.
  In the limit $R/L \to \infty $ or $\alpha\to0$, i.e. for the  Wheeler-deWitt patch of AdS,  the norm is   $|\zeta|_{\text{WdW}}= \kappa L/\cosh s$ and the proper time is   $\tau_{\text{WdW}} = 2 \kappa  L \arctan[\tanh(s/2)]$. For causal diamonds in flat space the norm is $|\zeta|_{\text{flat}} =  \kappa R/(\cosh s + \cosh x)$ and the proper time is $\tau_{\text{flat}} = 2 \kappa R \arctanh[\tanh(s/2)\tanh(x/2)]/\sinh x$ (see also appendix A of \cite{Visser:2019muv}).} Thus from the definitions of $\tau$ and $u$ we find $\zeta \cdot d \tau = \sqrt{-\zeta \cdot \zeta}$, which is identical to  the second equality in \eqref{eq:hammassrel}  since $\zeta \cdot d \tau = \partial \tau / \partial s$.  
 
The unique conformal Killing vector whose flow preserves a spherically symmetric causal diamond in   AdS reads in terms of the standard $t$ and $r$ coordinates~\cite{Jacobson:2018ahi}
\begin{equation} \label{eq:ckv}
\zeta = - \frac{2\pi L^2}{R} \left [   \left (1- \frac{\sqrt{1 + (R/L)^2 }}{ \sqrt{1 + (r/L)^2} } \cos (t/L)   \right) \partial_t + \frac{R}{L} \sqrt{(1 + (R/L)^2)(1+ (r/L)^2)} \sin (t/L) \partial_r \right ] .
\end{equation}
 The vector $\zeta$ generates a timelike flow within the causal diamond, it acts as a null flow on the null boundaries and it vanishes at the edge  of the diamond and the future and past tips.
We   normalized the conformal Killing vector such that its surface gravity  is  $\kappa = 2\pi$, in contrast to \cite{Jacobson:2018ahi} where the normalization of $\zeta$ was chosen such that $\kappa = 1$. The definition of surface gravity   used   in this setup is:   $\nabla_\mu (\zeta \cdot \zeta) = - 2 \kappa \zeta_\mu$, where both sides are evaluated  on the (future) null boundary of the diamond, which is a conformal Killing horizon  since $\zeta$ is tangent to its null generators. This definition of surface gravity is Weyl invariant~\cite{Jacobson:1993pf} and constant on any bifurcate conformal Killing horizon (see appendix  C of  \cite{Jacobson:2018ahi}). The value $\kappa = 2\pi$ is convenient, 
since  the Hawking temperature $T_{\text{H}}=\kappa/2\pi$ is equal to one for this normalization. Therefore, the conjugate quantities to the differential entropy and   holographic complexity simplify considerably in the CFT first law, cf.  \eqref{eq:conju1}.\footnote{Another convenient normalization of $\zeta$   is to set   $\sqrt{-\zeta\cdot\zeta} =1$ at  $O$. With this normalization the matter Hamiltonian is equal to the DJH mass, $H_\zeta^{\text{mat}}=m$, and the surface gravity becomes a function of $R$, or equivalently of $\alpha$: $\kappa =  \frac{R/L^2}{  \sqrt{1+(R/L)^2} -1}=\frac{\cos \alpha}{L (1- \sin \alpha)}.$ For the Wheeler-deWitt patch (i.e. $R/L \to \infty$) the surface gravity simplifies to $\kappa_{\text{WdW}} = 1/L.$} 

The norm of the conformal Killing vector evaluated at the center of the diamond is\footnote{This expression for the norm  also  follows from equation \eqref{eq:norm3} by setting $s=x=0$ and $\kappa=2\pi.$}
\begin{equation}
\begin{aligned} \label{eq:normckv}
   \sqrt{- \zeta \cdot \zeta}  \Big |_{O} & = \frac{2\pi L^2}{R} \left (  \sqrt{1 + (R/L)^2}  - 1 \right) = 2\pi L f(\alpha)  \\
   \text{with}  \quad   f(\alpha)  &=  \left (  \frac{1}{\cos \alpha} -\frac{ \sin \alpha }{\cos \alpha} \right) ,
  \end{aligned}
\end{equation}
where the factor $2\pi$ arises due to our choice of $\kappa$. The variation of the point particle Hamiltonian now follows from inserting this norm and the variation of the DJH mass, $\delta m = \delta \epsilon / (4G)$, into equation   \eqref{eq:conicalhamvar}  
\begin{equation} \label{eq:matHamvar4}
\delta H_\zeta^{{\text{mat}}}  
= \frac{ \pi L}{ 2G} f(\alpha) \delta \epsilon .
\end{equation}
Note that  the gravitational constant is not being varied in this variational expression, since~$\epsilon$ vanishes in the background spacetime. The change in the matter Hamiltonian is, of course, positive since the norm of $
\zeta$ is positive and $\delta \epsilon >0$ by assumption. Using the holographic dictionary~\eqref{eq:holographicdictionary}   the Hamiltonian variation can be easily expressed purely in terms of   CFT quantities
\begin{equation} \label{eq:hamvarcft}
\delta H_\zeta^{\text{mat}} = 2 \pi f(\alpha) \delta \Delta . 
\end{equation}
We recognize here the CFT energy variation, defined in \eqref{eq:rescaledE}, in the first law for differential entropy.  
The matter Hamiltonian  itself  \eqref{eq:hammassrel} is given by
\begin{equation} \label{eq:newmatterham}
 H^{\text{mat}}_\zeta  = 2\pi     f(\alpha) m L=  2\pi f(\alpha)( c / 6)  \epsilon,
\end{equation}
which reads in terms of the conformal dimension of the twist operator  
\begin{equation} \label{eq:newblabla}
	H^{\text{mat}}_\zeta  = 2\pi f(\alpha)(c/6 )  ( 1- \sqrt{1- 12 \Delta / c}  ) . 
\end{equation}
 Therefore, the boundary energy $E$ dual to the matter Hamiltonian is   equal to this expression up to all orders in $\Delta/c$ (or $\epsilon$). Up to first order in $\Delta/c$ it is equal to $2\pi f(\alpha)\Delta$, which is the only term that is relevant for the first law studied in this paper. Note that since $\Delta = ML$, the difference between the leading order term and the full expression for the matter Hamiltonian precisely corresponds to the difference between the ADM mass $M$ and the DJH mass $m$, respectively. In other words, the   expression \eqref{eq:newblabla} for $H^{\text{mat}}_\zeta $ above also follows from inserting the relation~\eqref{eq:djhadmmass} between the DJH and the ADM mass into \eqref{eq:newmatterham}.

\subsubsection{Combining the variations}
\label{sec:combining2}

In the previous three sections we   computed the area variation, volume variation and matter Hamiltonian variation for disks in three-dimensional AdS space.  In this section we combine these three variations into one variational relation,   thereby  reproducing the first law of causal diamonds  applied to variations from pure AdS to conical AdS space. 

We proceed   by studying
  the following combination of variations  for the present set-up, which appears in the first law of causal diamonds \eqref{eq:firstlawcd},
\begin{equation} \label{eq:combvar}
  \delta A - k \delta V, \qquad \text{with} \qquad k = \frac{1}{R} \sqrt{1 + (R/L)^2} = \frac{1}{L \cos \alpha}
\end{equation}
being the trace of the outward extrinsic curvature of $\partial D$ as embedded in the disk $D$. Below we compute this combination   explicitly for  the   $  \alpha, \epsilon  $ and $L$ variations. That is, we define the area and volume change by equations \eqref{eq:areavar1} and \eqref{eq:volumevardef},  respectively, and  insert   the results of the previous sections for the ($  \alpha,\epsilon, L$) variations.

Firstly, it follows from \eqref{eq:areachangealpha} and \eqref{eq:volalphavar} that the   combination of  variations above  for  variations which alter $\alpha$, but keep $\epsilon$ and $L$ fixed, vanishes
\begin{equation} \label{eq:alphavarvanishes}
\left ( \delta_\alpha  A - k \delta_\alpha   V \right) \Big |_{\epsilon, L}  = 0 . 
\end{equation}
This is because, by definition, the trace of the extrinsic curvature   is  equal to $k = ( \partial A / \partial V )_L.$ If the boundary opening angle varies, the volume change is given by $\delta_\alpha V \big |_{\epsilon, L}=(\partial_\alpha V)_L  \delta \alpha$, while the area change is   $\delta_\alpha A \big |_{\epsilon, L}=(\partial_\alpha A)_L  \delta \alpha   = k (\partial_\alpha V)_L  \delta \alpha  $, hence  equation \eqref{eq:alphavarvanishes} follows.
A more formal reason is that the variation induced by rescaling $\alpha$ is a diffeomorphism, and it was shown in section 3.3.2 of \cite{Jacobson:2018ahi} that the combination  \eqref{eq:combvar} vanishes for any  diffeo-induced variation. Below we can thus leave out the restriction of fixing $\alpha$ in the combination of variations under consideration. 

Secondly, for first-order variations of $\epsilon$,  at fixed   $L$,   the combination of area and volume variations becomes 
\begin{equation}
   \left ( \delta_\epsilon  A - k \delta_\epsilon   V  \right)\Big |_{ L}  =  - 2 \pi L \left ( \frac{1}{\cos \alpha} - \frac{\sin \alpha}{\cos \alpha} \right)   \delta \epsilon .
\end{equation}
Here we inserted expressions \eqref{eq:areavarepsilon} and \eqref{eq:volumevarepsilon}, respectively, for the area and volume variation induced by $\epsilon$. The right-hand side is related to minus   the matter Hamiltonian variation \eqref{eq:matHamvar4}, hence we find the following relation for $\epsilon$ induced variations
\begin{equation} \label{eq:areavolvarepsilonham}
  \frac{1}{4 G}  \left ( \delta_\epsilon  A - k \delta_\epsilon   V  \right)\Big |_{ L}  = -   \delta  H_\zeta^{\text{mat}} .
\end{equation}
The minus sign indicates that the area at fixed volume decreases due to the presence of a conical defect, while the volume  increases at fixed area.

Thirdly, for variations which change the AdS radius, at fixed $\epsilon$, the combination of variations   takes the form
\begin{equation} \label{eq:areavolvarads}
\left ( \delta_L A - k \delta_L V \right) \Big |_{ \epsilon} = -  2 \pi \left (   \frac{2}{\cos \alpha \sin \alpha} - \frac{2}{\cos \alpha} -\frac{\cos \alpha}{\sin \alpha} \right) \delta L .
\end{equation}
This follows from expressions \eqref{eq:areavarads} and \eqref{eq:volumevarads}, respectively, for the area and volume change under a variation  of $L$. The variation of the AdS radius is  related to the variation of the cosmological constant,   via $-2 \delta L /L = \delta \Lambda / \Lambda$. 
The conjugate quantity to $\delta L$  is therefore proportional to  the so-called `thermodynamic volume' $V_\zeta$, which  appears in extensions of the first law of black hole mechanics as the quantity conjugate to $\delta \Lambda$ \cite{Kastor:2009wy,Dolan:2010ha,Cvetic:2010jb}. The thermodynamic volume of maximally symmetric causal diamonds  is defined as the proper volume weighted by the norm of $\zeta$~\cite{Jacobson:2018ahi}
\begin{equation} \label{eq:thermovolume1}
V_\zeta = \int_D dV \sqrt{-\zeta \cdot \zeta}\,  . 
\end{equation}
It can   be computed explicitly for the conformal Killing vector  \eqref{eq:ckv} of  a causal diamond in pure AdS  
\begin{equation}\label{eq:vzeta}
\begin{aligned}
V_\zeta 
&= \frac{2\pi L^2}{R} \left ( \pi    R^2 - 2 \pi  L^2 \left(\sqrt{1+(R/L)^2 }-1\right)  \right) \\
&= 2 \pi^2 L^3 \left (  \frac{\sin \alpha}{\cos \alpha} + \frac{1}{\cos \alpha \sin \alpha} - \frac{2}{\cos \alpha}   \right) . 
\end{aligned}
\end{equation}
Since the (positive) functions of $\alpha$ within   parenthesis in equations \eqref{eq:areavolvarads} and \eqref{eq:vzeta} are the same, we can rewrite the former relation   as 
\begin{equation} \label{eq:ccvariationareavol}
 \left ( \delta_L  A - k \delta_L   V \right) \Big |_{ \epsilon} = -   \frac{V_\zeta}{2 \pi } \frac{2}{L^3} \delta L  = - \frac{V_\zeta}{2\pi} \delta \Lambda. 
\end{equation}
From  this variational identity   and  the scaling  properties $A \sim L$ and $V \sim L^2$ one can easily obtain a Smarr-like relation between the area, volume, cosmological constant and their conjugate quantities
\begin{equation} \label{eq:smarr3d}
   A - 2   k V  = \frac{V_\zeta}{2\pi }  2 \Lambda . 
\end{equation}
This is indeed the Smarr formula in three spacetime dimensions for maximally symmetric causal diamonds with $\kappa=2\pi$, derived in \cite{Jacobson:2018ahi}.

Finally, by combining the $\alpha$, $L$ and $\epsilon$ variations \eqref{eq:alphavarvanishes}, \eqref{eq:areavolvarepsilonham} and \eqref{eq:ccvariationareavol}, respectively, we arrive at the variational identity which relates the variations of the area, volume, matter Hamiltonian and cosmological constant
\begin{equation}  \label{eq:firstlawcd3}
\frac{1}{4G}\left (     \delta A -   k \delta V\right ) = - \delta H_\zeta^{\text{mat}}    - \frac{V_\zeta}{8 \pi G} \delta \Lambda  .
\end{equation}
This agrees with the first law of causal diamonds \eqref{eq:firstlawcd} if we make the identification $\kappa = 2\pi$. 

Note that   in the first law of causal diamonds only   the variation of the cosmological constant  of the Lagrangian theory appears, and not of Newton's constant. The absence of the  variation  of Newton's constant in \eqref{eq:firstlawcd3} is crucial for the comparison of the gravitational bulk first law with the microscopic boundary first law (see section \ref{subsec:firstlawcftfromads} below). Ultimately, the reason for this absence is that  the area, volume and matter Hamiltonian do not depend on the gravitational constant, whereas the first two quantities do scale with the cosmological constant. 
For completeness, in appendix~\ref{app:couplings} we  provide a careful analysis of the variations of the coupling constants in the   covariant phase space formalism,  and we prove that the terms proportional to the variation  of $G$ cancel out in the first law of causal diamonds.   For    variation of $ \Lambda$ such an analysis was done in \cite{Jacobson:2018ahi}   by treating  the cosmological constant as a perfect fluid, and thus as  part  of the fluid stress-energy tensor. However, variations of~$G$ cannot be computed in this way and hence an independent analysis is needed. 

As the gravitational constant is   a coupling constant of the Lagrangian theory,   it could in principle be varied in the space of gravitational theories (see \cite{Kastor:2010gq,Kastor:2014dra,Caceres:2016xjz,Kastor:2016bph} for other references where variations of all gravitational constants are   taken into account in  extended first laws). 
For example, in contrast to the extended first law of causal diamonds, in the extended version of the first law for AdS-Schwarzschild black holes   the variations of   $\Lambda$  and~$G$ could \emph{both} contribute\footnote{In comparison, in the extended first law for AdS-Rindler horizons only variations of $\Lambda$ appear, and not variations of $G$ \cite{Kastor:2014dra} (see equation \eqref{eq:firstlawadsrindler} and below for a further discussion).} 
\begin{equation} \label{eq:firstlawschw}
\delta M =\frac{1}{8 \pi G} \left ( \kappa \delta A + \bar V_\chi \delta \Lambda \right) - M\, \frac{\delta G}{G}.
\end{equation} 
Here  $\bar V_\chi $ is the background subtracted thermodynamic volume (denoted by $\Theta$ in \cite{Kastor:2009wy}), and $\chi$ represents the timelike Killing field of AdS-Schwarzschild. The variations $\delta \Lambda$ and~$\delta G$ could both  appear in the first law,  because the mass $M$ depends on     $\Lambda$ and $G$  via the Smarr formula, which is given by     \cite{Kastor:2009wy}
\begin{equation}
\frac{d-3}{d-2}M = \frac{1}{8\pi G} \left (  \kappa A     - \frac{ 2}{d-2} \bar V_\chi \Lambda   \right) 
\end{equation}
in $d  $ spacetime   dimensions. The $\delta G$ term in the first law \eqref{eq:firstlawschw}, which    is perhaps unfamiliar to the reader, arises  simply due to the scaling $M \sim G^{-1}$ in the Smarr formula.  Although we take the variations of the couplings $\Lambda$ and $G$ into account in the first law, in this paper we do not view them as thermodynamic variables in the bulk;    rather the number of degrees of freedom $N_{\text{dof}}\sim L^{d-2}/G$ plays the role of a thermodynamic variable in a holographic CFT. The  variations of $\Lambda$ and $G$ in the first law of AdS black holes correspond  to a chemical potential term for varying the number of degrees of freedom in the dual  CFT   first law (i.e.~$\mu \delta N_{\text{dof}}$ term  with $\mu$   the chemical potential), but only if the  CFT first law is expressed in terms of the dimensionless energy   $E = ML$   \cite{Visser}.  In the next section, we investigate how such a  $\mu \delta N_{\text{dof}}$ term arises in the CFT dual of the first law of causal diamonds.

\section{Matching the boundary and bulk first laws}
\label{sec:matching}

In  section \ref{sec:bdryfirstlaw}   we derived a  new variational relation in   two-dimensional CFTs  for differential entropy and holographic complexity, and in section \ref{sec:bulkderiv}   we analyzed the first law of causal diamonds for metric perturbations from vacuum  AdS$_3$   to  the conical defect geometry. In the present section we show that the boundary   first law of section \ref{sec:bdryfirstlaw}  is the holographic dual of the bulk first law of section \ref{sec:bulkderiv}.  In section \ref{subsec:firstlawcftfromads} we derive the boundary first law from the bulk first law, using the holographic dictionary in AdS$_3$/CFT$_2$. In section \ref{subsec:extensionhigherd} we generalize this argument to arbitrary dimensions -- assuming the dictionary between differential entropy and the bulk area generalizes to higher dimensions -- thereby obtaining a new variational identity for higher-dimensional holographic CFTs. 

\subsection{First law in holographic CFT$_2$ from first law in AdS$_3$}
\label{subsec:firstlawcftfromads}

We want to express the first law of causal diamonds   in terms of variations of the differential entropy and holographic complexity, for which we repeat here the holographic dictionary  used in this paper  
\begin{equation} \label{eq:entropyandcomplexity}
S_{\text{diff}} = \frac{A}{4 G} \qquad \text{and} \qquad \mathcal C = \frac{V}{4 G L} .
\end{equation}
Since we  allow for variations of Newton's constant $G$ and the AdS radius $L$,  the  variations of the differential entropy and complexity are not proportional to the variations of the area and proper volume, respectively, which appear in the   first law of causal diamonds. For example,  the inclusion of a factor $1/G$  inside the variation of the area has to be compensated by a term involving the variation of Newton's constant, i.e. $(\delta A) / G = \delta (A/G) - A \delta (1/G)$. Therefore, the area and volume variations are, respectively, related to  the variations of differential entropy and holographic complexity via 
\begin{align}
\frac{1}{4G} \delta A &=  \delta \left ( \frac{A}{4G} \right) + \frac{A}{4G}  \frac{\delta G}{G} , \label{eq:areavardiff}\\
\frac{1}{4G} \delta V &=  L\, \delta \! \left ( \frac{V}{4G L} \right) +   \frac{V}{4G }  \frac{\delta G}{G} +   \frac{V}{4G}  \frac{\delta L}{L}  .  \label{eq:volumevarcomplex}
\end{align}
If we insert these relations into the first law, we find a variational relation which now does involve a term proportional to~$\delta G$, in addition to a new term proportional to~$\delta L$. The   term involving $\delta L$ in \eqref{eq:volumevarcomplex} can be combined with the $\delta \Lambda$ term in the first law, by using the relation $ - 2 \delta L / L=\delta \Lambda / \Lambda $ and the Smarr formula \eqref{eq:smarr3d}. We thus obtain a new form of the first law of causal diamonds \eqref{eq:firstlawcd3}, in which the variations $\delta L$ and $\delta G$   share a common prefactor and  appear in a particular combination,
\begin{equation} \label{eq:newfirstlaw}
  \delta \! \left (  \frac{A}{4G} \right)  -   k L \, \delta \! \left (\frac{V}{4G L} \right)  =  \left ( \frac{A}{4G} -   k L  \frac{V}{4 G L} \right)  \left (  \frac{\delta L}{L} - \frac{\delta G}{G} \right) - \delta H_\zeta^{\text{mat}} . 
\end{equation}
This form of the first law can be easily translated into a boundary first law by using the holographic dictionary of the AdS$_3$/CFT$_2$ correspondence. On the left-hand side  we find the variations of the differential entropy and holographic complexity \eqref{eq:entropyandcomplexity}. The conjugate quantity $k L$ to the variation of the complexity is   a dimensionless function of the boundary opening angle, i.e. $1/\cos \alpha$, since $k$ is given by \eqref{eq:combvar}. On the right-hand side, the  combination of variations $\delta L / L - \delta G / G $ is equal to $\delta (L/G)/ (L/G)$ and hence to $\delta c/ c$, where $c$ is  the central charge of a dual $2d$ CFT. The boundary first law which  follows from the bulk first law is therefore
\begin{equation} \label{eq:bndyfirstlaw3}
 \delta S_{\text{diff}} - \frac{1}{\cos \alpha} \delta \mathcal C=   \left ( S_{\text{diff}} - \frac{1}{\cos \alpha} \mathcal C \right) \frac{\delta c}{c} - 2 \pi \left (  \frac{1}{\cos \alpha} -\frac{ \sin \alpha }{\cos \alpha} \right) \delta \Delta ,
\end{equation}
where we also replaced the matter Hamiltonian variation by   expression \eqref{eq:hamvarcft} involving the variation of the scaling dimension. This result agrees, of course, with the   boundary first law derived  in section  \ref{subsec:combining1}. The reason behind the appearance of   $\delta L / L - \delta G / G$ (or~$\delta c/ c$) in the first law is that  both   differential entropy and our definition of holographic complexity are proportional to the fraction $L/G$ (or $c$). The fact that differential entropy and complexity are proportional to a (generalized) central charge  in the CFT, and hence that the  variations of $L$ and~$G$ in the first law combine into   the    variation of the central charge, generalizes to higher dimensions, as we will see in the next section. 

Before moving on, we stress that the validity of the CFT dual of the first law of causal diamonds depends on the holographic dictionary for the area and   volume variation. Since the dictionary for holographic complexity has not been   established yet, it could be that   $\mathcal C = V/(4G L)$ is incorrect. As an example of a different expression, we consider  topological complexity \eqref{eq:topcomplexity1}, $\mathcal C_{\text{top}}:=-\frac{1}{2}\int_D   dV \mathcal R  =V/L^2$, which is     motivated from the Gauss-Bonnet theorem \eqref{eq:gaussbonnet} \cite{Abt:2017pmf}. We note that in this case the first law of causal diamonds can     be organized as   
\begin{equation}
\delta\!\left( \frac{A}{4G}\right) - \frac{k L}{6} \frac{3L }{2G} \delta \left (\frac{V}{L^2} \right) = \frac{A}{4G}  \left (  \frac{\delta L}{L} - \frac{\delta G}{G} \right)  - \delta H_\zeta^{\text{mat}}.
\end{equation}
This    suggests a   CFT first law for   differential entropy and topological complexity of the form 
\begin{equation} \label{eq:topocomplfirstlaw}
\delta S_{\text{diff}} - \frac{c}{6 \cos \alpha} \delta \mathcal C_{\text{top}} = S_{\text{diff}} \frac{\delta c}{c} - 2 \pi \left (  \frac{1}{\cos \alpha} -\frac{ \sin \alpha }{\cos \alpha} \right) \delta \Delta.
\end{equation}
The conjugate quantities to the complexity variation and to the central charge variation are then slightly different compared to the CFT first law \eqref{eq:bndyfirstlaw3}.  This signifies how important it is to establish the correct dictionary for the volume. In higher    dimensions, however, the dictionary for holographic complexity used in this paper seems to be more appropriate   than topological complexity, since the latter is not a dimensionless quantity for spacetime dimensions $d>3$ (because $V \sim L^{d-1}$ and $\mathcal R=2 \Lambda  \sim 1/L^2$ for static slices of pure AdS).

\subsection{Extension of the boundary first law to higher dimensions}
\label{subsec:extensionhigherd}

The  bulk first law  in the form  \eqref{eq:firstlawcd3} applies to causal diamonds in AdS space in arbitrary dimensions \cite{Jacobson:2018ahi}. Following the same procedure as in the previous section, we can thus derive a boundary first law in higher dimensions from the  bulk  first law. This translation procedure, however, is highly dependent  on the holographic dictionary \eqref{eq:entropyandcomplexity} for differential entropy and complexity, and the question is whether this dictionary generalizes   to higher dimensions. Since the original `complexity=volume'  conjecture \cite{Susskind:2014rva,Stanford:2014jda} is not restricted to any specific dimension, we can safely assume the dictionary between holographic complexity (for cutoff CFTs) and proper volume in higher dimensions. 

Furthermore, differential entropy has been extended to higher dimensions for certain  symmetric gravitational backgrounds by \cite{Myers:2014jia,Czech:2014wka} and for  general convex bounded   regions in the bulk  by \cite{Balasubramanian:2018uus}. The latter proposal for differential entropy is in terms of  an   integral of shape derivatives of the entanglement entropy of ball shaped regions.
 For the validity of the boundary dual of the first law of causal diamonds   it is necessary  that    differential entropy is related to the area of a bulk surface both in the background AdS spacetime and in the perturbed geometry with AdS asymptotics.  In this regard, it is satisfying that the extensions of differential entropy to higher dimensions are well defined for  arbitrary states in holographic CFTs, hence not only the vacuum state, and  are conjectured to be related to the area  of   bulk surfaces  for general   backgrounds.

In this section we assume that   the proposals for differential entropy and holographic complexity in higher dimensions are correct.  Given this assumption,  let us now derive a boundary dual of the   first law of causal diamonds for arbitrary dimensions. The formulas for the area and volume variations \eqref{eq:areavardiff} and \eqref{eq:volumevarcomplex}, respectively,  still hold in  higher dimensions,  but the Smarr formula  \eqref{eq:smarr3d} contains dimension dependent factors \cite{Jacobson:2018ahi}
\begin{equation} \label{eq:smarrgeneral}
(d-2)   A - (d-1)   k V =  \frac{ V_\zeta}{2\pi}2 \Lambda ,
\end{equation}
where $d$ is the number of bulk spacetime dimensions. 
As a result, the form \eqref{eq:newfirstlaw} of the    first law  generalizes to higher dimensions as \begin{equation} \label{eq:bulkfirstlawgendim}
  \delta \left (  \frac{A}{4G} \right)  -   k L \delta \left (\frac{V}{4G L} \right)  =  \left ( \frac{A}{4G} -   k L \frac{V}{4 G L} \right)  \left ( (d-2) \frac{\delta L}{L} - \frac{\delta G}{G} \right) - \delta H_\zeta^{\text{mat}} .
\end{equation}
For $d=3$   we related the combination of   variations of $L$ and $G$, on the right-hand side of the equation,  to the variation of the central charge $c$ in the dual two-dimensional CFT. In arbitrary $d$ bulk dimensions we should also relate this combination of variations to a central charge in the  holographic $(d-1)$-dimensional CFT. However, the standard  central charges parametrizing   the trace anomaly   $\langle {T^\mu}_\mu \rangle$ in a curved background exist only for even dimensions.
Two other candidates   for a generalized central charge, which are  also defined in odd dimensions, are the parameters $C_T$ and~$a^*$~\cite{Myers:2010tj}. The first parameter $C_T$ is defined as the overall normalization of the two-point function of the CFT stress tensor \cite{Osborn:1993cr}. 
The second parameter $a^*$ is the universal coefficient in the vacuum entanglement entropy for ball-shaped regions. In even dimensions $a^* $ is equal  to the coefficient $A$ of the Euler density in the trace anomaly, e.g. for two-dimensional CFTs we have $a^* = c/12$. Since $a^*$ evolves monotonically under the renormalization group flow, it can be thought of as counting the number of degrees of freedom in the CFT \cite{Myers:2010xs,Myers:2010tj,Casini:2017vbe}. Now for Einstein gravity the two CFT parameters are related to $L$ and $G$ via\footnote{The central charge in   the stress tensor two-point function is also sometimes defined with a different normalization, $\tilde C_T := \frac{\pi^{d-1} (d-2)}{\Gamma(d+1)}C_T$, such  that  $ a^*=\tilde C_T$ for   CFTs dual to Einstein gravity \cite{Hung:2011nu,Faulkner:2017tkh}.}
 \begin{equation} \label{eq:twoCFTparameters}
 	a^* =\frac{\pi^{d-1} (d-2)}{ \Gamma (d+1)}  C_T = \frac{\Omega_{d-2}L^{d-2}}{16\pi G},
 \end{equation}
where   $\Omega_{d-2}:= 2 \pi^{\frac{d-1}{2}}/ \Gamma\!\left(\frac{d-1}{2}\right)$  denotes the volume of a $(d-2)$-dimensional unit sphere. For the purpose of expressing the first law in terms of CFT quantities, however, only the scaling of the central charge with $L$ and $G$ is important and the proportionality factor is irrelevant. This is because the  combination of variations $(d-2) \delta L / L- \delta G /G$ in \eqref{eq:bulkfirstlawgendim} is equal to $\delta( L^{d-2}/ G)/(L^{d-2}/ G)$. Therefore, in terms of  the generic number of field theoretic   degrees of freedom  of the CFT\footnote{This `area law' for $N_\text{dof} $ is the ultimate reason why the   AdS/CFT correspondence implements the holographic principle  \cite{Susskind:1998dq}. Let us recap this area law for the canonical example of   AdS/CFT:     $\mathcal N=4$ $SU(N)$ super-Yang-Mills   theory dual to  type IIB string theory on $AdS_5 \times S^5$  \cite{Maldacena:1997re}.  The central charges in the trace anomaly for  $4d$ $\mathcal N=4$ super-Yang-Mills theory are the same and equal to  $c =( N^2 -1) / 4$, where we can drop the factor minus one for large $N$. The holographic dictionary states $(L/l_\text{s})^4 = g_{\text{YM}}^2 N$ and $g_{\text{s}}= g_{\text{YM}}^2 / (4\pi)$, and the ten- and five-dimensional Newton's constants are given by $16 \pi G_{10} = (2\pi)^7 g_{\text{s}}^2   l_{\text{s}}^8$ and $G_{5} = G_{10}/V_{S^5}=G_{10}/( \pi^3 L^5)$. Combining these equations yields  $N^2 = \pi L^3 /( 2 G_5)$  and hence the central charge is dual to $c = \Omega_3 L^3 / (16 \pi G_5)$, since $\Omega_3 = 2\pi^2$, consistent with the dictionary in \eqref{eq:twoCFTparameters}.\label{footnote:YMdof}} 
\begin{equation} \label{eq:dof}
N_{\text{dof}} \sim \frac{L^{d-2}}{G}  = \frac{L^{d-2}}{l_{\text{P}}^{d-2}}
\end{equation} 
 the particular combination of variations of $L$ and $G$ in the first law  becomes   $\delta N_{\text{dof}}/ N_{\text{dof}}$.
 We keep the notation $N_{\text{dof}}$ below, instead of   $C_T$ or $a^*$, because it is not entirely clear  which central charge should appear in the boundary first law. According to the definition of  \cite{Balasubramanian:2018uus} the differential entropy of the vacuum   is proportional to $C_T$ for three-dimensional CFTs, since this parameter appears in the second  shape derivative of the vacuum entanglement entropy of a ball \cite{Mezei:2014zla,Faulkner:2015csl}, but it is an open question whether the relation $S_{\text{diff}}
\sim C_T$ extends to higher-dimensional CFTs. Furthermore, it is not known  on which central charge     holographic complexity  depends, since it is not even   established what the precise definition of   complexity is for holographic CFTs.\footnote{The problem of extracting universal quantities from holographic complexity was also raised in \cite{Carmi:2016wjl}.} From \eqref{eq:twoCFTparameters} we see that this issue  does not matter  for Einstein gravity since $a^*\sim C_T$, but for CFTs dual to higher curvature gravity the   central charges are no longer proportional (see e.g. \cite{Myers:2010jv,Myers:2010tj}).\footnote{In   \cite{Bueno:2016gnv}  the first law of causal diamonds was extended to an arbitrary higher derivative theory of gravity.} It would be interesting to figure out which   central charge features in the CFT first law.  

Regarding the boundary dual of the term $k L$ in the first law, we note that the expression \eqref{eq:combvar} for the extrinsic trace generalizes to $k = \frac{d-2}{R}\sqrt{1 + (R/L)^2}$ in higher dimensions. The relation between $R$ and $\alpha$, i.e. $R=L \cot \alpha$, remains the same in higher dimensions, if we single out an angular coordinate on the boundary sphere:
$d \Omega_{d-2}^2 = d \theta^2 + \sin^2 \theta d \Omega^2_{d-3}$. Hence, we find in higher dimensions
\begin{equation} \label{eq:generalk}
	k L = \frac{d-2}{\cos \alpha}.
\end{equation}
Further, the matter Hamiltonian variation $\delta H_\zeta^{\text{mat}}$  in the gravitational first law can also  be replaced by   equation \eqref{eq:hamvarcft} in arbitrary dimensions, since the expression \eqref{eq:normckv}  for the norm of the conformal Killing vector is valid in any dimension, and the variational relation $\delta \Delta =L \delta m $ between  a point mass $m$ in AdS and the scaling dimension of $\Delta$ of the dual CFT  operator  generalizes to higher dimensions.  

Finally, inserting the holographic  dictionary \eqref{eq:hamvarcft}, \eqref{eq:entropyandcomplexity}, \eqref{eq:dof} and \eqref{eq:generalk}     into the bulk first law \eqref{eq:bulkfirstlawgendim}  yields the  boundary variational relation
\begin{equation} \label{eq:verygeneralfirstlaw}
 \delta   S_{\text{diff}} -   \frac{d-2}{\cos \alpha} \, \delta \mathcal C = \left ( S_{\text{diff}}-   \frac{d-2}{\cos \alpha}  \,  \mathcal C \right) \frac{\delta N_{\text{dof}}}{N_{\text{dof}}} - 2 \pi \left (  \frac{1}{\cos \alpha} -\frac{ \sin \alpha }{\cos \alpha} \right) \delta \Delta.
\end{equation}
Again the $\delta N_{\text{dof}}$ term can be easily computed from the left-hand side by using the proportionality of differential entropy and holographic complexity with the number of degrees of freedom 
$S_{\text{diff}} \sim N_{\text{dof}}$ and $ \mathcal C \sim N_{\text{dof}}$. These proportionalities follow from the holographic dictionary \eqref{eq:entropyandcomplexity}, together with the scaling of the area and volume in pure AdS with the curvature radius $A \sim L^{d-2}$ and $V \sim L^{d-1}.$

The CFT first law in higher dimensions can also be written as  
\begin{equation}
\begin{aligned} \label{eq:finalboundaryfirstlaw}
 \delta E  &=  T \delta S_{\text{diff}} + \nu \delta \mathcal C + \mu \delta N_{\text{dof}}, \qquad \text{with} \\
T&= -1, \quad \nu   =  \frac{d-2}{\cos \alpha} \, ,   \quad \mu   =  \frac{1}{N_{\text{dof}}}  \left ( S_{\text{diff}}-  \nu \,  \mathcal C \right). 
\end{aligned}
\end{equation}
The conjugate quantity $\mu$ to the number of degrees of freedom  is a chemical potential,   and the conjugate  quantity $\nu$ to the complexity is the energy cost of a changing the   complexity. The chemical potential is a positive, decreasing function of $\alpha$, which is maximal   at $\alpha=0$ and it vanishes at $\alpha=\pi/2$ for any $d$.  

\subsubsection{Comparison with extended first law of entanglement}\label{subsec:comparefirstlaws}

Finally, we  compare the CFT dual of the first law of causal diamonds to the   first law of entanglement for ball-shaped regions, which is the boundary dual  of the    first law for AdS-Rindler space \cite{Faulkner:2013ica}. 
The  extended   first law for   AdS-Rindler space, which includes a variation of the cosmological constant, is given by  \cite{Kastor:2014dra}
\begin{equation} \label{eq:firstlawadsrindler}
\delta \bar E_\xi = \frac{1}{4G} \left ( \delta A + \frac{\bar V_\chi}{2\pi} \delta \Lambda \right), \qquad \text{with} \qquad (d-2)A =   \frac{\bar V_\chi}{2\pi }2 \Lambda.
\end{equation}
Here $\chi$  is the boost   Killing vector of   the AdS-Rindler wedge, whose surface gravity is normalized to $\kappa = 2\pi$, and   $\xi = \lim_{r \to \infty} \chi$ is the   conformal Killing vector that preserves a diamond on the boundary.\footnote{See appendix \ref{app:kill} for a derivation of the boost Killing vector in several coordinate systems for AdS (there we set $\kappa=1$).}  Further, $A$ is the area of the bifurcation surface of the AdS-Rindler horizon, and $\bar V_\chi$ is the (background subtracted) thermodynamic volume of the codimension-one region between the bifurcation surface and the asymptotic boundary.   The bar on $\bar E_\xi$ and $\bar V_\chi$ indicates the implementation of  background subtraction in order to  cancel   divergences (see also \cite{Jacobson:2018ahi}).   By inserting     \eqref{eq:areavardiff} and $\delta \Lambda / \Lambda = -2 \delta L /L$  the AdS-Rindler first law can be repackaged as \cite{Caceres:2016xjz} 
\begin{equation} \label{eq:boundaryfirstlawent}
\delta \bar E_\xi = \delta \!  \left (  \frac{A}{4G}   \right) - \frac{A}{4G} \left ( (d-2) \frac{\delta L}{L} - \frac{\delta G}{G} \right) .
\end{equation}
We see that the same combination of variations of $L$ and $G$ appears here as in the first law of causal diamonds \eqref{eq:bulkfirstlawgendim}. The extended first law for AdS-Rindler is dual to an extended first  law of entanglement in the CFT,  where the extension involves a variation of the number of degrees of  freedom~\cite{Kastor:2014dra}
\begin{equation} \label{eq:CFTfirstlawextendedent}
\delta \langle K_\xi \rangle = \delta S  - \frac{S}{N_{\text{dof}}} \delta N_{\text{dof}}.
\end{equation} 
We identified $\delta \bar E_\xi  =\delta \langle K_\xi \rangle$, where $K_\xi$ is the modular Hamiltonian that generates the flow of $\xi$ on the boundary, and  employed the RT formula $S=A/(4G)$  to relate the area of the bifurcation surface of the AdS-Rindler horizon to   the  entanglement entropy    of the boundary region homologous to the bifurcation surface.  
If we fix the conformal frame at the asymptotic boundary of AdS such that the CFT lives in Minkowski space, then the   boundary of the AdS-Rindler wedge is the causal diamond of a  ball-shaped region in flat space and  $S$ is the vacuum entanglement entropy  of that ball with its complement~\cite{Casini:2011kv,Blanco:2013joa}. In this case we have $S\sim a^*$ and hence  $N_{\text{dof}}= a^*$ in \eqref{eq:CFTfirstlawextendedent}, where $a^*$ is the universal coefficient of the vacuum entanglement entropy of ball-shaped regions \cite{Myers:2010xs,Myers:2010tj}.\footnote{The parameter $a^*$ is not only the coefficient of   the universal (i.e. UV cutoff independent) contribution to the vacuum entanglement entropy of ball-shaped regions, but it is also   an overall coefficient to $S$. This follows at least from the holographic computations   in~\cite{Myers:2010tj}, where the authors found $S \sim a^*$ for the holographic entanglement entropy (i.e. Wald entropy) of the bifurcation surface of   AdS-Rindler horizons.}  
  In the CFT the  standard first law of entanglement   follows from the positivity of relative entropy~\cite{Blanco:2013joa}, and the extension to varying the number of degrees of freedom can be derived from the proportionality of $S$ with $a^*$, which implies $\delta_{a^*} S = (S/ a^* )\delta a^*$ (note   that $\delta_{a^*}  \langle K_\xi \rangle=~0$).
  The parameter~$a^*$ also appears in the extended first law for CFTs in  background geometries  which are Weyl equivalent to the causal diamond of a ball in flat space,  such as Rindler space or hyperbolic space times time, and for    CFTs  dual to  any  higher derivative theory of  gravity \cite{Kastor:2016bph,Caceres:2016xjz,Rosso:2020zkk}.

A striking difference between the  first law for AdS-Rindler space and the first law of causal diamonds is that only the latter  involves a variation of the proper volume, whereas the former does not. In the CFT this translates into the fact that the first law of entanglement does not include a variation of the holographic complexity, whereas the   first law  of differential entropy does.   It would be interesting to understand this fact purely from CFT considerations. This would require a derivation of the first law of differential entropy from first principles, in analogy to the derivations of the first law of entanglement in \cite{Blanco:2013joa,Faulkner:2013ica}.

Interestingly, the chemical potential $\mu_{\text{ent}} = - S/ N_{\text{dof}}$  that follows from \eqref{eq:CFTfirstlawextendedent} is quite similar to the chemical potential~\eqref{eq:finalboundaryfirstlaw} in the CFT dual of the first law of causal diamonds, except for the overall sign and the complexity term.  
 On the one hand, the   chemical potential in the extended first law of entanglement is negative since, at fixed entanglement entropy, the modular energy decreases if $N_{\text{dof}}$ increases. On the other hand, the   chemical potential in~\eqref{eq:finalboundaryfirstlaw} is positive since, at fixed differential entropy and complexity, the energy increases if $N_{\text{dof}}$  increases. Note that for both chemical potentials, at fixed energy, the entropy increases if $N_{\text{dof}}$   increases, as is usual in thermodynamics.

As far as we are aware,  it has not been fully appreciated in the literature     that the   two variations $(d-2)\delta L / L - \delta G / G$ combine  into a single variation $\delta N_{\text{dof}}/N_{\text{dof}}$, even if neither $L$ nor $G$ is kept fixed. 
In \cite{Caceres:2016xjz} the form \eqref{eq:boundaryfirstlawent} of the first law was known including the variations of $L$ and $G$, but $L$ was   kept fixed to arrive at \eqref{eq:CFTfirstlawextendedent}, which is unnecessary in our opinion.\footnote{The only exception is     AdS$_3$/CFT$_2$, in which case it was recognized in \cite{Caceres:2016xjz} that   $
\delta L / L - \delta G/ G = \delta c/c$. This relation did not appear in \cite{Kastor:2014dra}, since there $G$ was kept fixed in $3d$ gravity.} Their reason for keeping the AdS radius fixed is that   a variation of $L$ in the bulk entails both a variation of $N_{\text{dof}}$ and   of the  curvature radius  of the  boundary metric, in the   conformal frame where   the radius of the boundary cylinder is equal to the AdS radius \cite{Karch:2015rpa}. 
This means   if one is solely interested in  varying $N_{\text{dof}}$  on the boundary, while fixing the curvature radius, then $G$ should be varied and $L$  kept  fixed  in the bulk. These considerations are highly dependent though on the particular conformal frame where the curvature radius of the boundary cylinder is equal to $L$. However, the  extended  first law  of entanglement is valid for any conformal frame \cite{Rosso:2020zkk}. In a different conformal frame, where 
 the   curvature radius $L_{\text{bndy}}$ of the boundary metric is not related to the AdS radius $L$,   there is no need to fix~$L$, since there is a one-to-one correspondence between  varying both $G$ and $L$ in the bulk  and varying $N_{\text{dof}}$ on the boundary. We do note  that it is   possible, in general, to   vary  the boundary curvature radius, in addition to the number of degrees of freedom, but this will give an extra term on the right-hand side of the extended first law~\eqref{eq:CFTfirstlawextendedent}, given by  $-\delta_{L_{\text{bndy}}} S =- (\partial S^{\text{vac}}/\partial L_{\text{bndy}})\delta L_{\text{bndy}}$ (since $\delta_{L_{\text{bndy}}} \langle K_\xi\rangle=0$). So the  boundary curvature radius is kept fixed in the current form of the extended first law of entanglement. 

In contrast, in the   original work \cite{Kastor:2014dra}   in the  $AdS_5 \times S^5$ example,   it was realized that the term  involving  $\delta N_{\text{dof}}$ can be derived  from the combination of   variations of $L$ and $G_5$, the five-dimensional Newton's constant.\footnote{There is a typo  in   equation (3.39) of the published version of \cite{Kastor:2014dra}, since in comparison   to our equation~\eqref{eq:areavardiff} a factor of $1/(4G_5)$ is missing in the final term of their equation.}  But their actual derivation of the extended first law depends  on   a ten-dimensional perspective, in the sense that the ten-dimensional Newton's constant is kept fixed, such that the variation of $L$ is directly related to the variation of the rank $N$ of the gauge group or, equivalently, to $\delta N_{\text{dof}}$  (see  also   \cite{Johnson:2014yja,Dolan:2014cja} and  our footnote~\ref{footnote:YMdof}). In our view,  however, the ten-dimensional perspective   overcomplicates the derivation  and reference to the (five-dimensional) AdS space is sufficient to derive  the extended first law of entanglement.\footnote{Even if we take a ten-dimensional perspective, it is still unnecessary to fix the ten-dimensional Newton's constant. This is because  in the $AdS_5\times S^5$ example, the combination of variations of $L$ and $G_5$ becomes: $3 \delta L / L - \delta G_5 / G_5 =8\delta L / L - \delta G_{10} / G_{10} = \delta N^2 / N^2$, where we used $G_5 = G_{10} / (\pi^3 L^5)$ and $N^2 = \pi^4L^8/ (2 G_{10})$.} The derivation is arguably more transparent and   straightforward if the   extended AdS-Rindler first law is written as \eqref{eq:boundaryfirstlawent},  in terms of     the AdS radius and (five-dimensional) Newton's constant. A simple, but crucial step in the derivation  of \eqref{eq:CFTfirstlawextendedent} from \eqref{eq:boundaryfirstlawent} is to realize that  $(d-2)\delta L / L - \delta G / G =  \delta (L^{d-2}/ G) /(L^{d-2}/G)$, which seems to have been overlooked  for $d>3$ in previous work.


\section{Conclusion and outlook}\label{sec:conclude}

In this paper we found a   first law-like   relation in     CFT$_2$ which is dual to the first law of causal diamonds in AdS$_3$. Using a fixed coordinate approach, we    obtained the bulk first law for the specific example of a disk  inside a time slice of AdS$_3$, where the perturbed geometry is AdS$_3$ with a point mass. This complements the derivation of the first law of causal diamonds from the  covariant phase space formalism in~\cite{Jacobson:2018ahi}.
In our search for the boundary first law  we  considered three types of  independent variations:   a change of state, a change of subregion size and a change of the central charge of the~CFT. The resulting boundary first law relates the variations of the   differential entropy, holographic complexity, the central charge and the scaling dimension of the perturbed state. Remarkably,  there is no term proportional to the  variation of $\alpha$ in the first law, although the variations of the differential entropy and complexity separately do depend on the interval size. This is because the $\alpha$ variation cancels between the two terms   due to a natural choice of relative coefficient.  In AdS  it is related to the vanishing of the combination $\delta_\chi A - k \delta_\chi V$ for the maximal slice of   spherical causal diamonds   for variations induced by a diffeomorphism  $\chi$. 

We emphasize that the first law of differential entropy is a new relation in holographic two-dimensional CFTs. A similar first law in quantum information theory has   been widely studied in the AdS/CFT literature, the `first law of entanglement',  which relates the variation of the entanglement entropy to the variation of the expectation value of the modular Hamiltonian.  
The first law of entanglement has many applications, both in quantum field theory and in AdS/CFT, and perhaps the first law of differential entropy could find a similar wide applicability. Both first laws are not standard thermodynamic relations, since the entropy and energy that appear in both variational relations  are not standard thermodynamic quantities (except for   special subsystems, such as  spherical subregions   in a global vacuum CFT, which are thermodynamic systems). The first law of entanglement entropy is a quantum generalization of the first law of thermodynamics for density matrices. Similarly,   the first law of differential entropy  can perhaps be formulated as  a        variational relation in   quantum information theory, although this requires further study since differential entropy has not been investigated   for non-holographic CFTs (or  QFTs).

There are some similarities between the two   first laws, e.g. they can be extended by adding a chemical potential term associated to the variation of the central charge.   At least in part, the first law of differential entropy can be obtained from the first law of entanglement, since differential entropy is a derived notation from entanglement entropy. However, we would also like to mention three differences. Firstly, an obvious difference is that differential entropy is a global property of the CFT associated to  time strips,  whereas entanglement entropy is associated to   spatial subregions.
Secondly, a striking difference  between   our first law of differential entropy and the first law of entanglement is that the latter  does not involve the variation of  complexity. Assuming the `complexity=volume' proposal,    this corresponds in AdS to the fact that the volume variation   is absent in the   first law of AdS-Rindler space, which is dual to the first law of entanglement, whereas it does appear in the first law of causal diamonds, dual to the first law of differential entropy. The reason for this is that the variation of the  gravitational Hamiltonian vanishes  along the flow of   the boost Killing vector of AdS-Rindler space, whereas it is nonvanishing and proportional to the volume variation of the maximal slice along the flow of the diamond conformal Killing vector. Finally, another difference is that the formal `temperature' in the first law of differential entropy is negative, if the internal energy and differential entropy are positively defined, whereas it is positive   in the first law of entanglement. 

Arguably  the most interesting application of  the first law of entanglement in AdS/CFT is the derivation of the linearized gravitational equations \cite{Faulkner:2013ica}, assuming the Ryu-Takayangi formula for holographic entanglement entropy (see also \cite{Faulkner:2017tkh} for a derivation of the   second-order nonlinear equations).  In contrast, 
in previous work the first law of causal diamonds has   been used as a stepping stone to derive the   nonlinear   Einstein equation  from a \emph{local} thermodynamic argument, either by reinterpreting the (quantum corrected) first law as the stationarity of the generalized entropy at fixed volume in small local causal diamonds everywhere in spacetime (called `entanglement equilibrium')~\cite{Jacobson:2015hqa} or as the stationarity of free conformal   energy of small diamonds \cite{Jacobson:2018ahi,Jacobson:2019gco}. With the new dictionary between the bulk and boundary first law established in this paper, it might be possible to obtain   the entanglement equilibrium hypothesis (or the stationarity of free conformal energy) in the bulk from the    first law of differential entropy on the boundary. Since   entanglement equilibrium is the input in Jacobson's derivation of the nonlinear Einstein equation \cite{Jacobson:2015hqa}, perhaps one can even reformulate his derivation in (a local version of) AdS/CFT.\footnote{We thank  Maulik Parikh for discussions on this point.} To be more precise, suppose one assumes the first law of differential entropy in a cutoff CFT, which lives on the boundary of a small ball-shaped region inside a timeslice of AdS. Assuming the AdS/CFT dictionary for the relevant quantities  (such as differential entropy and holographic complexity) one can obtain  the first law of causal diamonds in the small AdS ball from  the   first law of differential entropy. The (quantum corrected) bulk first law should  then be connected to the entanglement equilibrium hypothesis, which could   be used to derive the nonlinear Einstein equation. It would be interesting to   work this out in     detail. In particular, it would   require a better understanding of holography at sub-AdS scales (see,~however,~ e.g. \cite{Heemskerk:2009pn,vanLeuven:2018pwv} for some progress in this direction). 

Naturally, there are quite a few other avenues to pursue for future investigations. Although we have proposed   a higher dimensional generalization of the first law for differential entropy, inspired by the higher dimensional version of the first law of causal diamonds,  a more detailed study is required to check its validity  in the CFT. A promising future direction would be to apply the first law of entanglement to the higher dimensional definition of differential entropy, in terms of the shape derivatives of the entanglement entropy, for the vacuum state \cite{Balasubramanian:2018uus}. Other natural generalizations of the first law of differential entropy are for higher-order corrections to the variations, and for CFT setups dual to off-center circular bulk disks and more general bulk subregions of arbitrary shapes. 
Using the fixed coordinate approach this seems feasible, and one might learn how universal the first law of differential entropy is and how it encodes the shape dependence of the bulk region.  
As for   off-center circular disks, the first law of causal diamonds already applies in this setup, since it is a covariant relation, but our current boundary first law does not hold, since this geometric setup corresponds to a boundary interval size $\alpha$ which depends  on the angular coordinate~$\theta$,   whereas we considered constant $\alpha$ in this paper.
 We expect, however, that the   first law of differential entropy can be appropriately generalized to off-center bulk disks, which   is particularly interesting since the linearized Einstein equation around the AdS background can be derived from the   bulk first law for all circular disks and their associated diamonds. 


 Furthermore,  it would be interesting to generalize the first law of differential entropy to CFTs which are dual to higher derivative gravity. On the gravitational side  a generalization of the first law of causal diamonds to higher order gravities was already derived in \cite{Bueno:2016gnv}. A more non-trivial task would be to understand to what extent the first law of differential entropy is   applicable to a general non-holographic CFT. Even though our CFT derivation is partly based on the existence of a holographic bulk dual (especially regarding the `complexity=volume' dictionary), the quantities appearing in our first law such as differential entropy, complexity  and operator dimension   can all be defined in generic CFTs. 

Another important direction to pursue would be to investigate to what extent  the boundary first law applies to    general excited states, other than excited states   dual to a classical point particle in AdS. For instance, a perturbative excited state can  be prepared using the path integral,    or it can be  created by acting with a  local  conformal transformation on the CFT vacuum. The entanglement entropy has been studied for path integral states in \cite{Rosenhaus:2014woa,Rosenhaus:2014zza} and for generic vacuum excitations in \cite{Holzhey:1994we,deBoer:2016pqk}. Equivalently, one can compute   the differential entropy for these   states and take the difference with the differential entropy for the vacuum state.  This could lead, for example,   either to the inclusion of $1/c$   corrections in the first law or to  a higher-order variational relation for perturbations that create a black hole in the bulk. The $1/c$   corrections correspond to perturbative quantum corrections in the bulk, due to quantum fields living in a fixed AdS background. The bulk first law has already been   extended to this semiclassical regime in  \cite{Jacobson:2018ahi}, but for future work it would be especially interesting to find the dual CFT first law    including leading order $1/c$ corrections for perturbative excited states (see e.g.  \cite{Belin:2018juv,Agon:2020fqs}).


Finally, already within our CFT first law, there are a couple of aspects that require further study. Although we have argued that the finite bulk volume is dual to the  boundary complexity in a   cutoff CFT, this proposal needs a better understanding \cite{Chen:2020nlj}. For example, it would be interesting to  study the relation between \emph{finite} bulk volumes and circuit complexity in quantum field theory, developed in \cite{Jefferson:2017sdb,Chapman:2017rqy}. 
Alternatively, it is tempting to suggest that the bulk volume is a measure of the complexity of the boundary mixed state, which is dual to the bulk state $\rho_{D, \text{bulk}}$ obtained by tracing out the quantum gravitational degrees of freedom living in the complementary  region of the disk~$D$. A similar interpretation   has been put forward for differential entropy, namely as the entanglement entropy of  $\rho_{D, \text{bulk}}$ \cite{Balasubramanian:2013rqa}.    Further, the internal energy in the first law  does not yet have a    covariant CFT definition. We know that the energy is dual to the matter Hamiltonian in the bulk generating evolution along the diamond conformal Killing flow. Perhaps such bulk conformal Killing flows could be related to the conformal  Killing flows of causal diamonds on the boundary, somewhat along the lines of how differential entropy is related to entanglement entropy.  Thus, it is worth investigating whether the CFT energy is a (possibly complicated) function of the  conformal   Killing vector which preserves a boundary diamond.

\bigskip
\bigskip
\goodbreak
\centerline{\bf Acknowledgements}
\noindent
MV is grateful to Irfan Ilgin and Erik Verlinde for  their  initial collaboration at the early stages of this project, and for sharing    notes   on the variation  of differential entropy. 
Our section \ref{sec:vardiffentropy} is largely based on their  computations, which are also  presented in Chapter~5 of Ilgin's   PhD thesis \cite{Ilgin:2019jhw}.  
We would  further like to thank  Jan de Boer, Alex Belin, Bartek Czech, Ted Jacobson, Alex Kieft and Christian Northe  for useful discussions, and Ted Jacobson and Juan Pedraza for helpful comments on the draft.  DS acknowledges the Institute for Advanced Study at Tsinghua University for support and hospitality during the course of this project.  We thank the organizers of the 6th general meeting of the   National Centre of Competence in Research (NCCR)  SwissMAP    in Villars-sur-Ollon in Switzerland, where this project was reinitiated. The work of MV is funded  by the Republic and canton of Geneva and by   the Swiss National Science Foundation, through Project Grants 200020$\textunderscore$182513 and the NCCR 51NF40-141869 The Mathematics of Physics (SwissMAP).

\appendix

\section{Embedding formalism and coordinate systems for   AdS$_3$ geometries}\label{app:embed}

 In this appendix we review the   embedding formalism for locally AdS$_3$ geometries, in particular for pure AdS$_3$ and  AdS$_3$  with a conical defect. The conical defect   spacetime is a quotient space of   AdS$_3$, and can therefore be obtained from the same embedding space as~AdS$_3$. The embedding formalism is useful for   computing the length of  geodesics  (see appendix \ref{app:embed_chord_crofton}) and for deriving the  Killing vector fields of these spacetimes (see appendix~\ref{app:kill}). 

Locally three-dimensional AdS spaces can be embedded in $\mathbb R^{2,2}$, on which the coordinates are  ($T^1, T^2, X^1, X^2$) and the metric is
\begin{equation} \label{eq:embeddingmetric}
ds^2 = - (dT^1)^2 - (dT^2)^2 + (dX^1)^2 + (dX^2)^2.
\end{equation}
AdS$_3$ is realised as a hyperboloid in this embedding space
\begin{equation} \label{eq:hyperboloid}
 - (T^1)^2 - (T^2)^2  + (X^1)^2 + (X^2)^2= - L^2,
\end{equation}
where $L$ is the curvature radius of AdS. Note that the isometry group of AdS$_3$ is by construction $SO(2, 2)$,
since it corresponds to the symmetry group that preserves the hyperboloid in $\mathbb R^{2,2}$. The embedding space naturally induces a metric on the hyperboloid through \eqref{eq:embeddingmetric}.
Below we present various embedding coordinates and their corresponding induced metrics  for pure AdS$_3$ and conical AdS$_3$.

\subsection{Pure AdS}
Embedding coordinates which cover the entire AdS$_3$ manifold are 
\begin{equation}
\begin{aligned} \label{eq:embads2}
T^1 &=\sqrt{r^2 + L^2}  \cos  (t/L)  \qquad&X^1 &=  r \cos   \phi \\
T^2 &= \sqrt{r^2 + L^2}  \sin (t/L)   \qquad & X^2 &=r \sin  \phi ,
\end{aligned} 
\end{equation}
with $0 \le t  < 2 \pi L$, $0\le r < \infty$, and $0 \le \phi < 2 \pi$. However, in order to avoid closed timelike curves, we will ignore the periodicity of the time coordinate  and declare that it ranges from $- \infty$ to $\infty$ (which is formally called the covering space of AdS).  For these embedding coordinates the induced metric \eqref{eq:embeddingmetric} on the hyperboloid is
\begin{equation} \label{eq:pureadsmetric3}
ds^2 = - \left (  1 + \frac{r^2}{L^2} \right) dt^2 + \left ( 1+ \frac{r^2}{L^2} \right)^{-1} \!\!\!dr^2 + r^2 d \phi^2 . 
\end{equation}
This is the main coordinate system of the present paper.

Further, in terms of  the  dimensionless  radial coordinate $z =\frac{L}{r} (  - 1 + \sqrt{1 + (r/L)^2}) $ and time coordinate $\tau = t/L$ the embedding coordinates take the form  
\begin{equation}
\begin{aligned} \label{eq:embads3}
T^1 &= L \frac{1+z^2}{1-z^2}  \cos \tau   \qquad&X^1 &= L \frac{2 z}{1 - z^2} \cos \phi   \\
T^2 &= L \frac{1+z^2}{1-z^2} \sin \tau \qquad& X^2 &= L \frac{2 z}{1 - z^2} \sin \phi  ,
\end{aligned} 
\end{equation}
\noindent where $0 \le z < 1$.   
With this parametrization the induced metric   becomes
\begin{align} \label{eq:poincaredisk}
ds^2  =      L^2 \left [ - \left( \frac{1+z^2}{1-z^2}\right)^2 d\tau^2 + \frac{4 (dz^2 + z^2 d \phi^2)}{(1-z^2)^2}  \right].
\end{align}
At constant $\tau$ the metric between brackets describes the Poincar\'e disk, which is a stereographic projection of the two-dimensional hyperbolic plane. Because of the cylindrical shape of AdS in these coordinates, they are sometimes called sausage coordinates.  The advantage of this coordinate system is that the asymptotic timelike boundary lies at a finite coordinate distance $z=1$. 

Embedding coordinates which cover only part of the AdS manifold are
\begin{equation}
\begin{aligned} \label{eq:embads4}
T^1 &=   \sqrt{\varrho^2+ L^2} \cosh (u/L)  \qquad &X^1 &=    \varrho \cosh (\sigma /L) \\
T^2 &=\varrho\sinh (\sigma /L)  \qquad& X^2 &=  \sqrt{\varrho^2 + L^2}  \sinh (u/L).
\end{aligned} 
\end{equation}
This leads to the induced metric
 \begin{equation} \label{eq:adsrindlermetric}
 ds^2 = -\frac{\varrho^2}{L^2} d \sigma^2   + \left( 1+ \frac{ \varrho^2}{L^2}   \right)^{-1}  \!\!\!d   \varrho^2   + \left( 1+ \frac{\varrho^2}{L^2}   \right)   du^2 ,
 \end{equation}
with $-\infty < \sigma < \infty$, $0\le \varrho < \infty$,  and $- \infty < u < \infty$. This metric describes the Rindler wedge of AdS space. The dimensionful time $\sigma$ is the proper time of Rindler observers in AdS.  The AdS-Rindler horizon is located at $ \varrho =0$, and the conformal boundary at $\varrho = \infty$. 
 
Embedding coordinates which divide the AdS hyperboloid  into two charts, $T^1>-X^1$ ($\mathrm{z} > 0$) and $T^1 < -X^1$ ($\mathrm{z} < 0$), are
\begin{equation}
\begin{aligned}   \label{eq:embads5}
T^1 &= \frac{1}{2\mathrm{z}}  \left (  L^2 - \mathrm{t}^2 +\mathrm{x}^2 + \mathrm{z}^2 \right)    \qquad &X^1 &=    \frac{1}{2\mathrm{z}}  \left (  L^2 + \mathrm{t}^2 -\mathrm{x}^2 -\mathrm{z}^2 \right)  \\
T^2 &=   L \mathrm{t} / \mathrm{z}  \qquad& X^2 &=   L \mathrm{x} / \mathrm{z} .
\end{aligned} 
\end{equation}
This brings the induced metric in Poincar\'{e} form
\begin{equation}
ds^2 = \frac{L^2}{\mathrm{z}^2} \left (  -d \mathrm{t}^2 + d\mathrm{x}^2 + d\mathrm{z}^2  \right) .
\end{equation}
The coordinates $\mathrm{t}$ and $\mathrm{x}$ range from $-\infty$ to $\infty$ and the conformally flat boundary is   at~$\mathrm{z}=0$.

Finally, other embedding coordinates  which divide pure AdS$_3$ into two   charts are
\begin{equation}  
\begin{aligned} \label{eq:embads10}
T^1 &=  L \frac{\cos\hat \tau }{\sin \hat \phi \sin \psi}    \qquad &X^1 &= L \frac{ \cos \hat \phi}{\sin \hat\phi \sin \psi}   \\
T^2 &=  L \frac{\sin \hat\tau }{\sin\hat \phi \sin \psi}  \qquad& X^2 &=  L \cot \psi  .
\end{aligned} 
\end{equation}
 The induced metric is conformal to   the metric on $\mathbb R \times S^2$
\begin{equation} \label{eq:adsspherical}
ds^2 = \left (\frac{L }{\sin  \hat\phi \sin  \psi } \right)^2 \left [   -d \hat\tau^2 + d{\hat \phi}^2 + \sin^2\hat \phi  d \psi^2  \right],
\end{equation}
where $- \infty <\hat \tau < \infty$, $0 \le\hat \phi <2   \pi$ and $0 \le \psi \le\pi $. 
 By removing the conformal factor $1/(\sin \hat \phi  \sin \psi)^2$  and taking the asymptotic limit $\psi\to0$ or~$\pi$ (depending on the chart), the metric on the conformal boundary becomes that of a Lorentzian cylinder whose radius is equal to the AdS scale: $d s^2_{\text{bndy}} = L^2 [- d \hat\tau^2 + d\hat\phi^2]$.

\subsection{Conical AdS}

For AdS$_3$ with a conical defect a simple set of embedding coordinates is 
\begin{equation}
\begin{aligned}
T^1 &= L \cosh \rho \cos(\gamma \tau)  \qquad&X^1 &=   L \sinh \rho \cos (\gamma \phi)   \\
T^2 &= L \cosh \rho \sin (\gamma \tau) \qquad& X^2 &=L \sinh \rho \sin (\gamma \phi) . 
\end{aligned} 
\end{equation}
The  conical defect parameter $\gamma$ ranges from 1 (pure AdS) to 0 (massless BTZ). 
Note that these coordinates still satisfy the hyperboloid equation \eqref{eq:hyperboloid}, and hence the conical spacetime is locally AdS$_3$. The  induced metric is
\begin{equation} \label{eq:conicalglobalmetric}
ds^2 = L^2 \left [ - \gamma^2 \cosh^2\!\rho\,  d\tau^2 +  d\rho^2  + \gamma^2 \sinh^2\!\rho\,    d\phi^2 \right] .
\end{equation}
Under the coordinate transformation $\tau ' = \gamma \tau$ and $\phi' = \gamma \phi$ the metric turns into that  of pure AdS, but with a different range for the angular coordinate $0 \le \phi' < 2\pi \gamma.$
Thus, in these coordinates conical AdS$_3$ is represented as an infinite solid cylinder,   foliated by hyperbolic planes with   deficit angle $2 \pi (1- \gamma)$.

Further, in terms of  the dimensionful radial coordinate $r=L \gamma \sinh \rho$ and time coordinate $t = \tau L$  the embedding coordinates read
\begin{equation}
\begin{aligned} \label{eq:embcoordconical2}
T^1 &= \sqrt{(r/\gamma)^2 + L^2} \cos(\gamma t/L)  \qquad &X^1 &= (r/\gamma) \cos (\gamma \phi)   \\
T^2 &= \sqrt{(r/\gamma)^2 + L^2} \sin (\gamma t/L) \qquad& X^2 &=(r/\gamma)  \sin (\gamma \phi)  .
\end{aligned} 
\end{equation}
and the induced metric is
\begin{equation}
ds^2 = - \left (  \gamma^2 + \frac{r^2}{L^2} \right) dt^2 + \left (  \gamma^2 + \frac{r^2}{L^2} \right)^{-1} dr^2 + r^2 d \phi^2 .
\end{equation}
This is the analog of   \eqref{eq:embads2} for conical AdS.

Finally, in  terms of the  radial   coordinate $z$,     defined by  
\begin{align}
z^\gamma = \tanh(\rho/2)  =  \frac{- \gamma + \sqrt{\gamma^2 + (r/L)^2}}{r/L},
\end{align}
the embedding coordinates are
\begin{equation}
\begin{aligned}
T^1 &= L \frac{1 + z^{2\gamma}}{1 - z^{2\gamma}} \cos(\gamma \tau)  \qquad&X^1 &= L \frac{2  z   }{z^{1-\gamma}(1-z^{2\gamma})} \cos (\gamma \phi)   \\
T^2 &= L \frac{1 + z^{2\gamma}}{1 - z^{2\gamma}} \sin(\gamma \tau) \qquad& X^2 &=L \frac{2  z   }{z^{1-\gamma}(1-z^{2\gamma})}  \sin (\gamma \phi)  .
\end{aligned} 
\end{equation}
In terms of these coordinates the induced metric takes the form
\begin{align} 
ds^2 &= L^2 \left [ - \gamma^2 \left ( \frac{1+z^{2\gamma}}{1-z^{2\gamma}} \right)^2 d\tau^2 + \frac{4 \gamma^2 \left (  dz^2 + z^2 d \phi^2 \right) }{z^{2(1-\gamma)} (1 - z^{2\gamma})^2}  \right]    \\
&= L^2 \left [  - \gamma^2 \coth^2(\gamma \ln z ) \, d\tau^2 + \frac{\gamma^2(dz^2 + z^2 d\phi^2)}{z^2 \sinh^2 ( \gamma \ln z )}  \right]  \label{eq:DJsystem}.
\end{align}
This is  the Deser-Jackiw coordinate system for a point mass in AdS$_3$ \cite{Deser:1983nh}.\footnote{The main case of interest of \cite{Deser:1983nh} was  point particles in dS$_3$. The metric \eqref{eq:DJsystem}  for a point mass in AdS$_3$ follows, however,  from inserting  the shift function and conformal factor for AdS, given by equation~(3.2),  into the static metric ansatz (2.5) in the Deser-Jackiw paper. Moreover, the master function $V(z)$ and coordinate $\varsigma$ in their (3.2) are given, respectively, by their equations (3.5a) and (3.5c) for the simplest case of a single point particle in AdS$_3$. Their notation corresponds to ours as follows:  $\sqrt{\epsilon}c \to \gamma$,   $r \to z$,  $t \to L  c \tau$.}

\section{Geodesics in AdS$_3$ geometries}\label{app:embed_chord_crofton}

In this appendix we compute the  `chord' length of spacelike geodesics -- which is simply the geodesic distance -- in conical  AdS$_3$ geometries using two different approaches. On the one hand, we   describe the geodesics in embedding space and express the geodesic distance between two bulk points in terms of the inner product of embedding coordinates (see~e.g.~\cite{Bengtsson}). On the other hand, we compute the chord length from an integral in kinematic space over the Crofton form, following the approach in \cite{Czech:2015qta}.  We start the appendix by deriving the geodesic equation  for  AdS$_3$ with a conical defect.

\subsection{Geodesic equation for conical AdS} \label{app:geoeq}

\begin{figure}
\begin{center}
\includegraphics[width=0.5\textwidth]{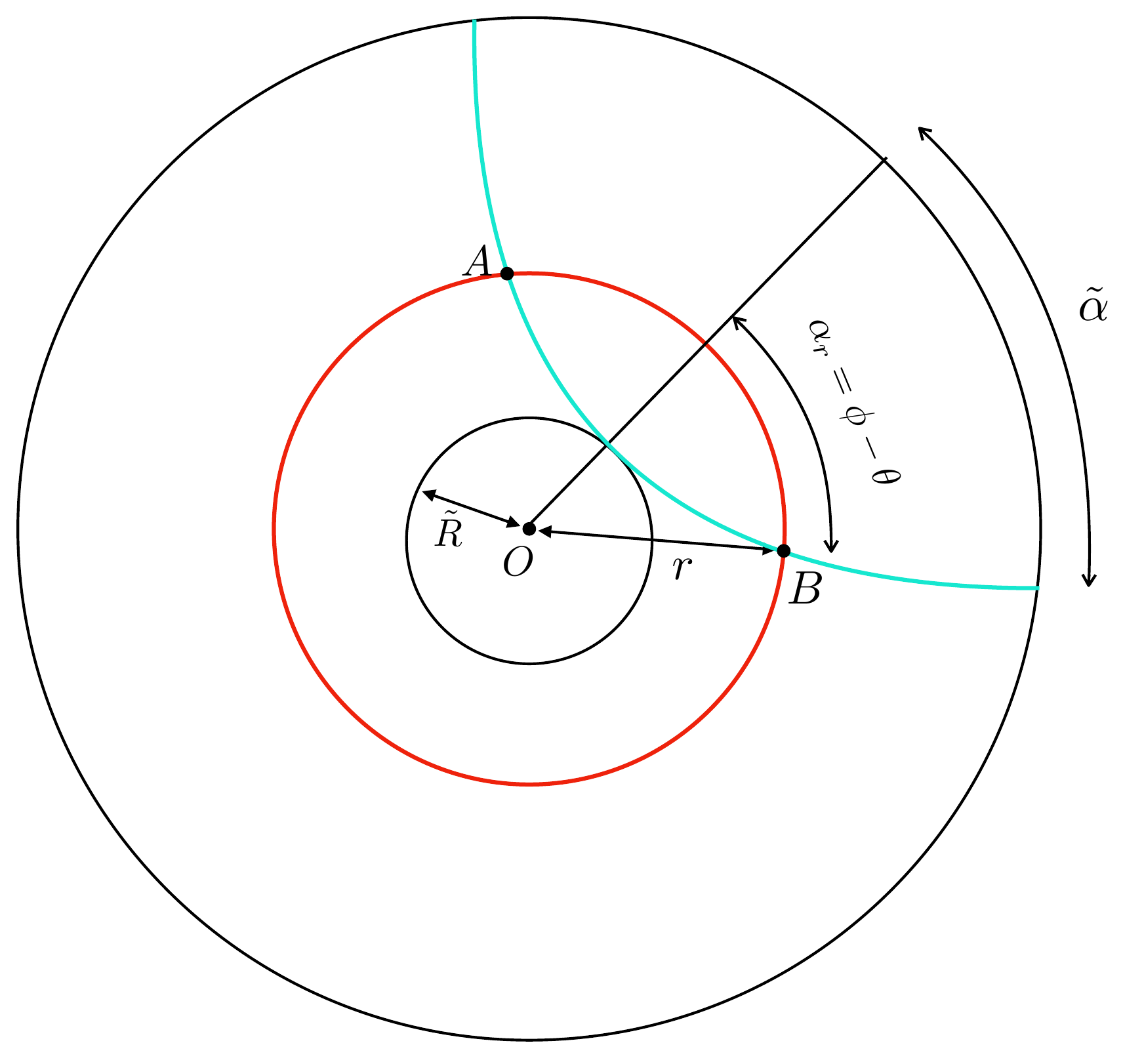}
\end{center}
\caption{A spacelike geodesic (in turquoise) at a time slice in conical or pure AdS   centered at boundary angular coordinate $\theta$ and with  a boundary opening angle $\tilde{\alpha}$. The geodesic is by construction tangent to a bulk disk of radius $\tilde{R}$. The geodesic distance between points $A$ and $B$, which both lie on a circle (in red) of radius $r$, is computed in appendix \ref{app:geoembed}.
\label{fig:bulkgeodesics}}
\end{figure}

Consider a spacelike geodesic  at a constant time slice in conical AdS \eqref{eq:conicalmetric2}, which is  centered at the boundary angular coordinate $\theta$ and which has     boundary opening angle $\tilde{\alpha}$ (see figure~\ref{fig:bulkgeodesics} for notational clarifications).  The geodesic distance functional  is 
\begin{equation}
I = \int ds = \int dr \sqrt{ \frac{1}{ \gamma^2 +  (r/L)^2 }+ r^2 \left ( \frac{d\phi}{dr}  \right)^2  } .
\end{equation}
Minimizing the geodesic distance yields
\begin{equation} \label{eq:minimalgeodesic1}
\frac{r^2 \dot \phi}{\sqrt{\left ( \gamma^2 + (r/L)^2  \right)^{-1} + r^2 \dot \phi^2} }=\text{constant},
\end{equation}
where the dot denotes differentiation with respect to the radial coordinate $r$. The constant is fixed by noting that the geodesic has a  turning point, at $r=\tilde{R}$, where the derivative diverges, i.e. $\dot \phi \to \infty$ as $r\to \tilde{R}$. Plugging the  resulting constant, $\tilde{R}$,   into \eqref{eq:minimalgeodesic1} leads to  the following differential equation
\begin{equation}
\frac{d\phi}{dr} = \frac{\tilde{R} L}{r\sqrt{( r^2 - \tilde{R}^2 ) \left ( r^2 + \gamma^2 L^2 \right)}} \, . 
\end{equation}
By integrating this equation between the turning point   $(r=\tilde{R}, \phi= \theta)$ and an arbitrary point on the geodesic $( r, \theta+ \alpha_r)$, where $\alpha_r$ is the opening angle in the bulk, we arrive at the following expression for the geodesics 
\begin{equation} \label{eq:geodesic1}
\tan^2(\gamma   \alpha_r) = \frac{ r^2  / \tilde{R}^2  - 1}{ r^2/(\gamma L)^2 +1   } \, . 
\end{equation}
Note that in  the limit $r \to \infty$ the bulk opening angle $\alpha_r$ becomes   the boundary opening angle $\tilde{\alpha}$ (both range from $0$ to $\pi/2$). Hence, by taking the limit $r \to \infty$ of the equation above, we find a relation between the radius of the disk and the boundary opening angle
\begin{equation} \label{eq:radiuscon2}
\tilde{R} = L \gamma \cot (\gamma \tilde{\alpha}) \, . 
\end{equation}
In terms of $\tilde{\alpha}$, instead of $\tilde{R}$, the geodesic equation reads\footnote{In terms of the other global coordinates \eqref{eq:conicalglobalmetric} the geodesic equation is given by  (with $\alpha_\rho := \phi - \theta$)
$$
\tan^2(\gamma   \alpha_\rho) =   \frac{\tanh^2  (\rho) }{\cos^2  (\gamma \tilde \alpha)} - 1    \qquad \text{or} \qquad \tanh( \rho) \cos (\gamma \alpha_\rho) = \cos (\gamma \tilde \alpha) . 
$$}
\begin{equation} \label{eq:finalgeodesic}
\tan^2(\gamma   \alpha_r) = \frac{r^2 \tan^2 ( \gamma\tilde{\alpha}) - \gamma^2 L^2}{r^2 + \gamma^2 L^2} \qquad \text{or} \qquad \frac{r }{\sqrt{r^2 + \gamma^2 L^2   }} \cos (\gamma      \alpha_r) = \cos (\gamma   \tilde{\alpha}).
\end{equation}
These results agree with the expressions in \cite{Balasubramanian:2014sra}, which were derived by rescaling the coordinates in the  pure AdS case. According to \eqref{eq:rescalingcoord}  the coordinate transformation from pure AdS to conical AdS is  $\phi'= \phi \gamma $ and $r' = r/\gamma$. Applying this transformation to the geodesic equation in pure AdS gives the required results.

In the main body of this paper we have used several special cases of this general set-up (also in pure AdS, with $\gamma =1$), such as $r=R>\tilde{R}$ (with $\alpha_r\to \alpha_R$) in equations  \eqref{eq:modRrrel} and \eqref{eq:geoequationconical},  and   $r=R=\tilde{R}$ (in which case $\tilde{\alpha}\to \alpha$ and $\alpha_R\to 0$) in equations \eqref{eq:radiusconalpha} and   \eqref{eq:radiuscon}.

\subsection{Chord length} \label{app:geoembed}

\textit{From the embedding formalism:} Given the embedding coordinates defined in appendix~\ref{app:embed}, one can derive an expression for the geodesic length. 
It is convenient to combine the embedding coordinates into a vector $X^\alpha = (T^1, T^2, X^1, X^2)$ and use the following notation for the inner product $X^2 = g_{\alpha\beta}X^\alpha X^\beta $ and $X_1 \cdot X_2 = g_{\alpha \beta}X_1^\alpha X_{2}^\beta,$ where  the embedding metric is given by \eqref{eq:embeddingmetric}. 

The Lagrangian in embedding space which describes geodesics in AdS is 
\begin{equation}
	\mathcal L = \frac{1}{2}   \dot X^2 + \mu (X^2 + L^2) ,
\end{equation}
where the dot indicates differentiation with respect to the proper distance $s$, and a Lagrange multiplier $\mu$ is introduced to ensure that the geodesics are confined to the hyperboloid~\eqref{eq:hyperboloid}. The Euler-Lagrange equation is $\ddot{X}^\alpha = 2 \mu X^\alpha.$ Combining this with the hyperboloid constraint $X^2 = - L^2$ yields an expression for the Lagrange multiplier $\mu =  \dot X^2 / (2L^2).$ Therefore, geodesics in AdS satisfy a simple equation in embedding space
\begin{equation}
	L^2 \ddot{X}^\alpha = \dot{X}^2 X^\alpha.
\end{equation}
The general solution for  spacelike geodesics ($\dot X^2 =1 $) is
\begin{equation}
	X^\alpha (s)= m^\alpha e^{s/L} + n^\alpha e^{-s/L},
\end{equation}
where $m^\alpha$ and $n^\alpha$ are constant vector that obey $m^2 =n^2 =0$ and $2 m \cdot n =-L^2.$ By taking the inner product between two points $X(s_1)$ and $X(s_2)$, we arrive at the following formula for the geodesic distance or chord length   $\lambda := s_2 - s_1$,
\begin{equation}\label{eq:mainforgeodesic}
	L^2\cosh\left(\frac{\lambda}{L}\right)= - X(s_1)\cdot X(s_2). 
\end{equation}
We can now compute the chord length for pure AdS and conical AdS by inserting   specific embedding coordinates. In the standard coordinates $(t,r,\phi)$ for pure AdS the geodesic length between the two bulk points $A=(0, r, \theta - \alpha_r)$ and $B=(0, r, \theta + \alpha_r)$    is (see~figure~\ref{fig:bulkgeodesics}) 
\begin{equation}
\begin{aligned}\label{eq:adsgeodesicembed}
	 \lambda_{\text{vac}}&=L\,\text{arccosh} \left[1+2\,(r/L)^2\,\sin^2(\alpha_r)\right] =2 L \arcsinh [(r/L) \sin (\alpha_r)]   \\
	&= 2L \arctanh \left [ \frac{(r/L)\sin( \alpha_r)}{\sqrt{1+(r/L)^2 \sin^2( \alpha_r)}}\right]\,. 
\end{aligned}
\end{equation}
A similar calculation  using the embedding coordinates \eqref{eq:embcoordconical2} for conical AdS shows 
\begin{equation}
\begin{aligned}\label{eq:conicalgeo}
	 \lambda_{\text{con}} 
	&=L\,\text{arccosh} \left[1+2\,r^2 / (\gamma L)^2\,\sin^2(\gamma\alpha_r)\right] = 2 L \arcsinh \left [  r/(\gamma L) \sin (\gamma \alpha_r)  \right]   \\
	&= 2 L \arctanh \left [ \frac{r/(\gamma L)\sin (\gamma \alpha_r)}{\sqrt{1+r^2/(\gamma L)^2 \sin^2 (\gamma \alpha_r)}}   \right] .  
\end{aligned}
\end{equation}
In other words, the conical AdS result for a disk of radius $r$ is obtained from the pure AdS case  by replacing $r\to r/ \gamma$ and $\alpha_r\to \gamma\alpha_r$.\\

\noindent \emph{From the kinematic space formalism:}
\begin{figure}
\begin{center}
\includegraphics[width=1\textwidth]{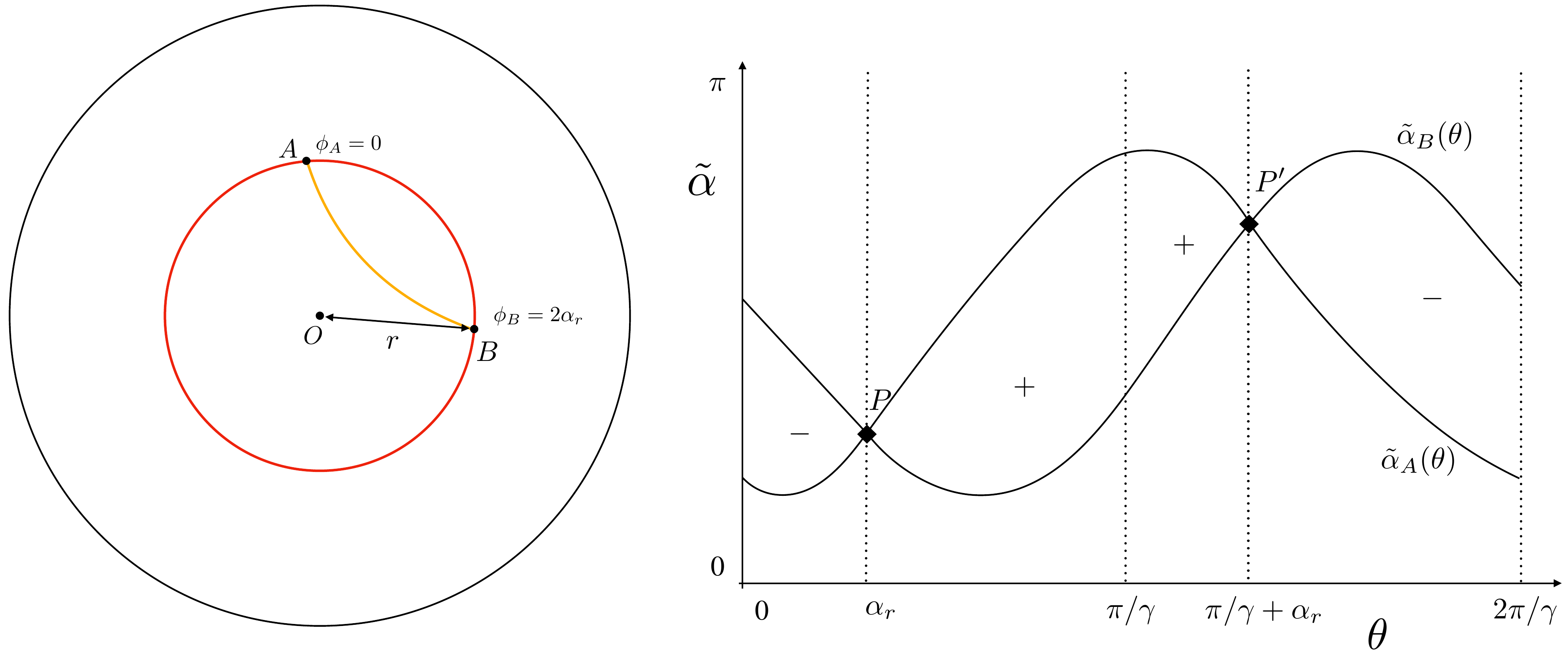}
\end{center}
\caption{  \emph{Left diagram}: geodesic arc (in yellow) on a constant time slice of conical AdS between two   points $A$ and $B$   on a circle (in red) of radius $r$ and with angular coordinates $\phi_A = 0$ and $\phi_B = 2 \alpha_r$. For simplicity, the conical parameter $\epsilon= 1- \gamma$ is taken to be much smaller than one, so the geodesics take a similar form as   those in pure AdS.   \emph{Right~diagram}: kinematic space $( \theta,\tilde \alpha)$ with   point curves $\tilde \alpha_A (\theta)$ and $\tilde \alpha_B (\theta)$ corresponding to the two bulk points in the left diagram. The geodesic distance between $A$ and $B$  is computed by an integral in kinematic space over the region $\Delta_{AB}$ enclosed by the two point curves. \label{fig:twofigappB2}} 
\end{figure}
An alternative way to derive the chord length between two bulk points $A$ and $B$ is from the integral of the Crofton form  over   kinematic space  \cite{Czech:2015qta}
\begin{equation}
	\frac{\lambda(A,B)}{4G}=\frac{1}{4}\int_{\Delta_{AB}}\omega(\theta, \tilde{\alpha}).
\end{equation}
In this appendix we parametrize kinematic space  with the pair $(\theta, \tilde{\alpha})$, where $\theta$ is the center of a spatial boundary region and $2\tilde \alpha$ is the angular size of the region (see figure \ref{fig:bulkgeodesics}). The Crofton form $\omega$ is the volume form on kinematic space, and  
the integration region $\Delta_{AB}$ is   the region in kinematic space between the two point curves  $\tilde \alpha_A(\theta)$ and $\tilde \alpha_B (\theta)$ associated to the bulk points $A$ and $B$, respectively. We recall that a point curve $\tilde \alpha_p (\theta)$   is formed by all geodesics on a constant time slice in AdS that intersect a bulk point $p$. The region $\Delta_{AB}$ corresponds  in the bulk to  the set of all geodesics which intersect the geodesic arc between the   points $A$ and~$B$ (see figure \ref{fig:twofigappB2}).
  
Using the expression for the Crofton form in terms of the entanglement entropy \eqref{eq:cf2}, and   $c=3L/ (2G)$, we can also write the chord length as  
\begin{equation}\label{eq:interm1}
	\lambda =-\frac{3L}{4c}\int_{0}^{2\pi}d\theta\,\partial_{\tilde\alpha} S(\tilde \alpha)\Big{|}_{\tilde \alpha_B(\theta)}^{\tilde\alpha_A(\theta)}.
\end{equation}
Here we    employed Stokes' theorem to remove   the integration over $\tilde{\alpha}$. 
We would like to compute the chord length between  two points $A$ and $B$  in conical  AdS space, which lie on a circle of radius $r$. The pure AdS case can be obtained by   setting $\gamma=1$ at every step. Suppose   the bulk angular coordinates are  given by  $\phi_A = 0$ and $\phi_B = 2 \alpha_r$. Then, it follows from~\eqref{eq:finalgeodesic} that the point curves of $A$ and $B$   satisfy the following equations 
\begin{equation}
\begin{aligned}
  \tilde  \alpha_A (\theta)  &=\frac{1}{\gamma}\arccos\left [ \frac{r}{\sqrt{r^2 + \gamma^2 L^2}} \cos (\gamma \theta) \right]  \\
     \tilde  \alpha_B (\theta)  &= \frac{1}{\gamma}\arccos\left[\frac{r}{\sqrt{r^2 + \gamma^2 L^2}} \cos [\gamma (\theta- 2 \alpha_r)] \right].
\end{aligned}
\end{equation}
The two point curves   intersect themselves, i.e. $\tilde \alpha_A (\theta)= \tilde \alpha_B (\theta)$, at   two points in kinematic space given by $P= \{\theta = \alpha_r \}$ and $P'=\{ \theta = \pi/\gamma + \alpha_r\}$. These two points  are depicted in the right diagram of figure~\ref{fig:twofigappB2} and they
  denote a unique geodesic in the bulk  passing through  both the   points $A$ and~$B$. The only  difference between $P$ and $P'$ is that the  orientation of the geodesic is opposite for these two points. 
    
The entanglement entropy of an excited state in a CFT dual to conical AdS is given by  
 $S^{\text{con}} (\tilde \alpha)= \frac{c}{3} \log \left [ 2L/(\mu \gamma) \sin (\gamma \tilde \alpha) \right]$, cf. equation \eqref{eq:conicalee}, where $\mu$ is a UV cutoff and $L$ is the radius of the cylinder, and hence its derivative is 
 \begin{equation}
 	\partial_{\tilde\alpha} S^{\text{con}}(\tilde \alpha)=\gamma\frac{c}{3}  \frac{\cos(\gamma \tilde \alpha)}{\sin (\gamma \tilde \alpha)} .
 \end{equation}
Plugging this into  \eqref{eq:interm1} yields that the contributions  from the two  point curves $\tilde \alpha_A$ and $\tilde \alpha_B$ are equal due to the circular symmetry of the setup. Further, to account for the orientation of the geodesics,   we  need to add appropriate signs for the  four different integration regions inside $\Delta_{AB}$ (see figure \ref{fig:twofigappB2} for our   sign convention).  The chord length thus consists of four different integrals  
\begin{equation}\label{eq:interm2}
		\lambda_{\text{con}} = - \frac{ \gamma L}{2 } \left[-\int_{0}^{\alpha_r} + \int_{\alpha_r}^{\pi/\gamma} + \int_{\pi/\gamma}^{\pi/\gamma + \alpha_r} - \int_{\pi/\gamma + \alpha_r}^{2\pi/\gamma}\right]d\theta \,  \frac{\cos(\gamma \tilde \alpha)}{\sin (\gamma \tilde \alpha)}\Bigg{|}_{\tilde \alpha_A(\theta)} .  
\end{equation} 
The transformation $\theta\to\theta+\pi/\gamma$ reverses the orientation of the geodesics, and it flips the sign of the integrand. This implies that the first and third integral, and the second and fourth integral, give the same result. The first and second integral are also the same, since the integral    vanishes for the  values $\theta =0$ and $\theta = \pi/ \gamma$.
The   four different integrals are therefore all equal, and hence after rewriting the integrand we find
\begin{equation}
	\lambda_{\text{con}} =  2 \gamma  L \int_0^{\alpha_r} d \theta \frac{x(\theta)}{\sqrt{1- x^2(\theta)}} \qquad \text{with} \qquad x(\theta) = \frac{r}{\sqrt{r^2 + \gamma^2 L^2}} \cos(\gamma \theta) .
\end{equation}
Finally,  this integral yields  the same expression  for the chord length    as \eqref{eq:conicalgeo} 
\begin{align}
	\lambda_{\text{con}}= 2 L \arctanh \left [ \frac{r/(\gamma L)\sin (\gamma \alpha_r)}{\sqrt{1+r^2/(\gamma L)^2 \sin^2 (\gamma \alpha_r)}}   \right] . 
\end{align}
 
\section{Conformal isometry of causal diamonds on the cylinder}\label{app:ckvs}

The conformal isometry of a causal diamond in Minkowski space is well studied in the literature \cite{Casini:2011kv,Faulkner:2013ica,Jacobson:2015hqa}.    However, Minkowski space corresponds to the conformal boundary of the Poincar\'e patch of AdS, whereas in the present paper we work in global AdS, whose conformal boundary is a (Lorentzian) cylinder. In this appendix we derive the conformal Killing vector generating the conformal isometry that  preserves a causal diamond on the two-dimensional cylinder in two distinct ways: from the generators of the conformal group and from the boundary limit of the boost Killing vector of Rindler-AdS$_3$.

\subsection{From the conformal group}\label{app:ckvbdry}

On the complex plane the  generators of the global conformal group are $\partial_z, \partial_{\bar z}$  which generate translations, $z \partial_z, \bar z  \partial_{\bar z}$  which generate dilatations and rotations, and $z^2 \partial_z, \bar z^2 \partial_{\bar z}$ which generate special conformal transformations. These generators can be mapped to the generators of the conformal group on the cylinder by the conformal transformation $\omega = i \log z$, where $\omega=\th+i \tau_E$ parametrizes the (Euclidean) cylinder.  The line element transforms as $dz  d \bar z = e^{i (\bar \omega - \omega)} d \omega d \bar \omega = e^{2\tau_E}(d\tau_E^2 + d \theta^2) $. Since a conformal generator remains a generator of the conformal group after a Weyl rescaling of the metric,   we can safely ignore the  conformal factor. The basis of conformal generators on the cylinder  is thus given by  
\begin{equation} \label{eq:basis1}
 \{e^{-i\omega}\partial_{\omega},\,\,\partial_{\omega},\,\,e^{i\omega}\partial_{\omega}\} \cup    \left \{  \omega \to \bar \omega\right\}.
\end{equation}
In Lorentzian signature the complex coordinate is $\omega = \theta -\tau$ and its complex conjugate is $\bar \omega = \theta + \tau$, where $\tau=-i \tau_E$ is the Lorentzian time. Together $\omega$ and $\bar \omega$ form a null coordinate system on the cylinder. In the following, however, we take the null coordinates to be the     retarded and advanced times $u = \tau-\theta $ and $v = \tau+ \theta$, since they are both increasing towards the future.
We write the basis of generators now in terms of  trigonometric functions of  these null coordinates
\begin{equation} \label{eq:basis2}
 \{ \partial_u,  \sin u \, \partial_u,   \cos u \,  \partial_u    \} \cup \left \{  u \to v\right\} . 
\end{equation}
The Killing vector fields $ \partial_u + \partial_v  $ and  $ - \partial_u+ \partial_v  $, respectively,  generate time translations and rotations on the cylinder, and the other four basis vectors are conformal Killing vectors which do not generate isometries of the metric $- du dv$. 

We put the origin of the null coordinate system, $u=v=0$, at the center of the causal diamond. The lines $u = \pm \alpha$ and $v = \pm \alpha$ are the null boundaries of the diamond. 
Since the diamond has a reflection symmetry across the $u=v$ line (the $t$-axis), the conformal isometry that preserves the diamond must be invariant under the exchange of $u$ and $v$.  
The conformal  Killing vector which generates this conformal isometry  therefore takes the general form
\begin{equation}
\begin{aligned}
&\xi =A(u) \partial_u +A(v) \partial_v , \\
  &\text{with}  \quad A(u) =  a + b \sin u + c \cos u.
\end{aligned}
\end{equation}  
To remain inside the diamond  the flow of $\xi$ must leave    the vertices $u=v= \pm \alpha$ and the edge $v=-u = \alpha$ of the diamond  fixed. This requirement yields $A(\pm \alpha)=0$, which determines the   function  up to a normalization: $A(u) = c \left ( \cos u - \cos \alpha  \right). $
The normalization is fixed by demanding that the surface gravity of $\xi$ is equal to  one, $\kappa = -   A'(\alpha)=   c \sin \alpha =1$,  at the future null boundary of the diamond. 
Thus, the conformal Killing vector whose flow preserves a causal diamond on the cylinder, and which has unit surface gravity, is  in terms of null coordinates
\begin{equation}
\begin{aligned} \label{eq:ckvcylinder1}
\xi  = \frac{1}{  \sin \alpha} \Big  [  \left ( \cos u - \cos \alpha  \right) \partial_u +  \left ( \cos v - \cos \alpha \right) \partial_v \Big ] . \\
\end{aligned}
\end{equation}
In terms of the $\tau$ and $\theta$ coordinates on the Lorentzian cylinder, $\xi$ becomes
\begin{equation} \label{eq:ckvcylinder2}
\xi =  \frac{1}{\sin \alpha} \Big [ \left (  \cos \tau \cos \theta - \cos \alpha \right)  \, \partial_\tau   - \sin \tau \sin \theta  \, \partial_\theta \Big].
\end{equation}
As a limiting case, note that for small diamonds, i.e.  $u, v, \alpha \ll 1$, $\xi$ reduces to the expression in flat space \cite{Faulkner:2013ica,Jacobson:2015hqa}
\begin{equation}
\begin{aligned}
\xi &= \frac{1}{2 \alpha}\left [   \left ( \alpha^2 - u^2  \right) \partial_u +  \left (  \alpha^2 - v^2  \right)   \partial_v \right]  . 
\end{aligned}
\end{equation} 
An illustration of $\xi$ is given in figure \ref{fig:xichi}, where we have also indicated the boost Killing   vector of the associated Rindler wedge in the bulk. We will discuss the latter in the following section. 
 
\subsection{From the boundary limit of the boost Killing vector}\label{app:kill}

\begin{figure}
\begin{center}
\includegraphics[width=.4\textwidth]{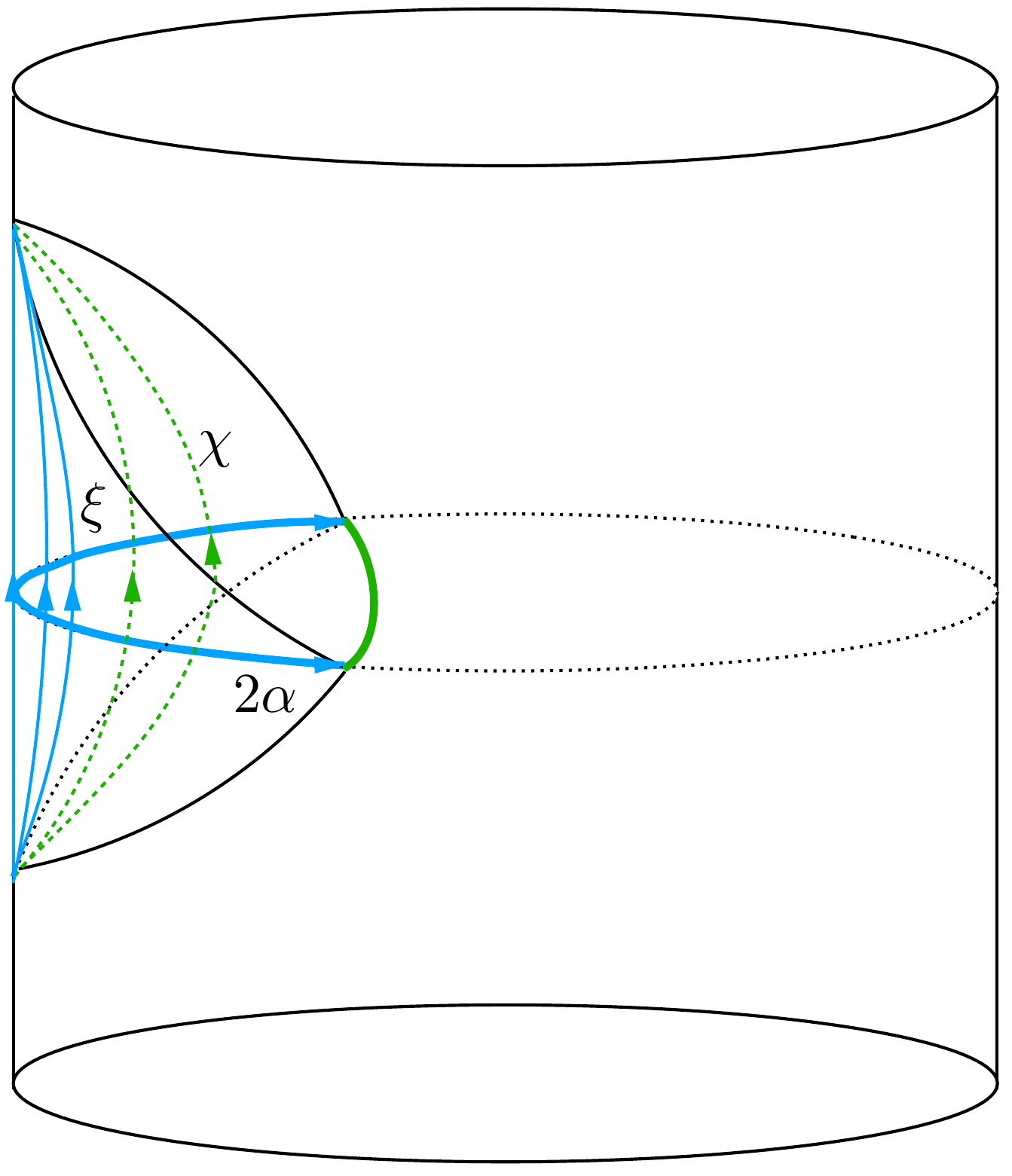}
\end{center}
\caption{The Rindler wedge of pure AdS$_3$  with associated boundary causal diamond (in black). The bifurcation surface (in green) of the AdS-Rindler horizon is the Ryu-Takayanagi surface associated to the entangling  region of angular size $2\alpha$ (in blue). The boost Killing vector $\chi$ (with flow in green) of   AdS-Rindler    asymptotes to the conformal Killing vector $\xi$, which generates a flow (in blue) inside the boundary causal diamond.\label{fig:xichi}}
\end{figure}

The conformal isometry that preserves a causal diamond on the Lorentzian cylinder can also be obtained from the boundary limit of a proper isometry of AdS space. This is because the conformal Killing vector $\xi$  of a causal diamond in flat space 
  extends to the  boost Killing vector $\chi$ of the  Rindler wedge of AdS space (see figure \ref{fig:xichi}).  The boost Killing vector generates proper  time translations  for uniformly accelerating (Rindler) observers in AdS with $a>L^{-1}$. 
  The vector field $\chi$ takes quite a  simple form  in Poincar\'{e} coordinates~\cite{Faulkner:2013ica}, but   is slightly more complicated in global AdS coordinates, as we will see below. We need the expression for $\chi$ in global coordinates, since its boundary limit yields the conformal Killing vector of a diamond on the     cylinder, which is the conformal boundary of global AdS.   
  
It is straightforward to derive the Killing vectors of AdS from the embedding space (see appendix \ref{app:embed}). The isometry group of $\text{AdS}_3$, $SO(2,2)$,   is generated by  six linearly independent Killing vectors in the embedding space~$\mathbb R^{2,2}$: two generators of rotations and four generators of boosts. The boost Killing vector  of the AdS-Rindler wedge    corresponds to the   Killing vector associated with a boost in the  $(T^2,X^1)$ plane of the embedding space. This is because the embedding coordinates $T^2$ and $X^1$ for AdS-Rindler space, in equation~\eqref{eq:embads4}, parametrize a Rindler observer (or hyperbola) in embedding space, with proper time $\sigma = L\, \text{arctanh} (T^2/X^1)$ and proper distance $\varrho = \sqrt{(X^1)^2 - (T^2)^2}$.   Rindler observers   in   embedding space are therefore in one-to-one correspondence with Rindler observers in AdS \cite{Deser:1997ri}. 
Indeed, the    Killing vector which generates a boost in the $(T^2,X^1)$ plane of embedding space,  $B=X^1 \partial_{T^2} +T^2 \partial_{X^1}$,  becomes the    generator of $\sigma$-time translations in AdS-Rindler space, i.e. $B=L \partial_{\sigma}$.   
  
Next, we compute the boost Killing vector in the global coordinates \eqref{eq:embads2} for AdS 
\begin{equation} \label{eq:adsboost1}
  B =    \!  \frac{r L}{\sqrt{L^2+r^2}} \cos  (t/L)\cos \phi     \partial_t +    \sqrt{L^2+ r^2} \sin (t/L) \cos \phi   \, \partial_r   -  \frac{\sqrt{L^2+r^2}}{r}  \sin  (t/L) \sin \phi   \, \partial_\phi  .
\end{equation}
The boost Killing vector becomes null on the AdS-Rindler horizon, which   in global coordinates is described by $\{\cos (t/L \pm \pi/2) = \frac{r}{\sqrt{r^2 + L^2}} \cos \phi \}$, and vanishes at the boundary vertices $\{ t= \pm \pi L/2, r =\infty, \phi =0 \}$ and at the straight line     $\{t=0, \phi = \pi/2, 3\pi/2 \}$. Note that the straight line, which is  the bifurcation surface of the horizon,  cuts the $t=0$ time slice of AdS in two wedges of equal size, since each   wedge subtends an angle $\pi$. In other words, the left and right AdS-Rindler wedges    both cover half of the boundary cylinder at $t=0$. This is also manifest from the boundary limit $r \to \infty$ of the boost Killing vector, which     equals   \eqref{eq:ckvcylinder2} only for $\alpha=\pi/2$. Hence, we have not yet found the extension into the bulk of the most general boundary conformal Killing vector $\xi$, which should hold for any~$\alpha$.  
  
Fortunately, we can move the bifurcation surface of the horizon  by an isometry of AdS to a new position that intersects the boundary at   $\phi=\alpha$ and $\phi = 2\pi - \alpha$. The new bifurcation surface subtends an angle $2\alpha$ at the boundary, instead of $\pi$ as in the previous case. A simple isometry that relates different bifurcation surfaces (and hence   different AdS-Rindler wedges) is a boost  in the $(T^1,X^1)$ plane of embedding space~\cite{Casini:2011kv}. This boost   transforms the embedding coordinates as
\begin{equation}
\begin{aligned}
\label{eq:explicitboost}
  (T^1)' &= \cosh \beta \, T^1 - \sinh \beta \, X^1, \\
  (X^1)' &= \cosh \beta \, X^1 - \sinh \beta \, T^1, 
\end{aligned}
\end{equation}
where $\beta$ is an arbitrary rapidity parameter which ranges from $-\infty$ to $\infty$. The   boost Killing vector of the transformed AdS-Rindler wedge is now given by the  boost generator   in the $(T^2,(X^1)')$ plane 
\begin{equation} 
\label{eq:boostkillingnew}
  \chi = (X^1)' \partial_{T^2} + T^2 \partial_{(X^1)'} . 
\end{equation}
By substituting $T^1$ and $X^1$ in  \eqref{eq:embads4} with $(T^1)'$ and $(X^1)'$, respectively,  
the induced metric for the new AdS-Rindler wedge is identical to the one   given by equation \eqref{eq:adsrindlermetric}, and  the boost Killing vector remains equal to 
\begin{equation} \label{eq:rindlerboostadskilling}
\chi = L \partial_\sigma
\end{equation}
in AdS-Rindler coordinates.  Relative to global coordinates, however, the AdS-Rindler wedge has been displaced by the boost \eqref{eq:explicitboost}, since we use the same   embedding coordinates  for global AdS as before performing the boost.
Hence in global coordinates the new boost Killing vector \eqref{eq:boostkillingnew}   takes a different form compared to the one in equation \eqref{eq:adsboost1}.
In order to derive this form, we first express the boost Killing vector  in terms of the unprimed embedding coordinates   
 \begin{align}
  \chi   &= \cosh \beta \,  B  - \sinh \beta \, H, \quad \text{with} \label{eq:boostkillingads2} \\
 B =  X^1 \partial_{T^2} +&T^2 \partial_{X^1}   \quad  \text{and} \quad H   = T^1 \partial_{ T^2} - T^2 \partial_{T^1}.  
  \end{align}  
Here $H$ generates rotations in the $(T^1,T^2)$ plane. 
We would like to express $\chi$, however, in terms of the  boundary opening angle~$\alpha$ instead of   the rapidity $\beta$. Their relation   can be derived as follows. By inverting the boost \eqref{eq:explicitboost} and plugging in  the embedding coordinates $(T^1)' = \sqrt{\varrho^2 + L^2} \cosh (u/L)$ and $(X^1)'= \varrho \cosh (\sigma /L)$, we find  
 \begin{equation}
 \begin{aligned}
T^1 &= \cosh \beta \, \sqrt{\varrho^2 + L^2} \cosh(u/L) + \sinh \beta \, \varrho \cosh (\sigma/L) , \\
X^1&=  \cosh \beta \, \varrho \cosh (\sigma/L) + \sinh \beta \, \sqrt{\varrho^2 +L^2} \cosh (u/L)  .
\end{aligned}
\end{equation}
The   transformation between the AdS-Rindler and the global coordinate system can be found by inserting the  embedding coordinates $T^1$ and $X^1$ for global AdS given by \eqref{eq:embads2} into these equations, and
  identifying the other two (not boosted) embedding coordinates $T^2$~and~$X^2$ for global AdS with those for AdS-Rindler space given by \eqref{eq:embads4}. The resulting  coordinate transformation is\footnote{For $\beta=0$ this is consistent with the coordinate transformation in equation (2.6) of \cite{Parikh:2012kg}.}
  \begin{align} \label{eq:coordrindlerglobal}
\tan (t/L) & = \frac{\varrho \sinh (\sigma /L)}{   \cosh \beta \, \sqrt{\varrho^2 + L^2} \cosh(u/L) + \sinh \beta \, \varrho \cosh (\sigma/L)   } , \nonumber \\
\tan \phi &= \frac{\sqrt{\varrho^2 + L^2} \sinh (u/L) }{   \cosh \beta \, \varrho \cosh (\sigma/L) + \sinh \beta \, \sqrt{\varrho^2 +L^2} \cosh (u/L)  } ,\\
r^2 &= \left (   \cosh \beta \, \varrho \cosh (\sigma/L) + \sinh \beta \, \sqrt{\varrho^2 +L^2} \cosh (u/L) \right)^2 + \left ( \varrho^2 + L^2 \right) \sinh^2(u/L) . \nonumber
\end{align}
The time slice $\sigma=0$ corresponds to $t=0$ in global coordinates. The bifurcation surface of the horizon lies inside that time slice and  intersects the asymptotic boundary at $u/L=\infty$ in AdS-Rindler coordinates, and at $\phi = \alpha$ and $\phi= 2\pi - \alpha$ in global coordinates. The relation between $\beta$ and $\alpha$ thus follows from evaluating the second equation  at   $\sigma =0$ and taking the limit  $\varrho/L \to \infty$ and, subsequently, $u/L\to \infty$, i.e. 
 \begin{equation} \label{eq:relationab}
\cosh \beta = \frac{1}{\sin \alpha} \qquad \text{and}\ \qquad \sinh \beta = \frac{1}{ \tan \alpha} \, ,
 \end{equation}
 where   $\beta$ now ranges from $0$ to $\infty$, and we still have $0\le \alpha \le \pi/2$.  
 As a function of $\alpha$ the boost Killing vector \eqref{eq:boostkillingads2} is hence given by
 \begin{equation}
 	\chi = \frac{1}{\sin \alpha} \big ( B - \cos \alpha\,  H \big).
 \end{equation}
 Note that for $\alpha=\pi/2$ (or $\beta=0$)   the boost Killing vector reduces to $\chi =B$.
 In  global AdS coordinates   the Killing vector  $H$  is simply the generator of time translations $
H= L \partial_t 
$, whereas  $B$ is  given by~\eqref{eq:adsboost1}. Therefore,  in global coordinates  we find
\begin{equation}
\begin{aligned}  \label{eq:adsboost7}
\chi = & \frac{1}{\sin \alpha}\! \Bigg [ \! \left (\frac{r L}{\sqrt{L^2+r^2}} \cos  (t/L)\cos \phi    - L \cos \alpha \right)  \partial_t +    \sqrt{L^2+ r^2} \sin (t/L) \cos \phi   \, \partial_r    \\
&\qquad  \,\, -  \frac{\sqrt{L^2+r^2}}{r}  \sin  (t/L) \sin \phi   \, \partial_\phi \Bigg]   .
\end{aligned}
\end{equation} 
One can readily verify that the vertices $\{t = \pm \alpha L, r=\infty, \phi=0 \}$ of the boundary causal diamond and  the extremal surface $\{ t= 0,  \cos \alpha= \frac{r}{\sqrt{r^2 + L^2}}  \cos \phi   \}$ are fixed points of the flow of $\chi$. The extremal surface is the bifurcation surface of  the horizon.  Further, $\chi$ becomes null on the past and future  Killing horizon, which are given by  $\{ \cos (t/L \pm \alpha)= \frac{r}{\sqrt{r^2 + L^2}}   \cos \phi \}$ in global coordinates or   $\varrho=0$ in AdS-Rindler coordinates. The normalization  of $\chi$ is  such that    the surface gravity is unity, $\kappa=1$, at the future  horizon.   Furthermore, let us stress that under the mapping \eqref{eq:coordrindlerglobal} the boost Killing field above  turns into $L \partial_\sigma.$

In terms of the dimensionless sausage  coordinates \eqref{eq:poincaredisk} the boost generator reads 
\begin{align}  \label{eq:boostkill2}
 \chi 
 = \frac{1}{\sin \alpha}\! \left [ \! \left (  \frac{2z}{1 + z^2}    \cos \tau \cos \phi    - \cos \alpha \right)  \partial_\tau +  \frac{1-z^2}{2} \sin \tau \cos \phi   \, \partial_z  - \frac{1+z^2}{2 z}   \sin \tau\sin \phi   \, \partial_\phi \right]   ,
\end{align}
and in terms of the spherical coordinates \eqref{eq:adsspherical} the boost Killing vector is simply\footnote{This expression agrees with the boost Killing vector in equation (2.32) of the recent paper \cite{Rosso:2020zkk}.} 
\begin{equation}
	\chi = \frac{1}{\sin \alpha}\! \left [ \Big ( \cos\hat\tau \cos\hat \phi - \cos \alpha \Big) \partial_{\hat\tau}  - \sin \hat \tau \sin \hat \phi \partial_{\hat \phi} \right]   .
\end{equation}
Surprisingly, this is independent of the bulk angular coordinate $\psi$, and hence   the expression does not change in the asymptotic limit $\psi \to 0$ or $\pi$. The   boundary limit $r\to \infty$ and $z\to 1$ of   the other   expressions for $\chi$, respectively \eqref{eq:adsboost7} and \eqref{eq:boostkill2}, yield the same result for the associated conformal Killing vector on the boundary cylinder
\be\label{eq:ckvbdryfrombulk}
\xi =\lim_{\text{bndy}}  \chi =\frac{1}{\sin \alpha} \Big [ \left (  \cos \tau \cos \theta - \cos \alpha \right)    \partial_\tau   - \sin \tau\sin \theta  \, \partial_\theta \Big].
\ee
As expected, this matches with   expression \eqref{eq:ckvcylinder2} for the conformal Killing  vector   which generates the conformal isometry of a causal diamond on the boundary cylinder. Note that we expressed $\xi$ in terms of the dimensionless boundary coordinates   $\tau$ and  $\theta$. Although we have focussed on   a three-dimensional bulk spacetime in this appendix, the
 equations above for the boost Killing vector of AdS-Rindler space (and its corresponding boundary limit) are also valid in higher dimensions. This is because the expression \eqref{eq:boostkillingads2} for $\chi$ in embedding coordinates remains the same. 

Finally, for completeness,     let us  check that the embedding expression  reproduces the    boost Killing vector in Poincar\'{e} coordinates \eqref{eq:embads5}. The rotation and boost generators are  
\begin{equation}
\begin{aligned}  \label{eq:poincaregenerators}
B &=   \frac{1}{2L} \left [  (   L^2  -        \mathrm{t}^2 - \mathrm{x}^2 - \mathrm{z}^2)\partial_\mathrm{t} - 2\mathrm{t}(\mathrm{x} \partial_\mathrm{x} + \mathrm{z} \partial_\mathrm{z})   \right] ,\\
 H &= \frac{1}{2L} \left [  (  L^2 +       \mathrm{t}^2 + \mathrm{x}^2 + \mathrm{z}^2)\partial_\mathrm{t} + 2\mathrm{t}(\mathrm{x} \partial_\mathrm{x} + \mathrm{z} \partial_\mathrm{z})   \right]  .
\end{aligned}
\end{equation}
The rapidity $\beta$ is in this case  related to the radius $R$ of a sphere at $\mathrm{t}=0$ in the flat boundary space via  $R=e^{-\beta} L$~\cite{Casini:2011kv}.\footnote{Comparing this  to \eqref{eq:relationab} we see that the radius of the sphere in flat space is given in terms of  the   opening angle on the cylinder by $R = L \tan (\alpha/2).$ } Combining this relation with \eqref{eq:boostkillingads2} and \eqref{eq:poincaregenerators}, we recover the known expression for  the boost Killing vector in Poincar\'e coordinates  \cite{Faulkner:2013ica} 
\begin{equation}
\begin{aligned} \label{eq:poincareboost}
\chi &= \frac{L}{2R} \left [ \left ( 1+ (R/L)^2  \right)B -  \left ( 1-(R/L)^2   \right) H   \right] \\
  &= \frac{1}{2R} \left [  \left (  R^2 -\mathrm{ t}^2 -\mathrm{x}^2 - \mathrm{z}^2  \right) \partial_\mathrm{t} - 2 \mathrm{t} \mathrm{x} \partial_\mathrm{x} - 2 \mathrm{t} \mathrm{ z} \partial_\mathrm{z}  \right] . 
\end{aligned}
\end{equation}
Note that the   term involving $H$ in the  first equation    is only nonzero if the radius of the   sphere   is   not equal to the AdS radius (see also appendix D in  \cite{Jacobson:2018ahi}). The   transformation between Poincar\'{e} coordinates and AdS-Rindler coordinates, which maps the boost Killing vector \eqref{eq:poincareboost} to the time translation generator \eqref{eq:rindlerboostadskilling}, can be obtained in a similar fashion as the transformation \eqref{eq:coordrindlerglobal}  for global coordinates above (see for instance   equation (80) in  \cite{Espindola:2018ozt}).

\addtocontents{toc}{\protect\enlargethispage{2\baselineskip}}

\section{Variation of coupling constants  in the first law of causal diamonds}
\label{app:couplings}
 
In this section we compute the contributions of  the variation   of gravitational coupling constants in  the   first law of causal diamonds in maximally symmetric spacetimes. We~consider variations of both the cosmological constant and Newton's constant, and    prove that terms proportional to the variation  of Newton's constant cancel out in the first law.    We employ the covariant phase space method \cite{Wald:1993nt,Iyer:1994ys}, which has been extended in \cite{Urano:2009xn,Caceres:2016xjz} to include variations of couplings, and we   follow the   notation of \cite{Jacobson:2018ahi}. 

Suppose  $L(\phi, \alpha_i)= \mathcal L(\phi, \alpha_i)  \epsilon$ is a diffeomorphism invariant Lagrangian $d$-form that depends on the dynamical fields $\phi$ and coupling constants $\alpha_i$. If one allows for variations of coupling constants $\alpha_i$, then the on-shell fundamental variational identity in the covariant phase space formalism becomes\footnote{See appendix A in \cite{Caceres:2016xjz} and section 2 in \cite{Urano:2009xn} for a derivation of this identity. }
 \begin{equation}
\delta H_\zeta =\oint_{\partial D}  \delta Q_\zeta   
 +\int_D \zeta \cdot E^{\alpha_i} \delta \alpha_i.
\end{equation}
Here    $\zeta$ is   the conformal Killing vector of a maximally symmetric causal diamond, $  H_\zeta$ is the Hamiltonian generating evolution along the flow of $\zeta$, $Q_\zeta$ is the associated Noether charge $(d-2)$-form,   and the $d$-form $E^{\alpha_i}=(\partial \mathcal L / \partial \alpha_i)\epsilon$ is the derivative of the Lagrangian   with respect to the coupling~$\alpha_i$. 

Assuming minimal coupling, the   Lagrangian uniquely splits into a gravitational   and a matter part: $L = L^\text{grav} + L^{\text{mat}}$. We consider the gravitational Lagrangian for general relativity plus a cosmological constant  
\begin{equation}
L^{\text{grav}} = \frac{R- 2 \Lambda}{16 \pi G } \epsilon .
\end{equation}
We  only take   variations of the gravitational  coupling constants  $\alpha_i=\{\Lambda, G\}$ into account in the first law, and not of the matter couplings. The derivatives of the gravitational Lagrangian with respect to $\Lambda$ and $G$ are
\begin{equation}
E^\Lambda = - \frac{\epsilon}{8 \pi G}  \qquad \text{and} \qquad E^G = -  \frac{R- 2 \Lambda}{16 \pi G^2 } \epsilon =  - \frac{\Lambda}{(d-2)4 \pi G^2} \epsilon,
\end{equation}
where we evaluated the Ricci scalar on the maximally symmetric background in the last equality. For Einstein gravity the fundamental identity thus takes the form
\begin{equation}
\delta H_\zeta = \oint_{\partial D} \delta Q_\zeta - \frac{1}{8 \pi G}  \left ( \delta \Lambda  + \frac{2 \Lambda }{d-2 } \frac{\delta G}{G } \right) \int_D \zeta \cdot \epsilon .
\end{equation}
The integral of $\zeta \cdot \epsilon$ over the disk can be identified with the thermodynamic volume $V_\zeta$~\eqref{eq:thermovolume1}. Further, the Noether charge variation is given by 
\begin{equation} \label{eq:noethercharge}
	\oint_{\partial D} \delta Q_\zeta = - \frac{\kappa}{8 \pi  } \delta \left ( \frac{A}{G} \right) ,
\end{equation}
where Newton's constant is included in the variation. The Hamiltonian variation splits into a gravitational and matter part, $\delta H_\zeta = \delta H_\zeta^{\text{grav}}+ \delta H_\zeta^{\text{mat}}$, with the gravitational part   defined as the symplectic form evaluated on the Lie derivative of the metric along $\zeta$
\begin{equation}
	\delta H_\zeta^{\text{grav}} = \int_D \omega (g, \delta g, \mathcal L_\zeta g) = \int_D   \left[\delta \theta (g, \mathcal L_\zeta g) - \mathcal L_\zeta \theta (g, \delta g) \right].
\end{equation} 
Here $\theta$ is the so-called symplectic potential $(d-1)$-form.
The Hamiltonian variation contains   contributions from both the variation of the metric and the variation of Newton's constant,  $\delta H_\zeta^{\text{grav}} = \delta_g H_\zeta^{\text{grav}} + \delta_G H_\zeta^{\text{grav}}$, which are given by
\begin{equation} \label{eq:hambothvar}
	 \delta_g H_\zeta^{\text{grav}} = -\frac{\kappa k}{8 \pi G} \delta V \quad \text{and} \quad  \delta_G H_\zeta^{\text{grav}}= \int_D \delta_G \theta (g, \mathcal L_\zeta g) =- \frac{d-1}{d-2}\,\frac{\kappa k }{8 \pi  }\, V \delta\!\left ( \frac{1}{G} \right). 
\end{equation}
This follows respectively from equations (3.35) and  (3.9) in Ref. \cite{Jacobson:2018ahi}. Therefore, plugging \eqref{eq:noethercharge} and \eqref{eq:hambothvar} into the fundamental identity, we find a first law which includes variations of both $\Lambda$ and $G$  
\begin{equation}
\delta H_\zeta^{\text{mat}} - \frac{\kappa k }{8 \pi G} \left (  \delta V - \frac{d-1}{d-2}\, V \,\frac{\delta G}{G }   \right) =- \frac{\kappa}{8 \pi G }\, \delta A + \frac{\kappa A}{8 \pi G}\, \frac{\delta G}{G} - \frac{V_\zeta}{8 \pi G}  \left ( \delta \Lambda  + \frac{2 \Lambda }{d-2 } \frac{\delta G}{G } \right) .
\end{equation}
However, we can deduce from the Smarr formula \eqref{eq:smarrgeneral},  
\begin{equation}
	 (d-1) \kappa k V =(d-2) \kappa A - 2 V_\zeta \Lambda,
\end{equation}
  that  the terms involving the variation  of  Newton's constant cancel each other. Thus,   the final form of the first law of causal diamonds is
\begin{equation}
\delta H_\zeta^{\text{mat}} - \frac{\kappa k }{8 \pi G}   \delta V =- \frac{\kappa}{8 \pi G } \delta A - \frac{V_\zeta}{8 \pi G}     \delta \Lambda   ,
\end{equation}
which agrees with the result   in \cite{Jacobson:2018ahi}.

\bibliography{Notes-new}

\end{document}